\documentclass[superscriptaddress,reprint,amsmath,amssymb,aps,prapplied,longbibliography]{revtex4-1}

\usepackage{amsmath,gensymb,textcomp,bm,dcolumn,eurosym,array,tabu,multirow,nicefrac,color,subfigure,graphicx,upgreek}
\usepackage[colorlinks, linkcolor=blue, citecolor=blue, urlcolor=blue, breaklinks=true]{hyperref}

\usepackage{mathpazo,times} 

\newcommand{\ket}[1]{\lvert #1 \rangle}
\newcommand{\braket}[2]{\langle #1 \vert #2 \rangle}

\usepackage[subsectionbib]{bibunits}
\usepackage{graphicx}
\usepackage{dcolumn}
\usepackage{bm,MnSymbol}
\usepackage{units}
\usepackage{psfrag}
\usepackage{float}
\usepackage{color}

\begin{document}
\title{Quantum Wiener-Khinchin theorem for spectral-domain optical coherence tomography}

\author{Yuanyuan Chen}
\email{chenyy@xmu.edu.cn}
\affiliation{Department of Physics and Collaborative Innovation Center for Optoelectronic Semiconductors and Efficient Devices, Xiamen University, Xiamen 361005, China}
\author{Lixiang Chen}
\email{chenlx@xmu.edu.cn}
\affiliation{Department of Physics and Collaborative Innovation Center for Optoelectronic Semiconductors and Efficient Devices, Xiamen University, Xiamen 361005, China}

\begin{abstract}
Wiener-Khinchin theorem, the fact that the autocorrelation function of a time process has a spectral decomposition given by its power spectrum intensity, can be used in many disciplines. However, the applications based on a quantum counterpart of Wiener-Khinchin theorem that provides a translation between time-energy degrees of freedom of biphoton wavefunction still remains relatively unexplored. Here, we use a quantum Wiener-Khinchin theorem (QWKT) to state that two-photon joint spectral intensity and the cross-correlation of two-photon temporal signal can be connected by making a Fourier transform. The mathematically-defined QWKT is experimentally demonstrated in frequency-entangled two-photon Hong-Ou-Mandel (HOM) interference with the assistance of spectrally-resolved detection. We apply this method to spectral-domain quantum optical coherence tomography that detects thickness-induced optical delays in a transparent sample, and show that our method suffices to achieve great advantages in measurement precision within a wide dynamic range and capturing time over the conventional HOM interferometric schemes. These results may significantly facilitate the use of QWKT for quantum information processing and quantum interferometric spectroscopy.
\end{abstract}
\maketitle

\section{Introduction}
Understanding how the strength of a signal is distributed in the frequency and time domains is essential for a variety of practical applications, as well as for studying fundamental physics. The Wiener-Khinchin theorem (WKT) relates between the power spectrum intensity and the autocorrelation of its time process \cite{wiener1930generalized,khinchin1934korrelationstheorie,kubo1985statistical}, which has been widely used in many disciplines, including statistics, signal analysis, applied mathematics, and optics. For example, based on optical version of WKT, the interferometric spectroscopy that can extract the longitudinal structural information by making a Fourier transform on its spectral-domain Mach-Zehnder or Michelson interference pattern has been well established \cite{davis2013fourier,griffiths1983fourier}. This Fourier transform spectroscopy at the infrared wavelength has been commercially used in clinical imaging, polymer testing, and pharmaceutical analysis \cite{yu2018application,adhi2013optical}.

In recent years, advances in quantum technologies have made possible an exciting breadth of applications in field such as quantum computation, secure communication and high-resolution imaging \cite{2019Quantum,lo1999Unconditional,manuel2018super}. As a necessary prerequisite towards quantum applications, quantum entanglement involves the correlations between paired photons such that the wavefunction in time domain refers to two-photon relative time signal, and in frequency domain refers to two-photon joint spectral intensity. This naturally gives rise to several questions: First, while WKT states the link between time process and spectral distribution, is it possible to establish a mathematical connection between two-photon relative time signal and two-photon joint spectral intensity? Second, how to translate this mathematically-defined theorem into experimental implementation? Third, are there any practical applications based on the connection between two-photon time and spectral signals?

Thus, a complete quantum counterpart of WKT that defines the connection between temporal and spectral distributions of biphton wavefunctions has long been hailed for inspiring new, even more efficient quantum tools, as required for meet the ever-increasing performance requirements of quantum optics experimentation and practical quantum information processing applications. To tackle these issues, a well-designed extended Wiener-Khinchin theorem (e-WKT) has been proposed to state that the difference-frequency distribution of the biphoton wavefunctions can be extracted by applying a Fourier transform on the time-domain HOM interference patterns \cite{jin2018extended}. However, the reverse process that extracts the relative temporal distribution of the biphoton wavefunctions from spectral-domain HOM interference patterns still remains relatively unexplored.

Here, we express the two-photon joint spectral intensity in terms of the cross-correlation functions of two-photon relative time delay by using a Fourier transform. Since HOM interferometry determines the time-time correlation of entangled photons \cite{hong1987measurement}, we exploit it to perform the Fourier transform on two-photon temporal signals, and observe the two-photon joint spectral intensity at the output of HOM interferometer with the assistance of spectrally-resolved coincidence detection. Thus, an elaborate design of frequency-entangled two-photon HOM interference is sufficient for the experimental demonstration of QWKT. We apply the QWKT to spectral-domain quantum optical coherence tomography (SD-QOCT) \cite{nasr2003demonstration,pablo2020spectrally}, wherein the thickness-induced optical delays in a transparent sample can be extracted from two-photon joint spectral intensity. This method is a crucial requisite towards the tomography for those photon-sensitive biological and chemical samples. Building on the measurement and estimation strategy by analyzing its Fisher information \cite{lyons2018attosecond}, we explore the sensitivity limit as a function of single-photon bandwidth and target delays. Compared to the conventional HOM interferometry, our SD-QOCT promises enhancements in precision and sensitivity within a wide dynamic range. Aside from this, this SD-QOCT provides an improvement in shortening the capturing time as two-photon joint spectral intensity can be obtained from single measurement and without the strict requirement of temporal scanning.

\begin{figure*}[!t]
\centering
\includegraphics[width=\linewidth]{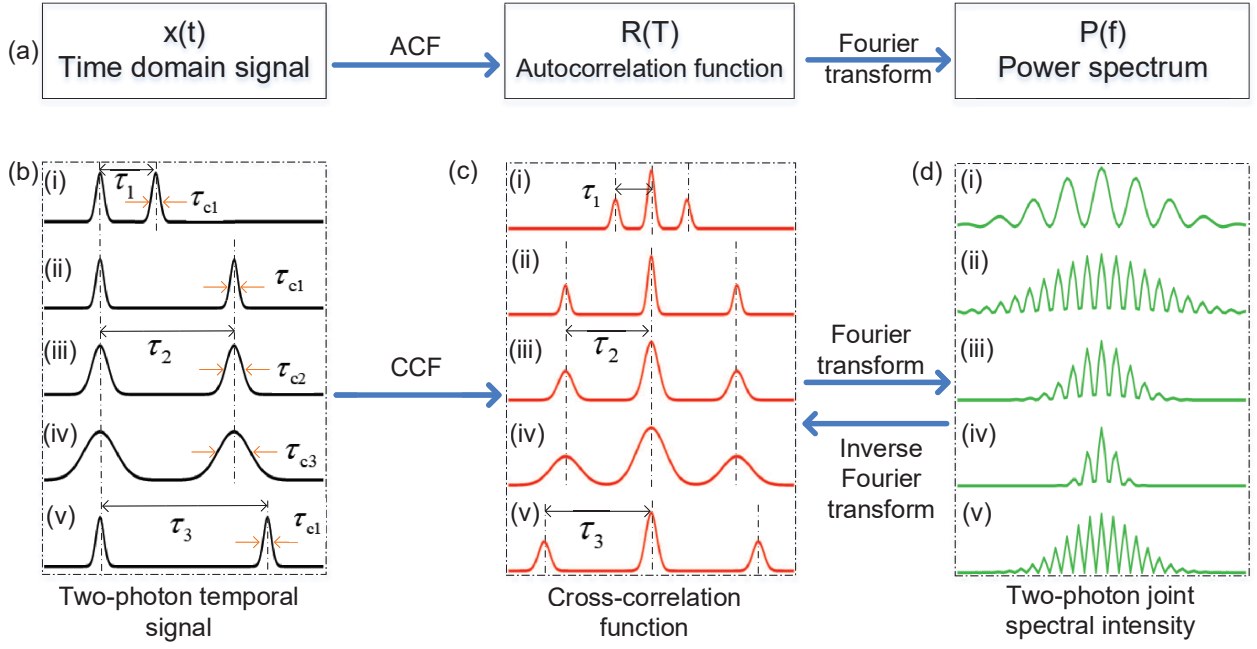}
\caption{(a) Demonstration of a classical WKT. In a QWKT, (b) the input is defined as two-photon relative delay as shown in \eqref{eq: idler time}. By making a Fourier transform on (c) its cross-correlation function, (d) the two-photon joint spectral intensity manifests itself as a sinusoidal oscillation within a typically Gaussian envelope.}
\label{figure_1}
\end{figure*}
These results show that QWKT provides a well-defined mathematical link between spectral and temporal degrees of freedom of entangled photons. An experimental demonstration of QWKT in frequency-entangled two-photon HOM interference facilitates its application in SD-QOCT and may also indicate a new direction towards fully harnessing QWKT in quantum information processing and quantum interferometric spectroscopy.

\section{Quantum Wiener-Khinchin theorem}
Typically, WKT states that the autocorrelation function of a wide-sense-stationary random process has a spectral decomposition given by the power spectrum of that process \cite{kubo1985statistical}. As shown in Fig. \textcolor{blue}{1(a)}, let us consider a random process $x(t)$ that represents a random variable that evolves in time. Its autocorrelation function can be calculated as
\begin{equation}
R(T)=\langle x(t)x^*(t+T) \rangle=\int_{-\infty}^{\infty}x(t)x^*(t+T)dt,
\end{equation}
where the bracket denotes averaging over an ensemble of realizations of the random variable, $*$ denotes the conjugate operator. This theorem indicates that the spectral power density is the Fourier transform of its autocorrelation function as
\begin{equation}
P(f)=\int_{-\infty}^{\infty} dT R(T) e^{-ifT}.
\end{equation}
In particular, if $x(t)$ is a discrete time series obtained from random sampling, its calculated spectral density is a periodic function in the frequency domain.

By comparison with spectral and temporal process in classical physics, quantum optics has many unique properties and intriguing phenomena. In particular, the spectral and temporal properties of quantum entanglement is determined by the relationship between entangled photons instead of single independent photons, namely the spectral and temporal processing of biphoton wavefunction. Here, we exploit the quantum counterpart of WKT that providing a mathematical connection between spectral and temporal domains of biphoton wavefunction. For a clearer picture, this theory is demonstrated in frequency-entangled two-photon HOM interference with the assistance of spectrally-resolved detection. We consider a generic case that the paired photons (signal and idler) are generated from spontaneous parametric down conversion (SPDC) process pumped by a strong laser \cite{pan2012multiphoton}. Since difference frequency of down converted photons exceeds that of the pump laser, frequency entanglement arises quite naturally as a consequence of energy conservation. The inherent property of simultaneous conversion makes the initial relative time delay between paired photons is exactly zero.

Let us first consider the two-photon temporal signal. When the signal photon arrives at time stamp $t$ and the idler photon arrives at time stamp $t+\tau$, we can define the discrete time-bin modes for two photons as
\begin{equation} \label{eq: signal time}
f_s(t)=exp(-\Delta^2t^2/2),
\end{equation}
\begin{equation} \label{eq: idler time}
f_i(t)=exp(-\Delta^2t^2/2)+exp(-\Delta^2(t+\tau)^2/2),
\end{equation}
where $\Delta$ is the RMS (root mean square) bandwidth in temporal domain, and $\tau$ is the relative time delay between paired photons. The synchronous time signal that corresponds to signal photon is also exhibited in the time signal of idler photon such that two-photon relative time delay can be demonstrated in a time process. We note that the synchronous time signal is necessary such that the two-photon relative temporal signal agrees well with the experimental results \cite{xie2015Harnessing,chen2021temporal}. As shown in Fig. \textcolor{blue}{1(b)}, we use two-photon temporal signals as the input processes of QWKT. Instead of an autocorrelation function, we use cross-correlation functions (CCFs) to demonstrate the temporal correlation between paired photons. Importantly, since the signal and idler photons after the balanced beam splitter in HOM interferometer are indistinguishable, the resultant two-photon temporal signal is expressed in the form of \cite{thompson2017analysis}
\begin{equation}
\begin{split}
R(T)&=(\langle f_s(t)f_i(t+T)\rangle+\langle f_i(t)f_s(t+T)\rangle)/2\\
    &=(\int_{-\infty}^{\infty}f_s(t)f_i^*(t+T)dt+\int_{-\infty}^{\infty}f_i(t)f_s^*(t+T)dt)/2,
\end{split}
\end{equation}
where the sum of commutative cross-correlation functions that denotes the indistinguishability can ensure the symmetry of $R(T)$. By substituting the specific temporal functions \eqref{eq: signal time} and \eqref{eq: idler time} into calculation, the resultant temporal signal reads
\begin{equation}\label{eq: crosscorrelation}
\begin{split}
R(T)\propto &exp(-\Delta^2T^2)+\frac{exp[-\Delta^2(T+\tau)^2]}{2}\\
           &+\frac{exp[-\Delta^2(T-\tau)^2]}{2}.
\end{split}
\end{equation}
The corresponding theoretical simulation for various two-photon time delays are demonstrated in Fig. \textcolor{blue}{1(c)}. Since the cross-correlation describes the periodicity of two time processes, our calculated CCFs manifest themselves as a main peak that accompanied by two bilateral symmetrical side peaks. Undoubtedly, the CCFs only depend on the difference of two-photons' arriving time $\tau$ but independent of start time, which determines the distance between main peak to side peak (see Fig. \textcolor{blue}{1(c.i,ii,v)}). The single-photon coherence time $\Delta$ still determines the distribution width (see Fig. \textcolor{blue}{1(c.ii,iii,iv)}).

Next let us consider the two-photon joint spectral intensity. The paired photons are prepared with central frequency $\omega_s$ (signal) and $\omega_i$ (idler) from a SPDC process. The two-photon state of interest can be expressed as
\begin{equation}\label{eq:probe}
\ket{\psi}=\int_0^\infty\int_0^\infty d\omega_sd\omega_if(\omega_s,\omega_i)\hat{a}_s^\dag(\omega_s)\hat{a}_i^\dag(\omega_i)\ket{0}.
\end{equation}
where $f(\omega_s,\omega_i)$ is the Gaussian spectral amplitude function that fulfills the normalized condition as $\int\int d\omega_sd\omega_i|f(\omega_s,\omega_i)|^2=1$. These paired photons are incident on the two arms of a HOM interferometer, where a time delay of interest is introduced that caused by imbalance between two arms of the interferometer. As a common case in HOM interferometry, the nonlocal interference pattern is observed by scanning the time of arrival of one of the paired photons incident on the beam splitter. The HOM interferometer transforms the incident two-photon state to a superposition of bunched and anti-bunched state as
\begin{equation}
\ket{\psi(\tau)}\rightarrow\ket{\psi_A(\tau)}+\ket{\psi_B(\tau)},
\end{equation}
where $\ket{\psi_A(\tau)}$ and $\ket{\psi_B(\tau)}$ correspond to the events that two photons anti-bunch into opposite and bunch into identical outports, respectively. Followed by detecting paired photons at opposite spatial modes, the normalized coincidence detection probability $P_c(\tau)=|\braket{\psi(\tau)}{\psi_A(\tau)}|^2$ is obtained as (see Appendix for more details)
\begin{equation}\label{eq: hom interference}
\begin{split}
P_c(\tau)=&\frac{1}{4}\int_0^\infty\int_0^\infty d\omega_1d\omega_2[|f(\omega_1,\omega_2)|^2+|f(\omega_2,\omega_1)|^2\\
&-2f(\omega_1,\omega_2)f(\omega_2,\omega_1)cos(\omega_1-\omega_2)\tau].
\end{split}
\end{equation}
Since the desired frequency-entangled state are prepared with broadband nondegenerate frequencies that are more than the spectral bandwidth of the pump laser, $f(\omega_1,\omega_2)=f(\omega_2,\omega_1)$ with the energy conservation as $\omega_1+\omega_2=\omega_p$, henceforth the normalized spectral distribution of two-photon state in opposite ports can be simplified from \eqref{eq: hom interference} as
\begin{equation}\label{eq:probability}
F(\omega)=\frac{exp(-\omega^2/8\sigma^2)}{2\sqrt{2\pi (2\sigma)^2}}[1-cos(\omega\tau+\phi)],
\end{equation}
where $\omega$ denotes the difference frequency $\omega=\omega_s-\omega_i$, $\sigma$ is the RMS bandwidth of the initial down-converted photons in spectral domain that is inversely proportional to temporal bandwidth $\Delta$, the term $exp(-\omega^2/8\sigma^2)/\sqrt{2\pi (2\sigma)^2}$ denotes the Gaussian spectral amplitude function, and $\phi$ is a phase factor. At the center wavelength ($\omega=\omega_s-\omega_i=0$), this wavelength indistinguishability decreases the HOM interference to zero, which agrees well with the fact that two identical photons impinge on a balanced beam splitter would leave through the same output port, i.e., the well-known HOM dip. The two-photon spectral signal as expressed in \eqref{eq:probability} reveals that the anti-bunched photons in opposite spatial modes manifest themselves as sinusoidal oscillation of the interference fringe with determined spectral separation that depends on the time delay $\tau$, where the single photon spectral bandwidth $\sigma$ determines the base-to-base envelope width (see Fig. \ref{figure_1}\textcolor{blue}{(d)}). Backed by these calculations, it is also revealed that if the input of QWKT is a discrete time process embedded with two well-separated time-bin modes, its joint spectral intensity manifests as a periodic oscillation with respect to single-photon frequencies.

It is obvious that the cross-correlation of two-photon temporal signal and two-photon joint spectral intensity can be connected by making a Fourier transform as
\begin{equation}
\begin{split}
&F(\omega)=\mathcal{F}[R(T)]=\frac{1}{2\pi}\int_{-\infty}^{\infty } R(T)e^{i\omega T}dT,\\
&R(T)=\mathcal{F}^{-1}[F(\omega)]=\int_{-\infty}^{\infty } F(\omega)e^{-i\omega T}d\omega.
\end{split}
\end{equation}
As a result, it indicates the statement of QWKT that the cross-correlation function of two-photon temporal signal and the two-photon joint spectral density are linked by a Fourier transform, which is in analogy to the classical WKT.

\begin{figure*}[!t]
\centering
\includegraphics[width=\linewidth]{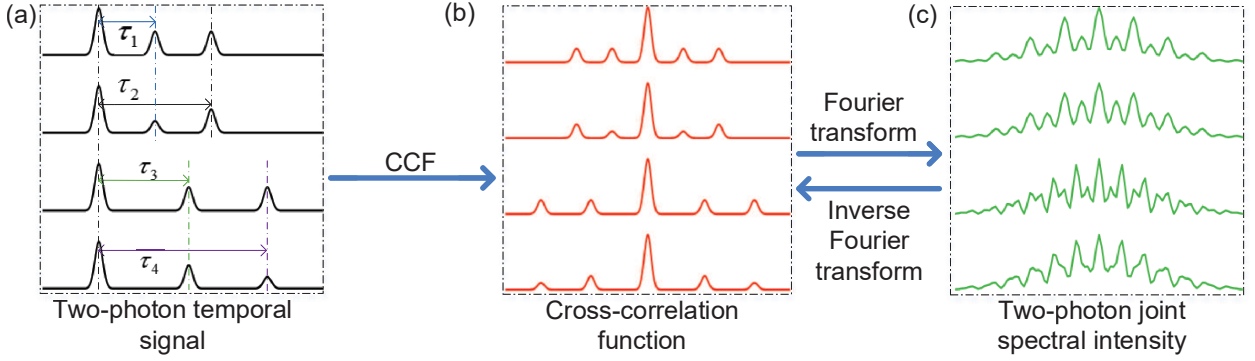} 
\caption{Quantum Wiener-Khinchin theorem for single photons that are superimposed in two temporal modes as shown in \eqref{eq: time superposition}.}
\label{figure_2}
\end{figure*}
\begin{figure*}[!t]
\centering
\subfigure[]{
\label{Fig3.sub.1}
\includegraphics[width=0.24\linewidth]{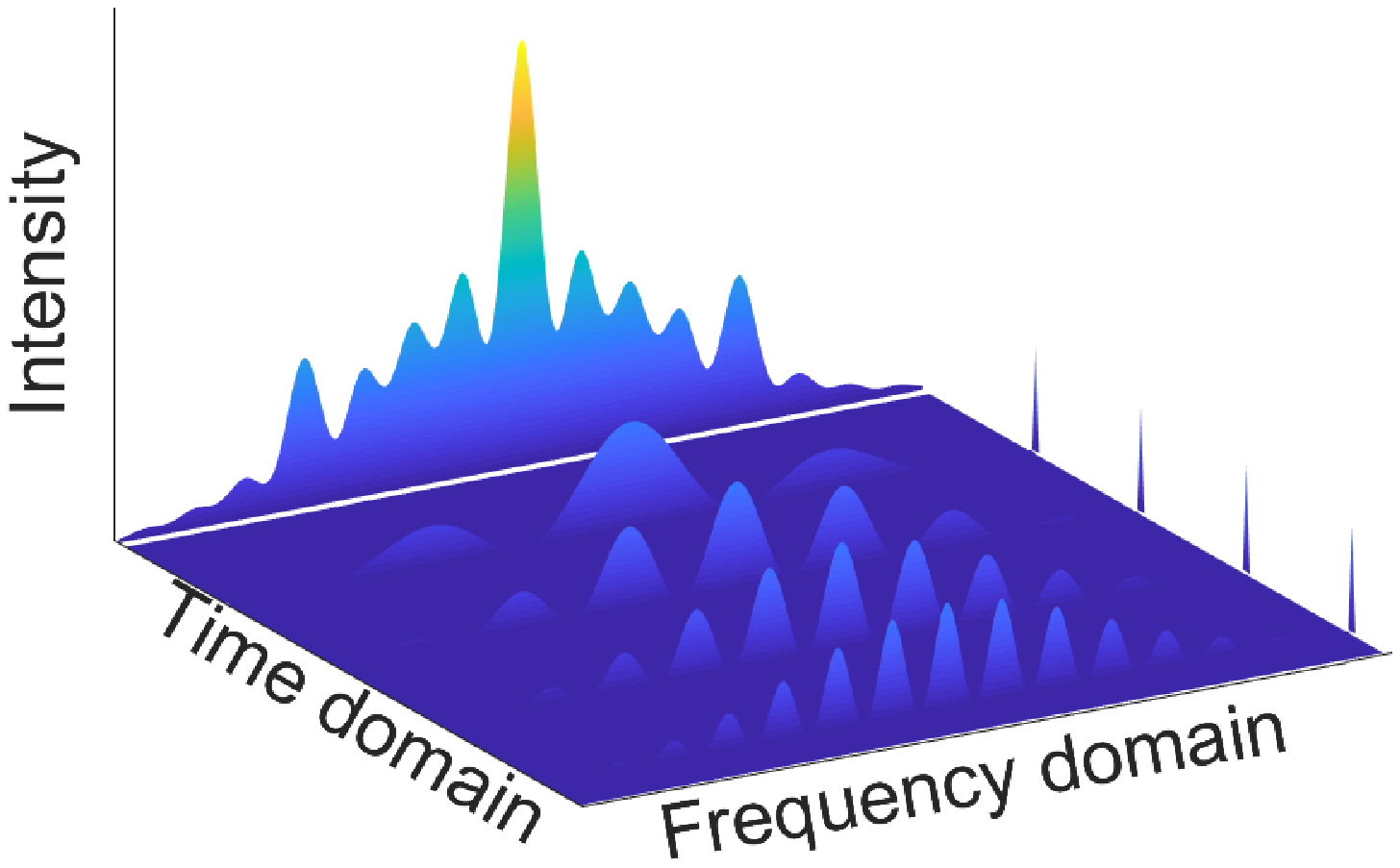}}
\subfigure[]{
\label{Fig3.sub.2}
\includegraphics[width=0.24\linewidth]{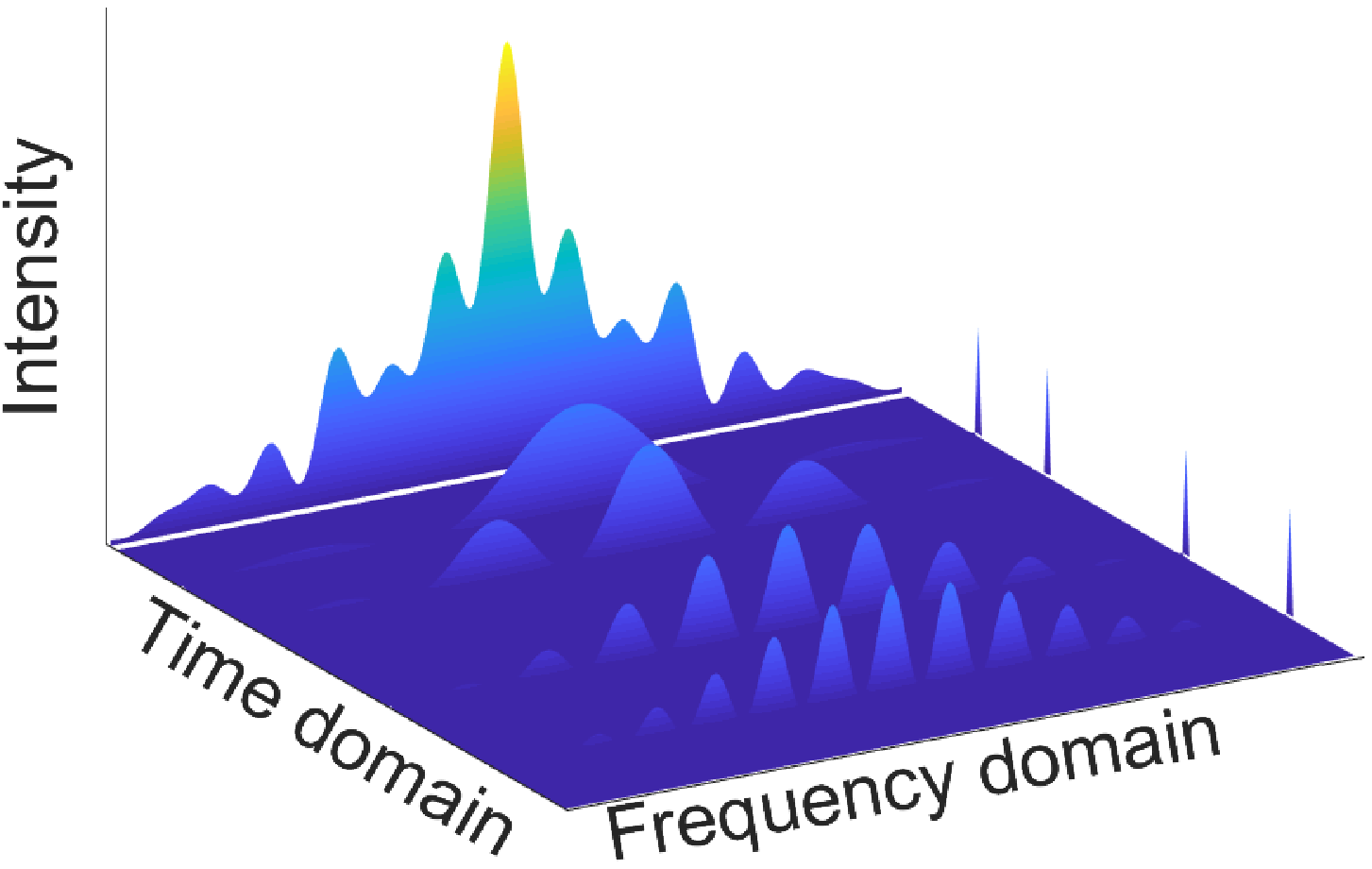}}
\subfigure[]{
\label{Fig3.sub.3}
\includegraphics[width=0.24\linewidth]{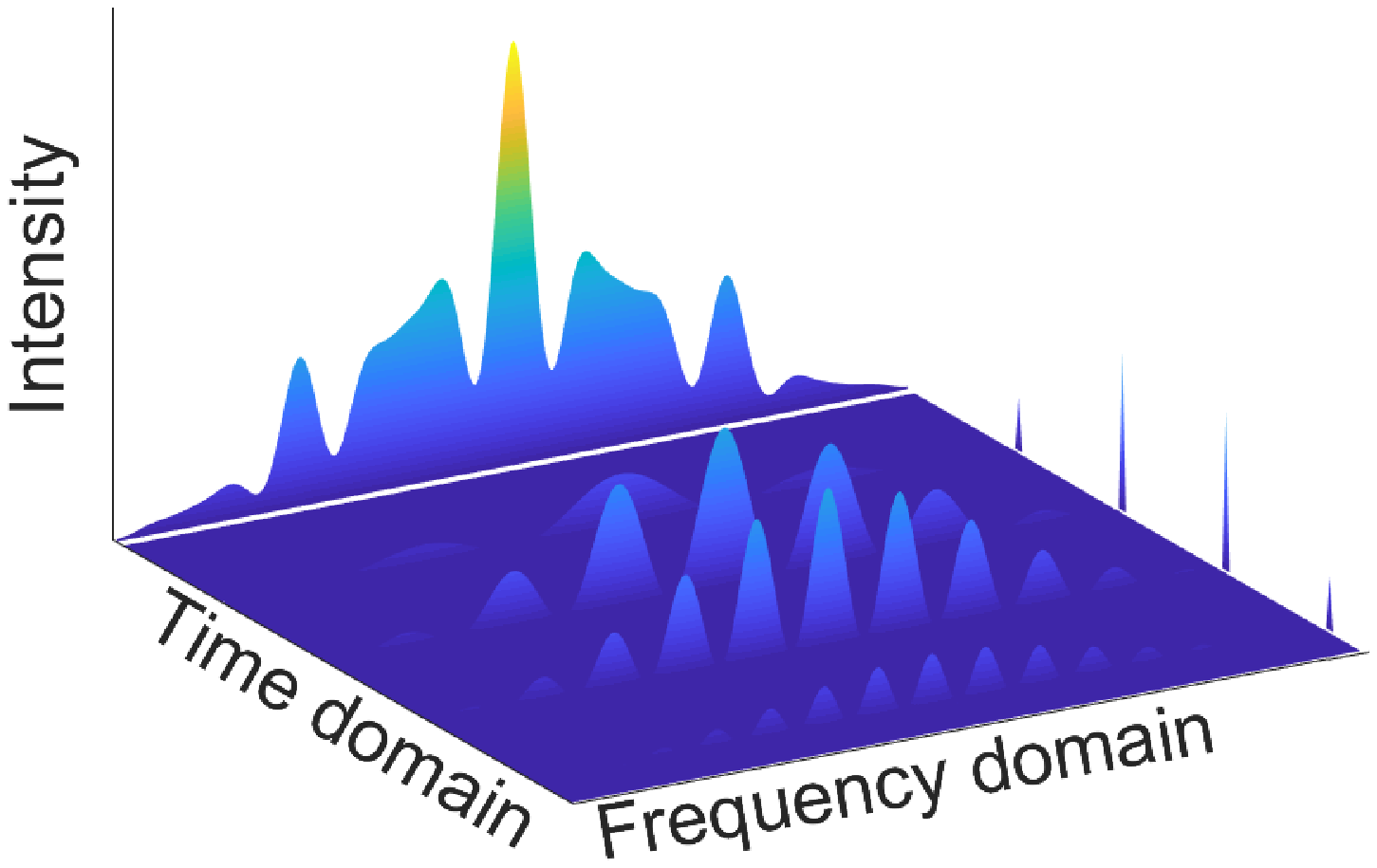}}
\subfigure[]{
\label{Fig3.sub.4}
\includegraphics[width=0.24\linewidth]{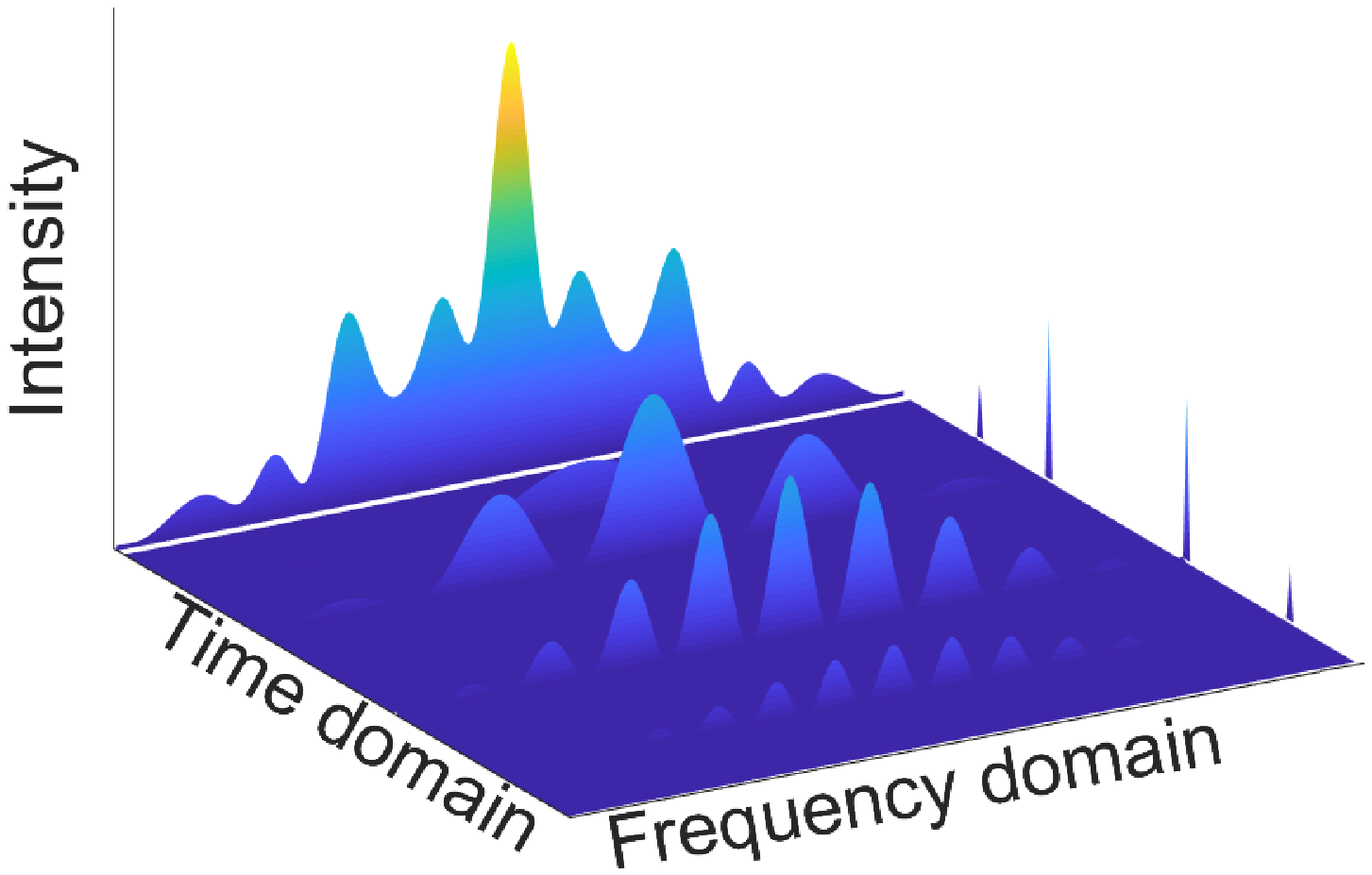}}
\caption{Spectral analysis can decompose the spectral distributions in Fig. \textcolor{blue}{2(c)} into a series of spectral intensities obtained from Fig. \textcolor{blue}{1(d)}. The relative heights of individual peaks in the time domain directly yield information about their relative weights of the superposition states.}
\label{figure_3}
\end{figure*}
For a more generic task that the light beam transmits through a material with non-uniform thickness in the transverse direction \cite{devaux2020imaging}, this material would introduce multiple time delays that is a superposition state as
\begin{equation}\label{eq: time superposition}
f_i(t)=exp(-\Delta^2t^2/2)+\sum_{i=1}^{n}A_iexp(-\Delta^2(t+\tau_i)^2/2),
\end{equation}
where $\tau_i$ represents the $i$-th two-photon relative time delay, $A_i$ denotes the corresponding probability amplitude that fulfills the normalized condition as $\Sigma_{i=1}^{n}A_i=1$ (see Fig. \textcolor{blue}{2(a)}). Their CCFs are calculated as shown in Fig. \textcolor{blue}{2(b)}, where the temporal signal also exhibits as a main peak but accompanied by multiple pairs of bilateral symmetrical side peaks, and the time delay $\tau_n$ determines the distance between main peak to $n$-th side peak, $a_n$ determines the normalized intensity of each side peak. Likewise, by a performing Fourier transform on the CCFs, two-photon joint spectral intensities can be obtained as shown in Fig. \textcolor{blue}{2(c)}. Since the resultant spectral pattern results from the superposition of multiple two-photon time delays, inversely, they can be decomposed into a superposition of multiple spectral patterns that obtained from single two-photon relative time delay as shown in Fig. \ref{figure_3}. In particular, we note that the weight of individual spectral pattern is directly determined by its corresponding probability amplitude of time-bin superposition states. For an instructive means of understanding, this approach can be considered as a quantum version of ``spectrum analysis'', which decomposes the complex periodic vibration into a series of simple harmonic motion.

Since this QWKT establishes the connection between the time-energy domains of biphoton wavefunction, it is allowed to extract temporal information from spectral pattern (i.e., quantum optical coherence tomography \cite{pablo2020spectrally,sylwia2020fourier}), or extract spectral information from temporal pattern (i.e., quantum interferometric spectroscopy \cite{chen2022entanglement}). Furthermore, three-dimensional quantum tomography is also achievable with the assistance of spatiotemporal quantum interference \cite{devaux2020imaging}.

\begin{figure*}[!t]
\centering
\includegraphics[width=\linewidth]{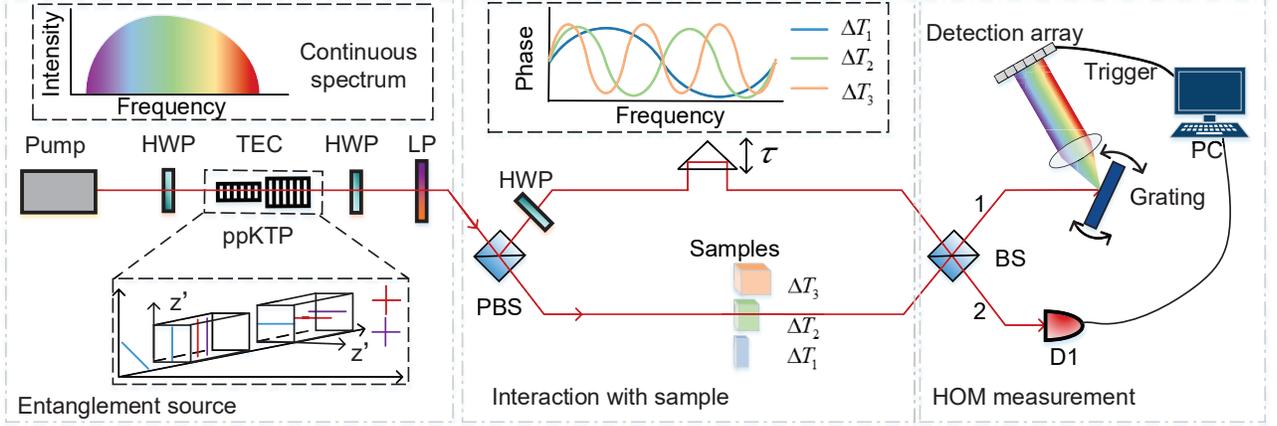}
\caption{Experimental demonstration of quantum Wiener-Khinchin theorem in frequency-entangled two-photon HOM interference. HWP, half-wave plate; TEC, temperature controller; PPKTP, periodically poled potassium titanyl phosphate crystal; LP, long pass filter; PBS, polarising beam splitter; BS: balanced beam splitter; sample: transparent sample with multi-layer structure, $D_1$, single photon detector; grating: reflective diffraction grating. The inset in the entanglement source part illustrates the horizontally and vertically orientated crystals, which ensures that the incident diagonally-polarized  photons can pump these two crystals with equal probability. The inset in the interaction with sample part illustrates the sample-induced phase shift as a function of frequency distribution of idler photons, wherein $\Delta T$ determines the oscillation period.}
\label{figure_4}
\end{figure*}
\section{Experimental demonstration of QWKT through HOM interference}
We experimentally confirm that the mathematically-defined QWKT can be implemented by using HOM interference \cite{chen2019Hong,xie2015Harnessing,giovannini2015spatially}. Since our method has the requirements of pairs of identical photons that are separated into two distinct spatial modes with broad spectral distribution, time-reversed HOM interference is used to prepare these entangled photons \cite{chen2018polarization}. Thereinto, polarization entanglement is used as an auxiliary degree of freedom for the spatial separation of indistinguishable photons. As shown in Fig. \ref{figure_4}, frequency-entangled photons are generated by SPDC process pumped with a continuous-wave grating-stabilized laser diode. A pair of crosssed PPKTP crystals are placed in sequence, whereby the optical axis of second crystal is rotated by $90^\degree$ with respect to the first crystal. Both of these two nonlinear optical crystals are designed for collinear type-0 phase matching such that the frequency bandwidth of down-converted photons are sufficiently broad for measuring the spectral pattern. A half wave plate in the pump beam is used to set a diagonal polarization state such that two mutually orthogonally oriented crystals are pump equally. Balanced pumping enables equal probability amplitudes for SPDC emission $\ket{H}\rightarrow\ket{HH}$ in the first crystal, and $\ket{V}\rightarrow\ket{VV}$ in the second crystal, where $\ket{H}$ and $\ket{V}$ represent horizontal and vertical polarizations. By rotating the polarization of down-converted photons to diagonal and anti-diagonal direction and setting the relative phase $\phi=\pi$, the resultant state reads as
\begin{equation}
(\ket{AA}-\ket{DD})/\sqrt{2}=(\ket{HV}+\ket{VH})/\sqrt{2}.
\end{equation}
Subsequently, a polarizing beam splitter (PBS) is used to deterministically route a pair of frequency-entangled photons into two distinct spatial modes \cite{chen2018polarization}. In our experimental realization of broadband frequency entanglement source, two mutually orthogonally oriented 10-mm-long ppKTP crystals are manufactured to provide collinear phase matching with pump (p), signal (s) and idler (i) photons at center wavelengths of $\lambda_p \approx \unit[405]{nm}$ and $\lambda_{s,i} \approx \unit[810]{nm}$. The single-photon bandwidth is $\sim \unit[20]{nm}$ at a temperature of $21^{\degree} C$. The wavelength-dependent phase shift is compensated by tilting a half wave plate. Without any bandpass filter, we detect a two-fold coincidence rate of $R_c = 8$ kcps. Then the signal photons pass through a translation stage, which can be used to scan the arriving time that incident on a balanced beam splitter. On the other hand, the idler photons interact with the test samples, and introduce a relative time delay to be estimated. These pairs of photons impinge on a balanced beam splitter from separated input modes, which constitutes a HOM interferometer \cite{chen2020verification,xie2015Harnessing}. Two-photon joint spectral intensity is identified in the opposite spatial modes when two photons arrive at the detectors within a coincidence window of $\sim\unit[1]{ns}$.

\begin{figure*}[!t]
\centering
\subfigure[]{
\label{Fig5.sub.1}
\includegraphics[width=0.24\linewidth]{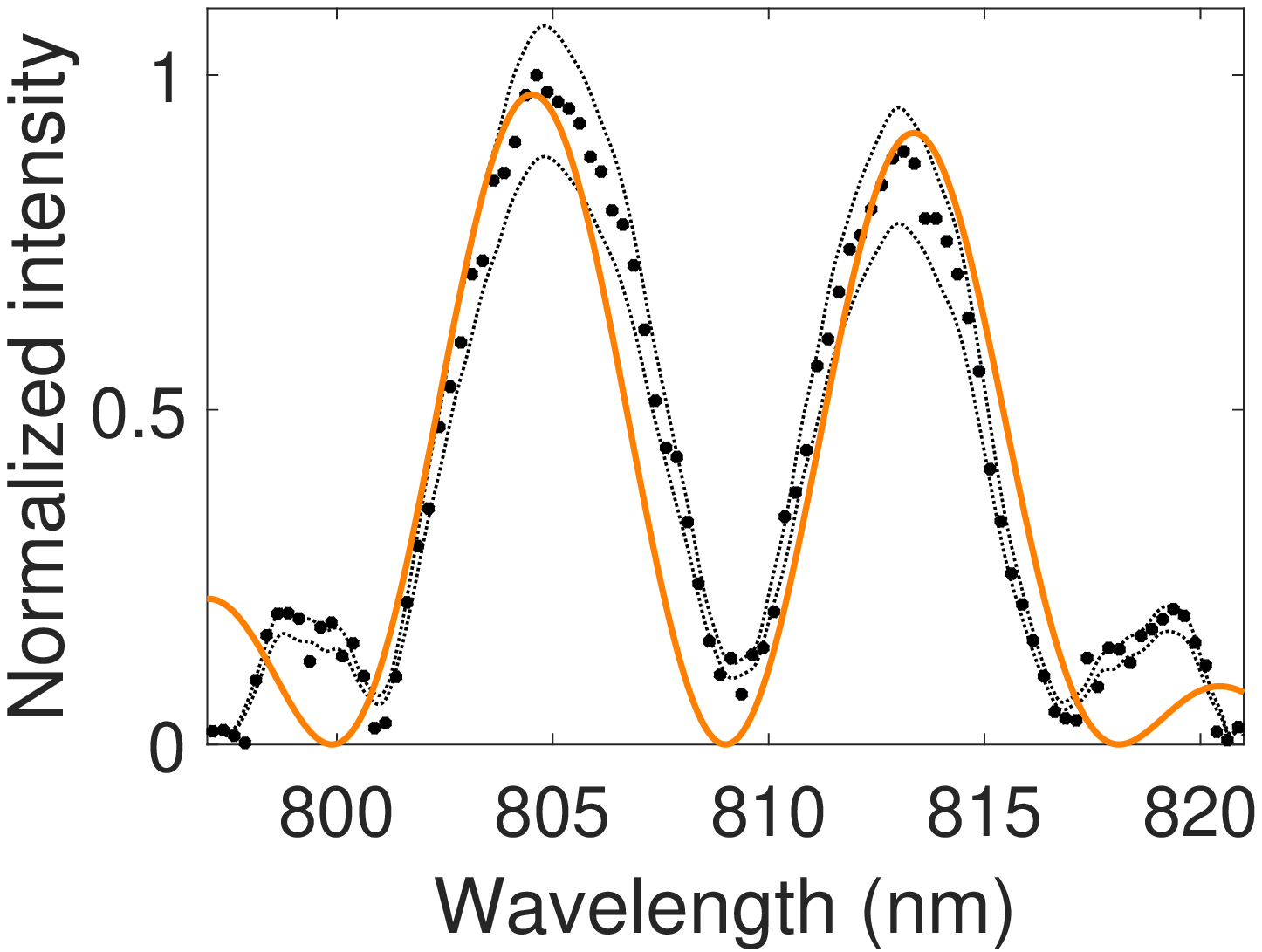}}
\subfigure[]{
\label{Fig5.sub.2}
\includegraphics[width=0.24\linewidth]{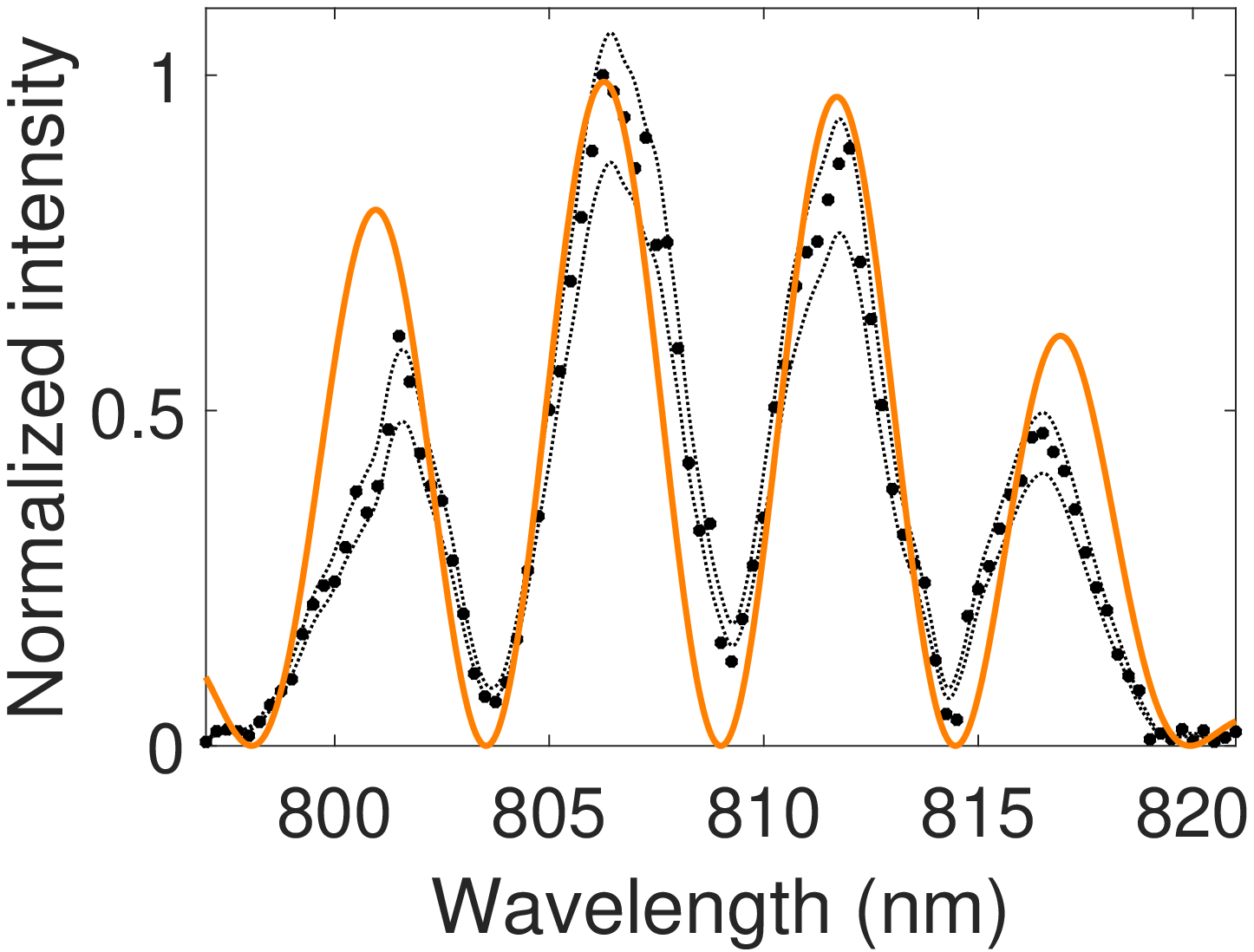}}
\subfigure[]{
\label{Fig5.sub.3}
\includegraphics[width=0.24\linewidth]{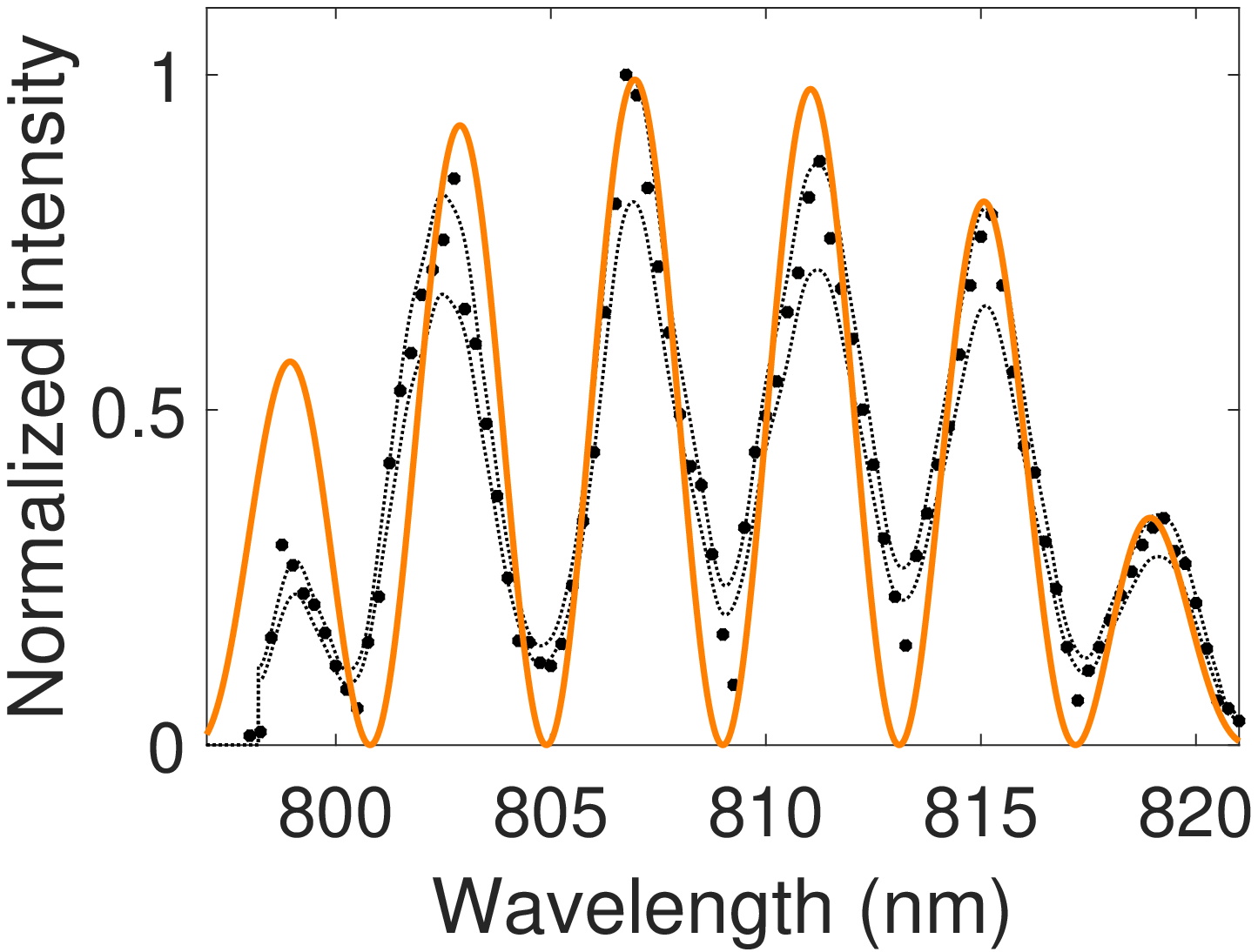}}
\subfigure[]{
\label{Fig5.sub.4}
\includegraphics[width=0.24\linewidth]{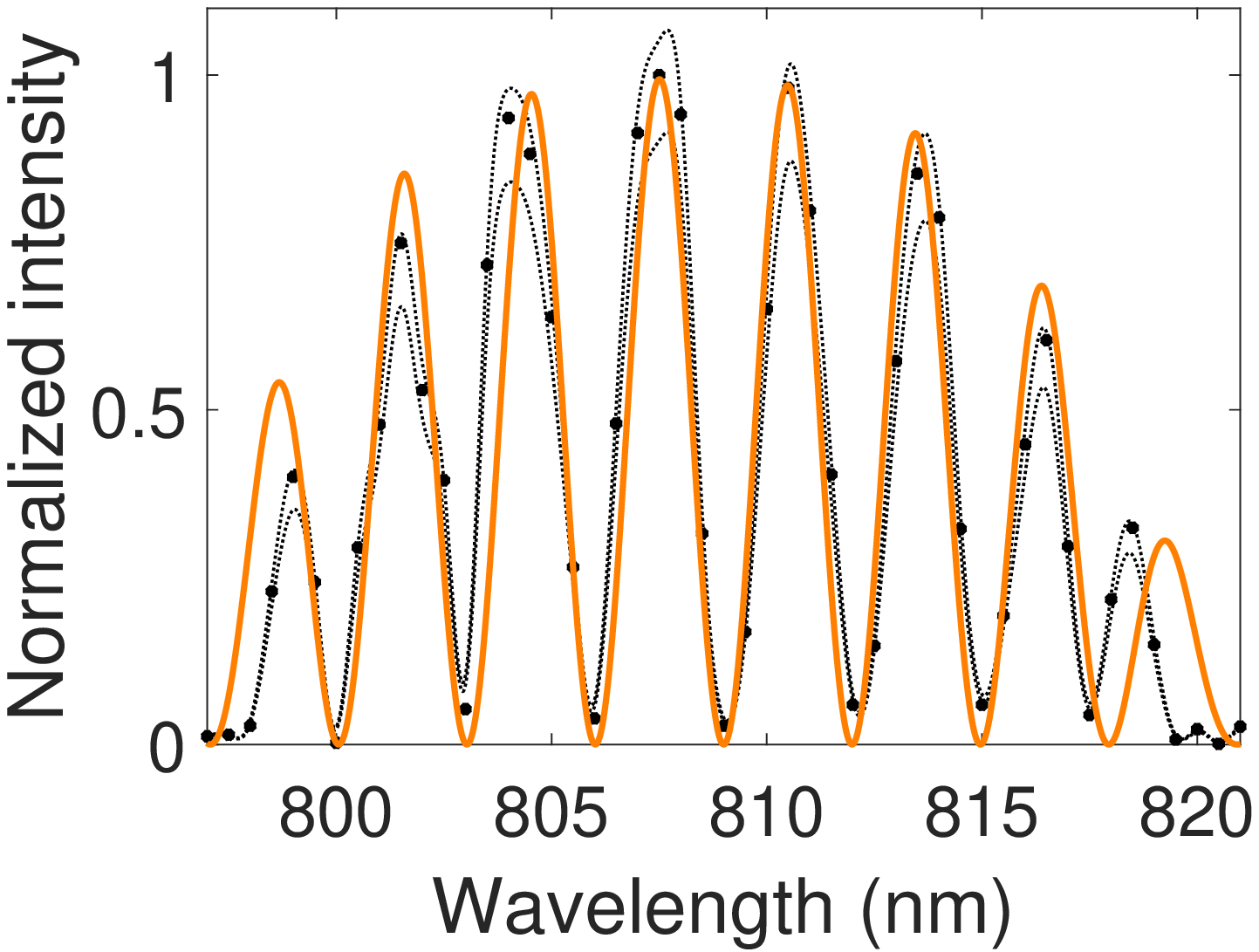}}
\subfigure[]{
\label{Fig5.sub.5}
\includegraphics[width=0.24\linewidth]{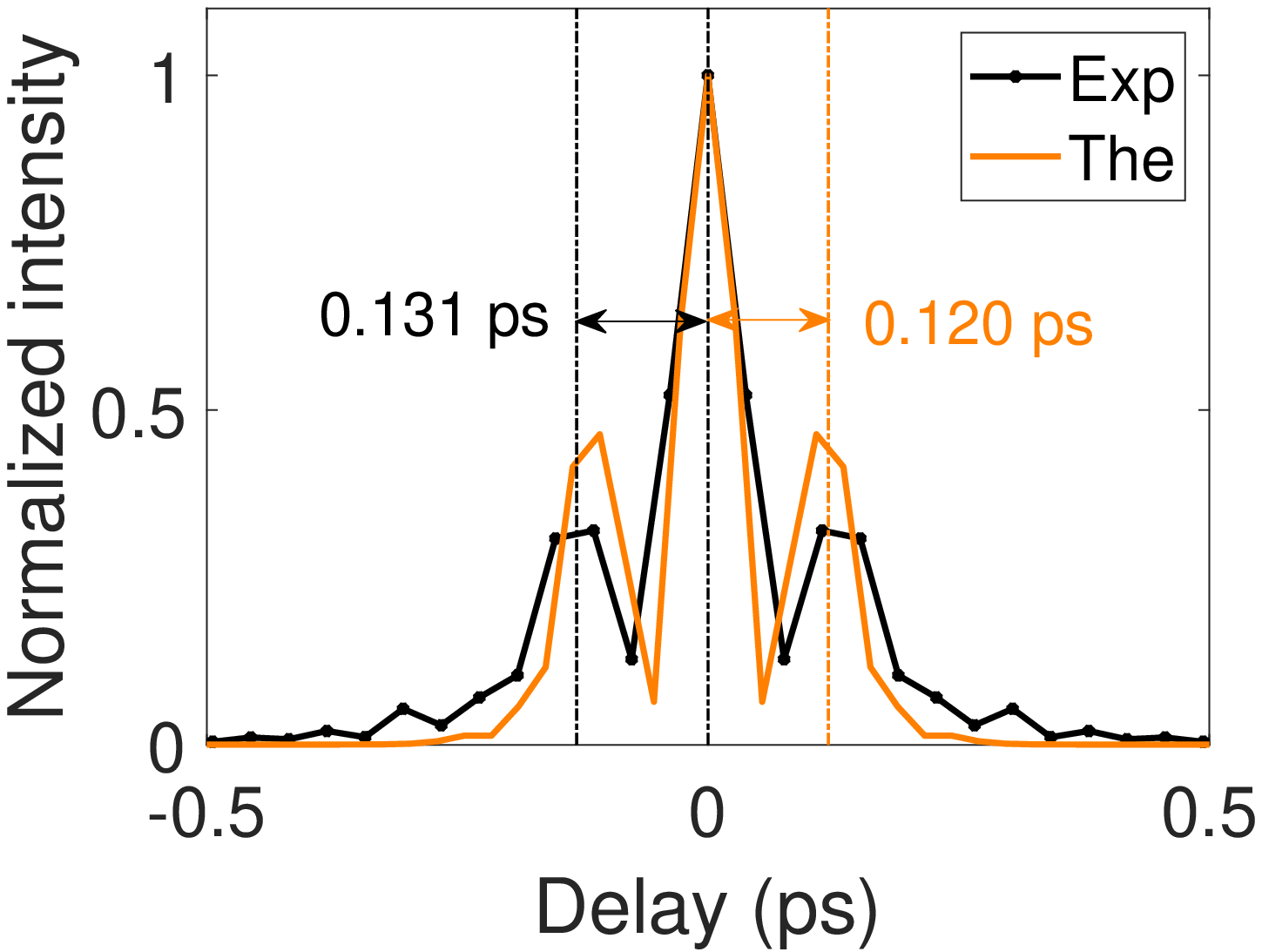}}
\subfigure[]{
\label{Fig5.sub.6}
\includegraphics[width=0.24\linewidth]{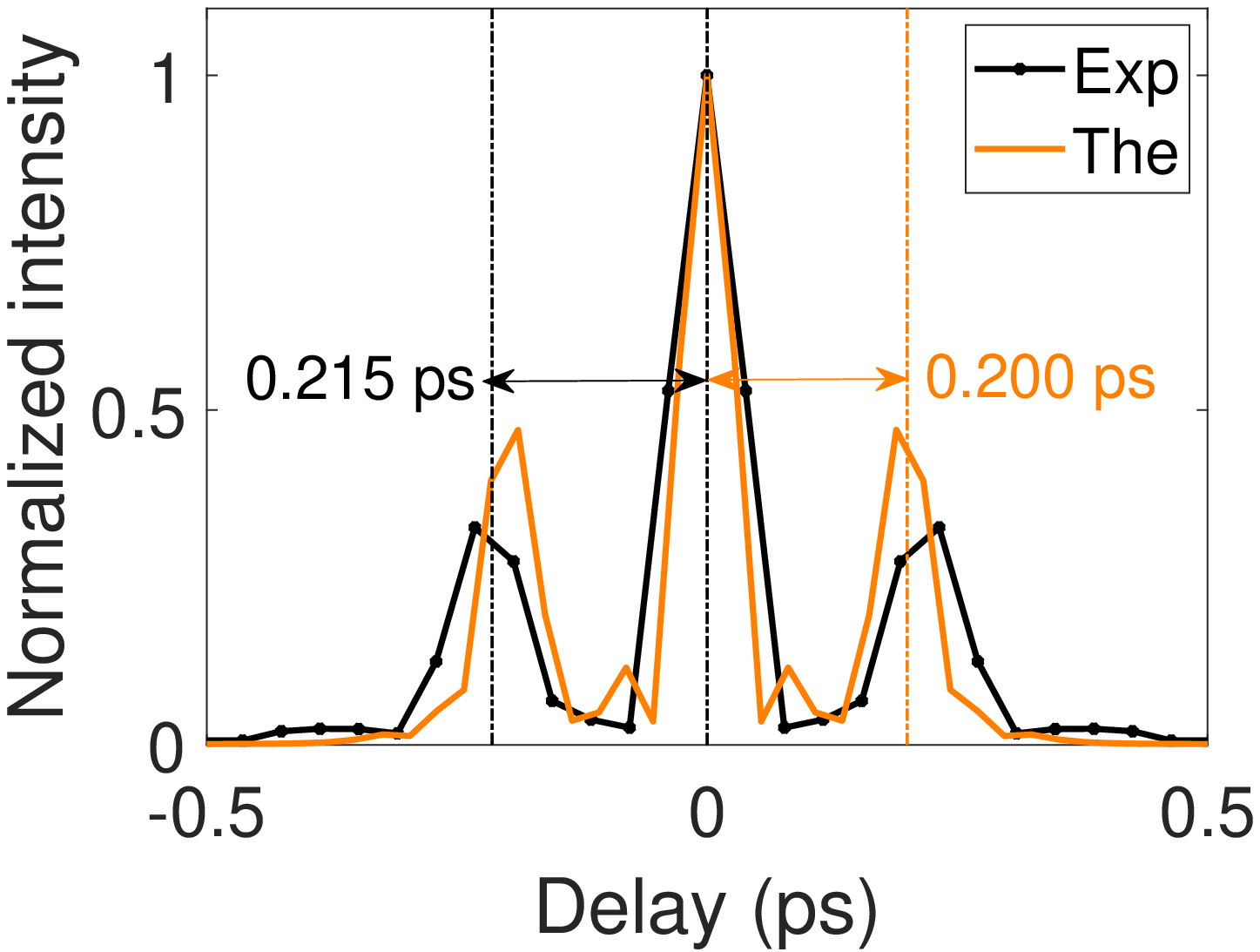}}
\subfigure[]{
\label{Fig5.sub.7}
\includegraphics[width=0.24\linewidth]{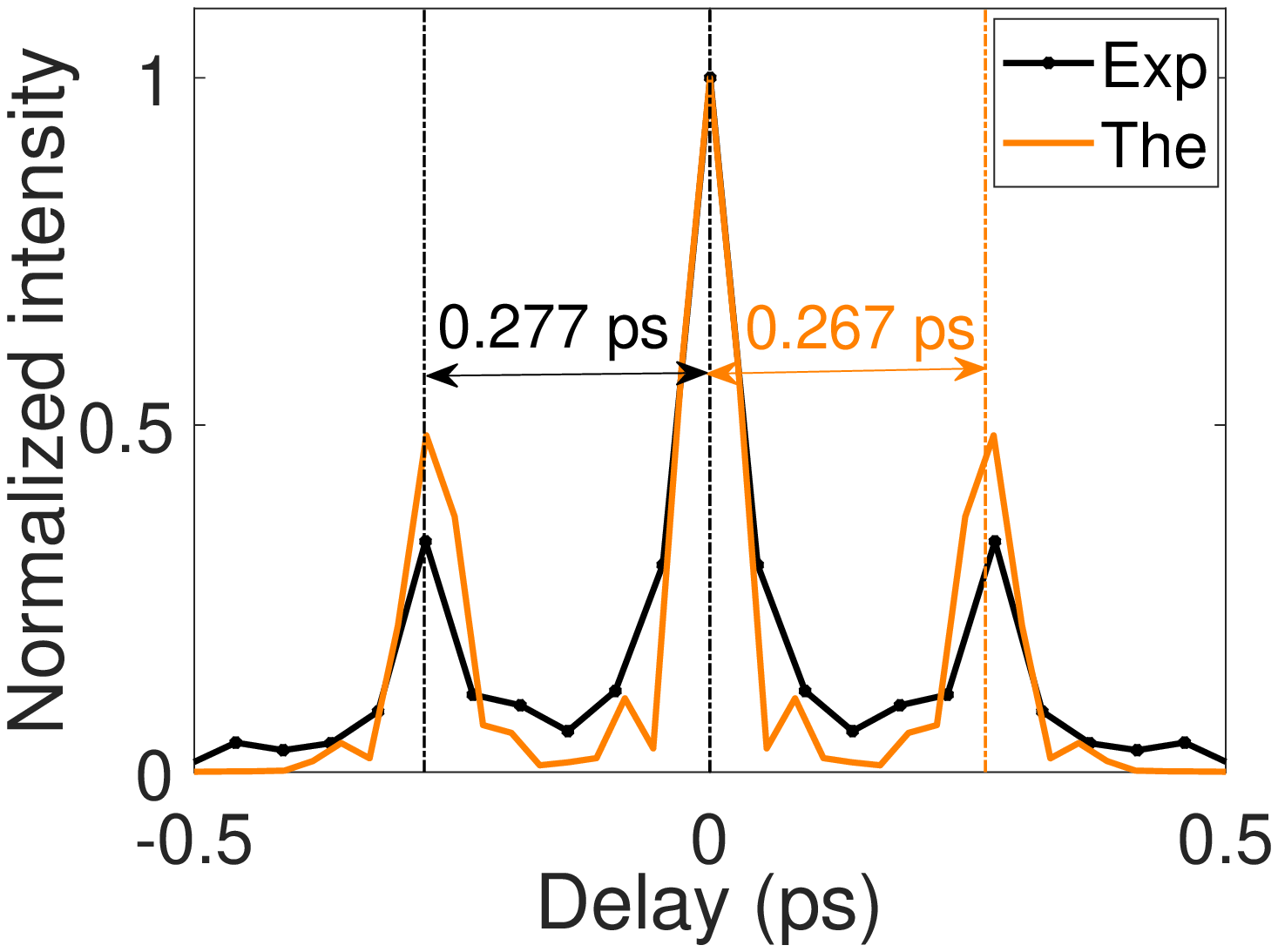}}
\subfigure[]{
\label{Fig5.sub.8}
\includegraphics[width=0.24\linewidth]{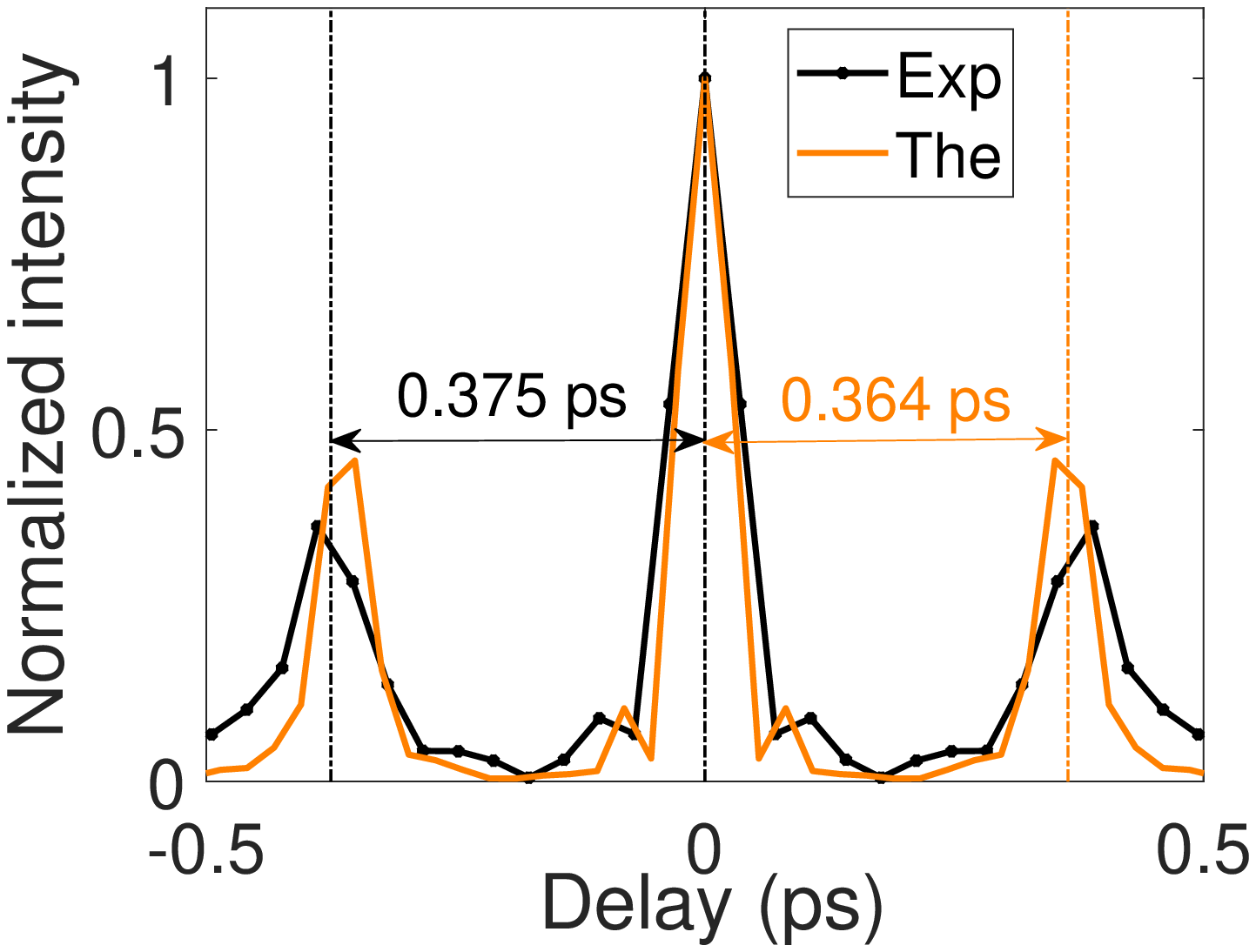}}
\subfigure[]{
\label{Fig5.sub.9}
\includegraphics[width=0.24\linewidth]{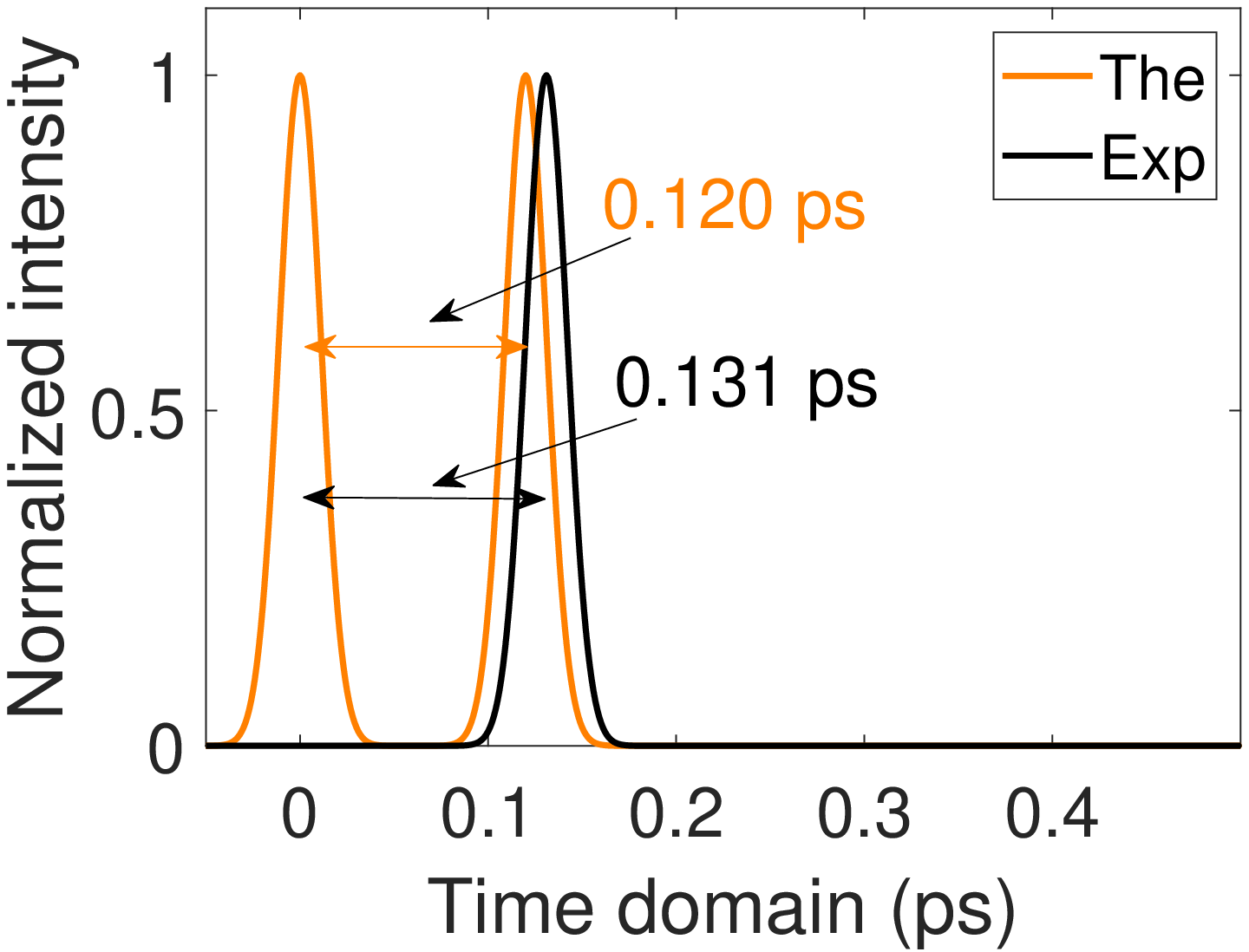}}
\subfigure[]{
\label{Fig5.sub.10}
\includegraphics[width=0.24\linewidth]{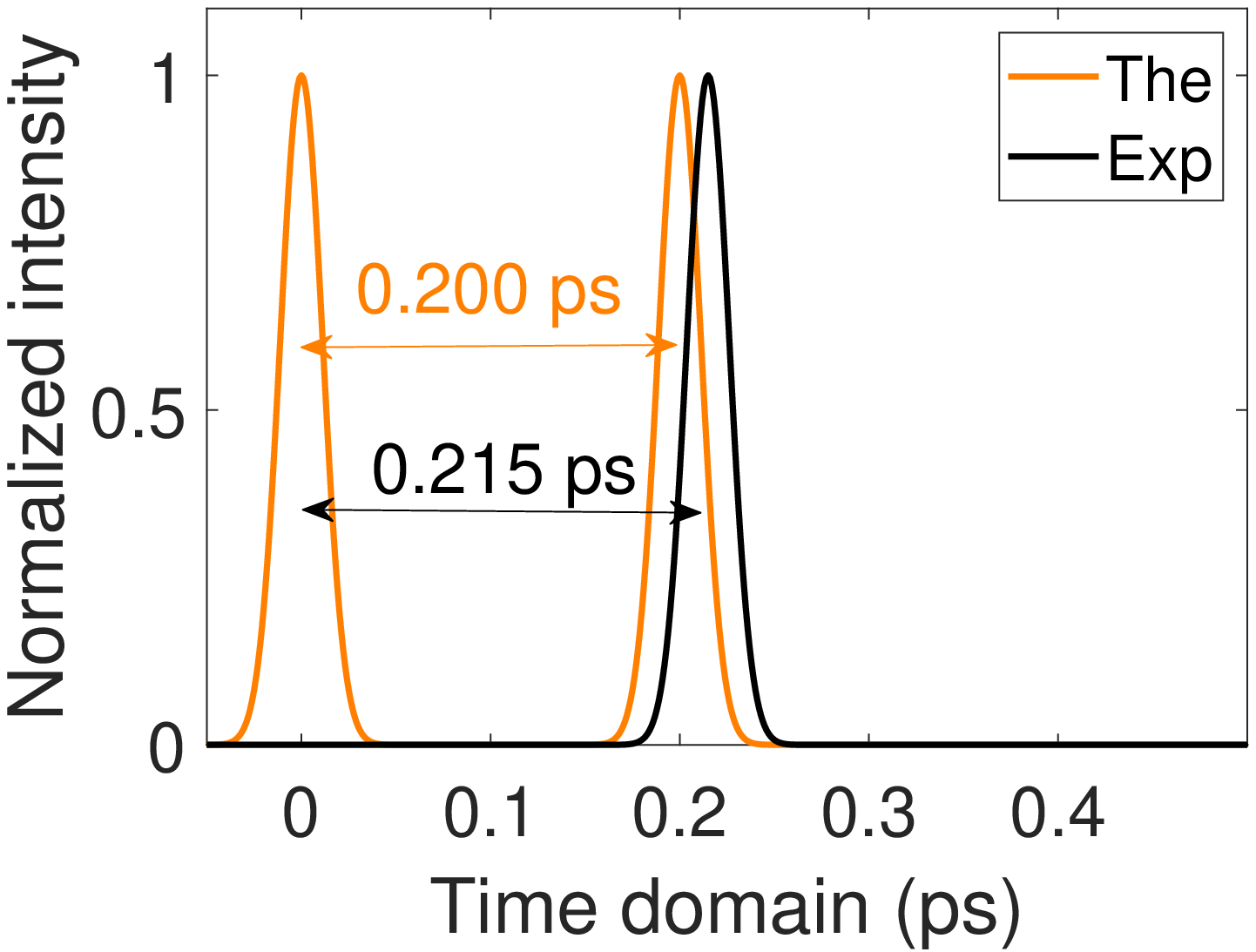}}
\subfigure[]{
\label{Fig5.sub.11}
\includegraphics[width=0.24\linewidth]{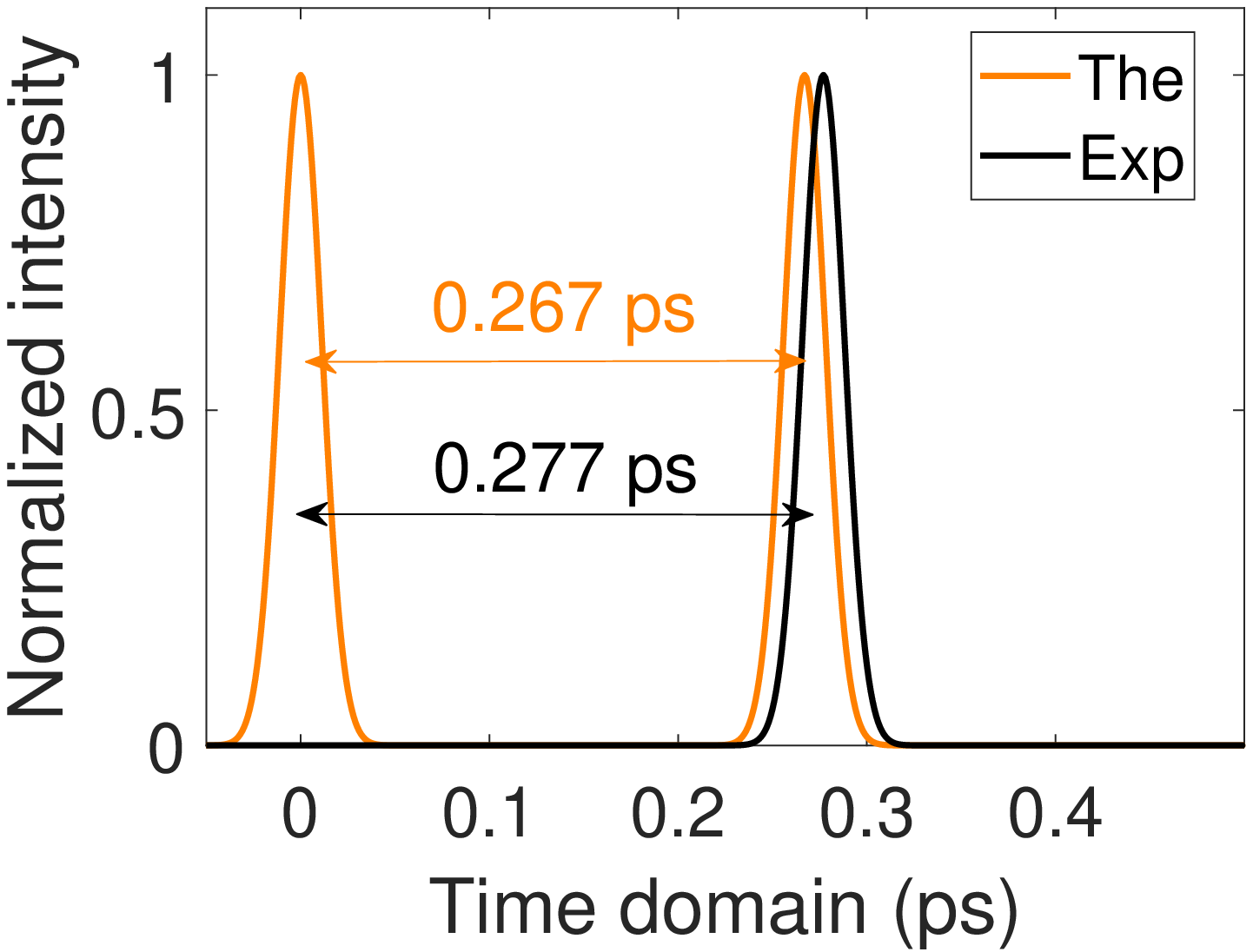}}
\subfigure[]{
\label{Fig5.sub.12}
\includegraphics[width=0.24\linewidth]{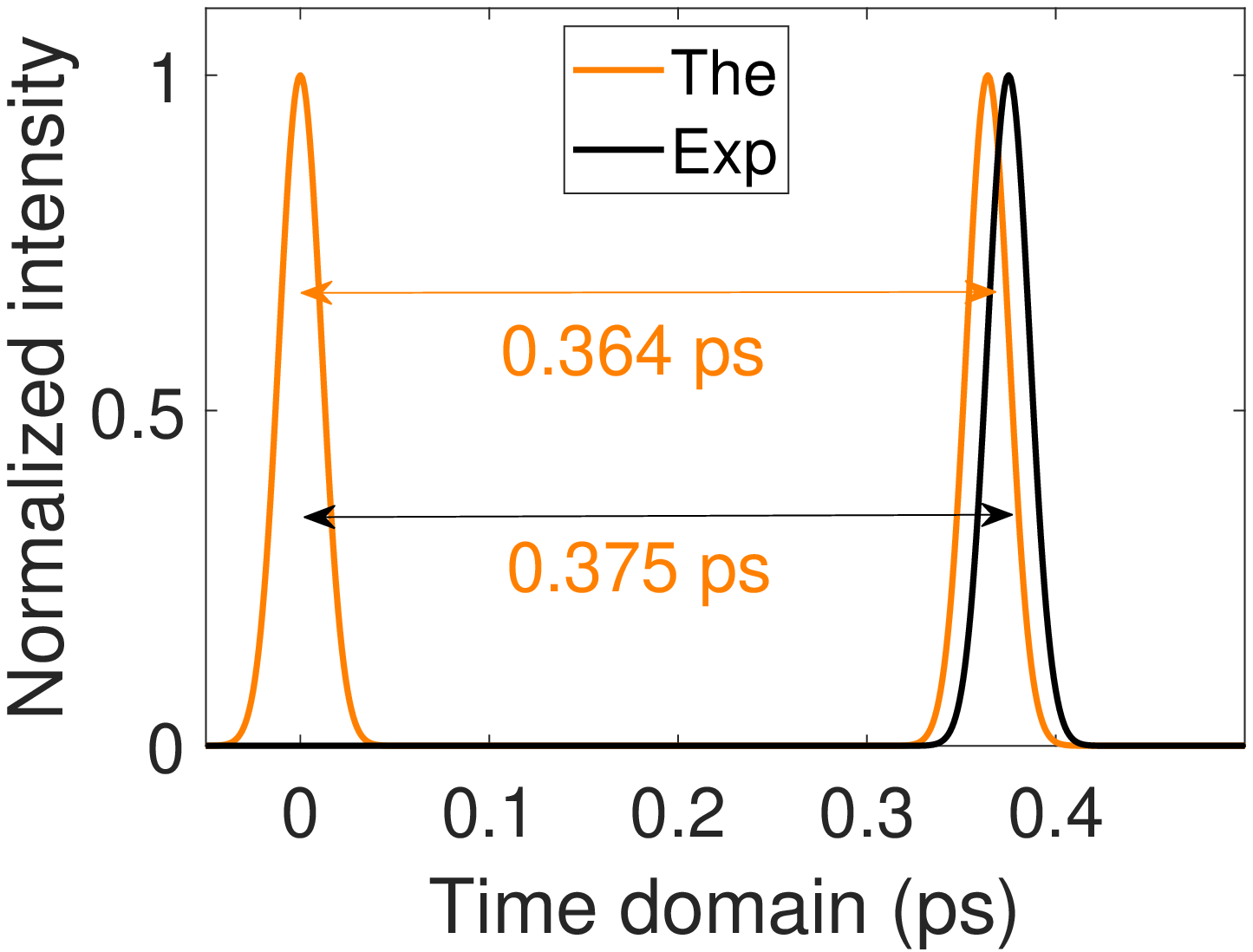}}
\caption{Experimental measurement and theoretical simulation of (a-d) two-photon joint spectral intensities, where the orange lines represent the theoretical predictions, and the black points represent the experimental results that are bounded by the standard deviation estimated by statistical methods assuming a Poisson distribution. (e-h) the cross-correlation functions by making an inverse Fourier transform, and (i-l) the predicted two-photon relative delays by setting $\Delta T$ as (a,e,i) \unit{0.120}{ps}, (b,f,j) \unit{0.200}{ps}, (c,g,k) \unit{0.267}{ps}, and (d,h,l) \unit{0.364}{ps}.}
\label{figure_5}
\end{figure*}
\begin{figure*}[!t]
\centering
\subfigure[]{
\label{Fig6.sub.1}
\includegraphics[width=0.24\linewidth]{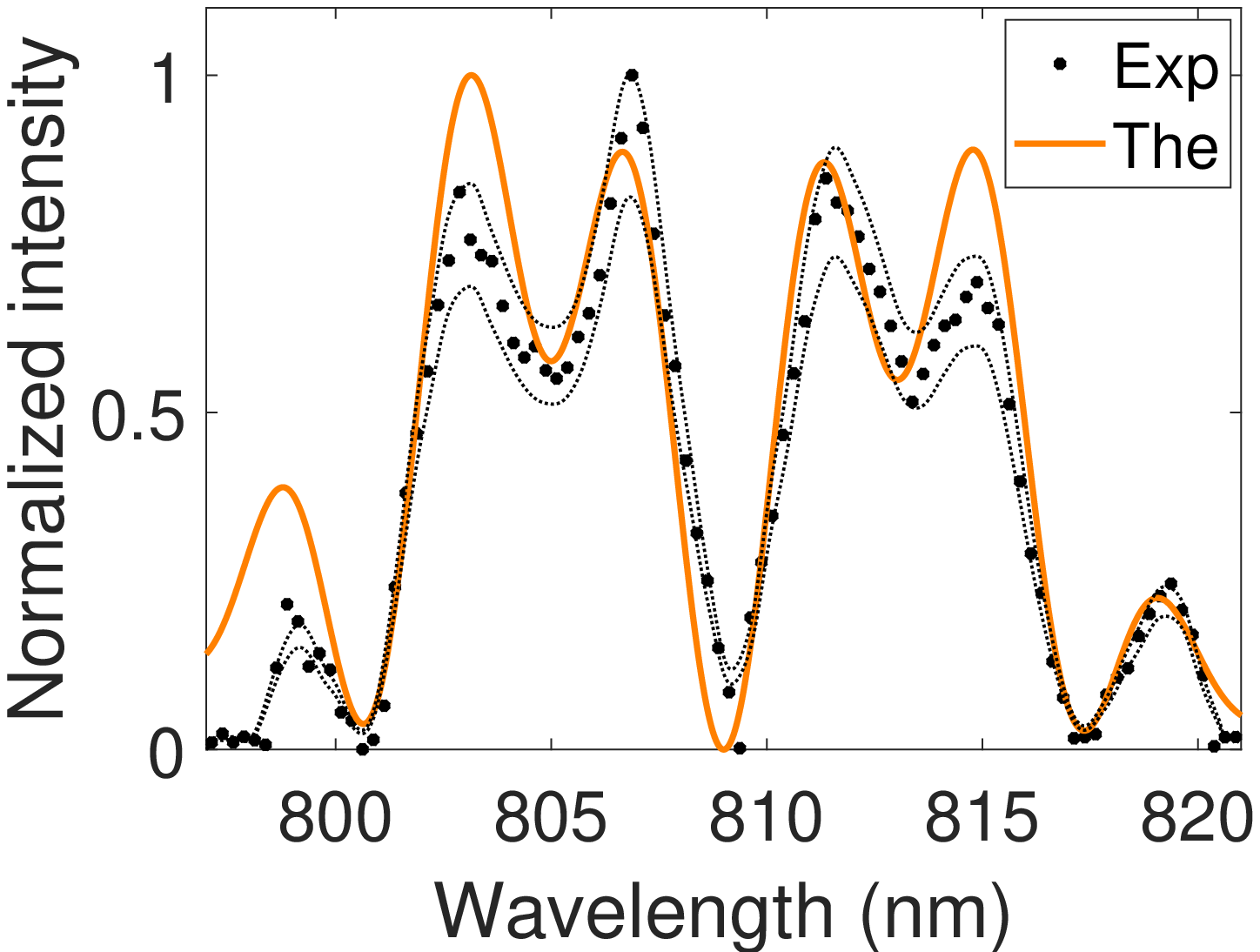}}
\subfigure[]{
\label{Fig6.sub.2}
\includegraphics[width=0.24\linewidth]{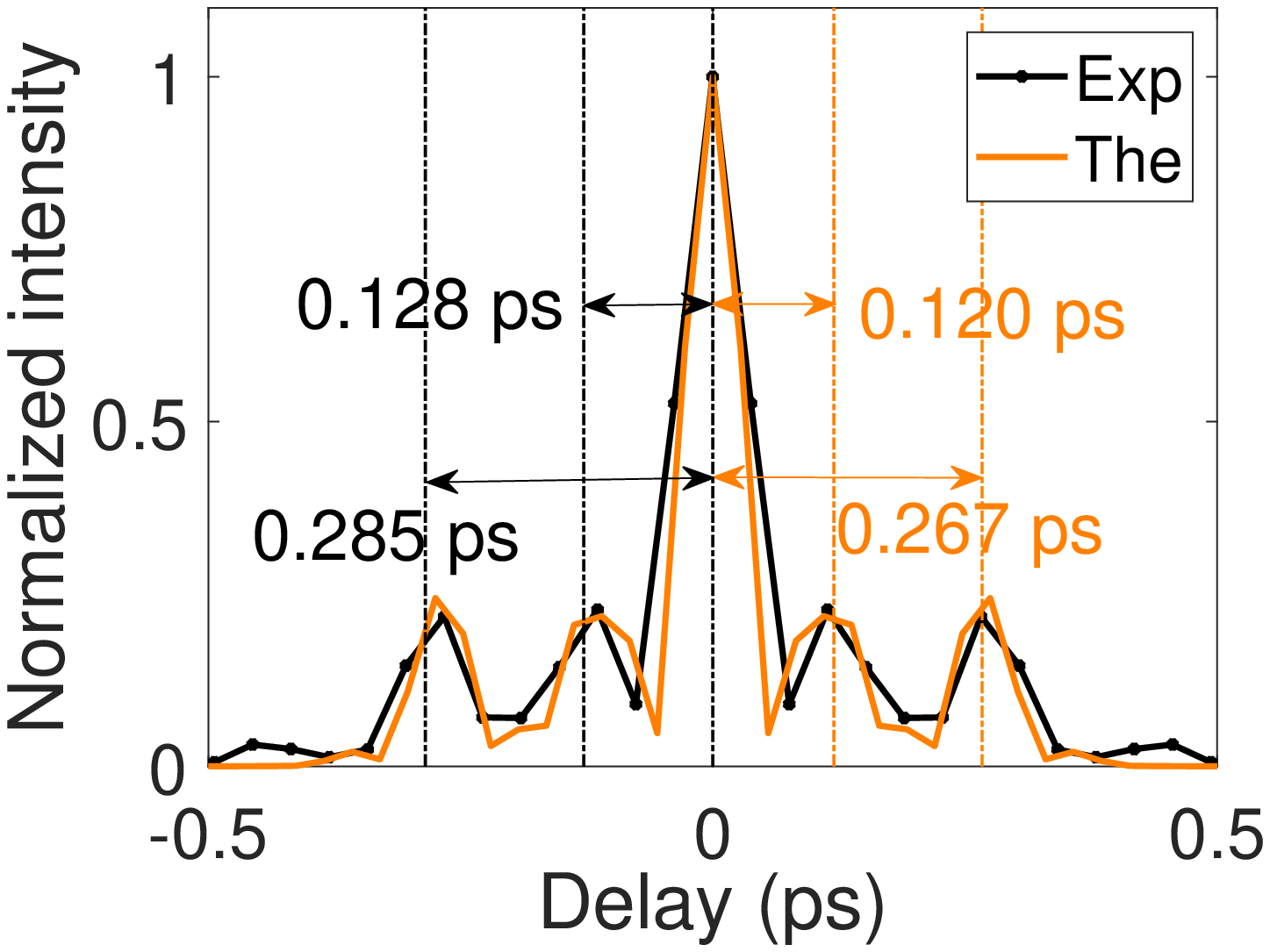}}
\subfigure[]{
\label{Fig6.sub.3}
\includegraphics[width=0.24\linewidth]{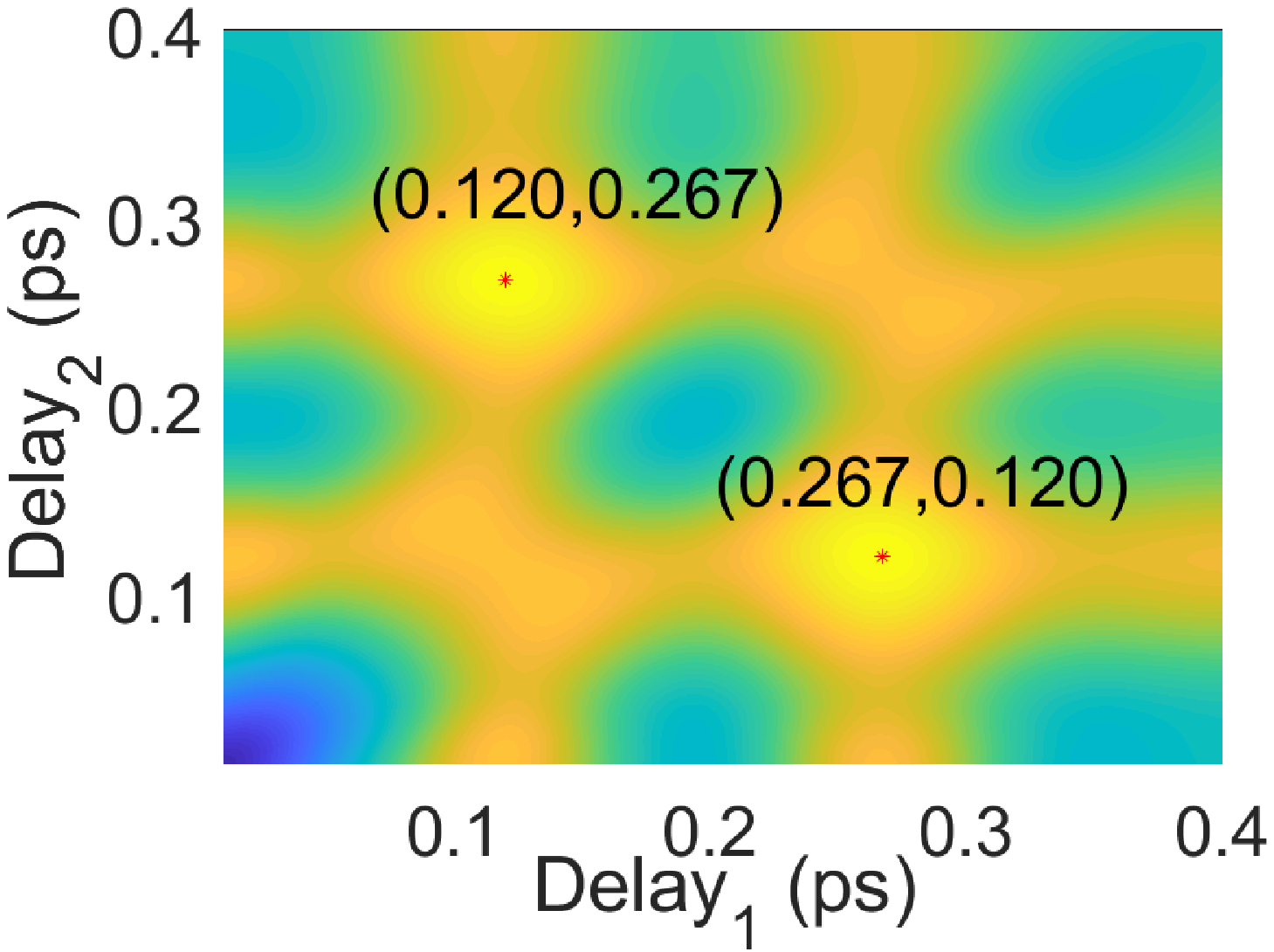}}
\subfigure[]{
\label{Fig6.sub.3}
\includegraphics[width=0.24\linewidth]{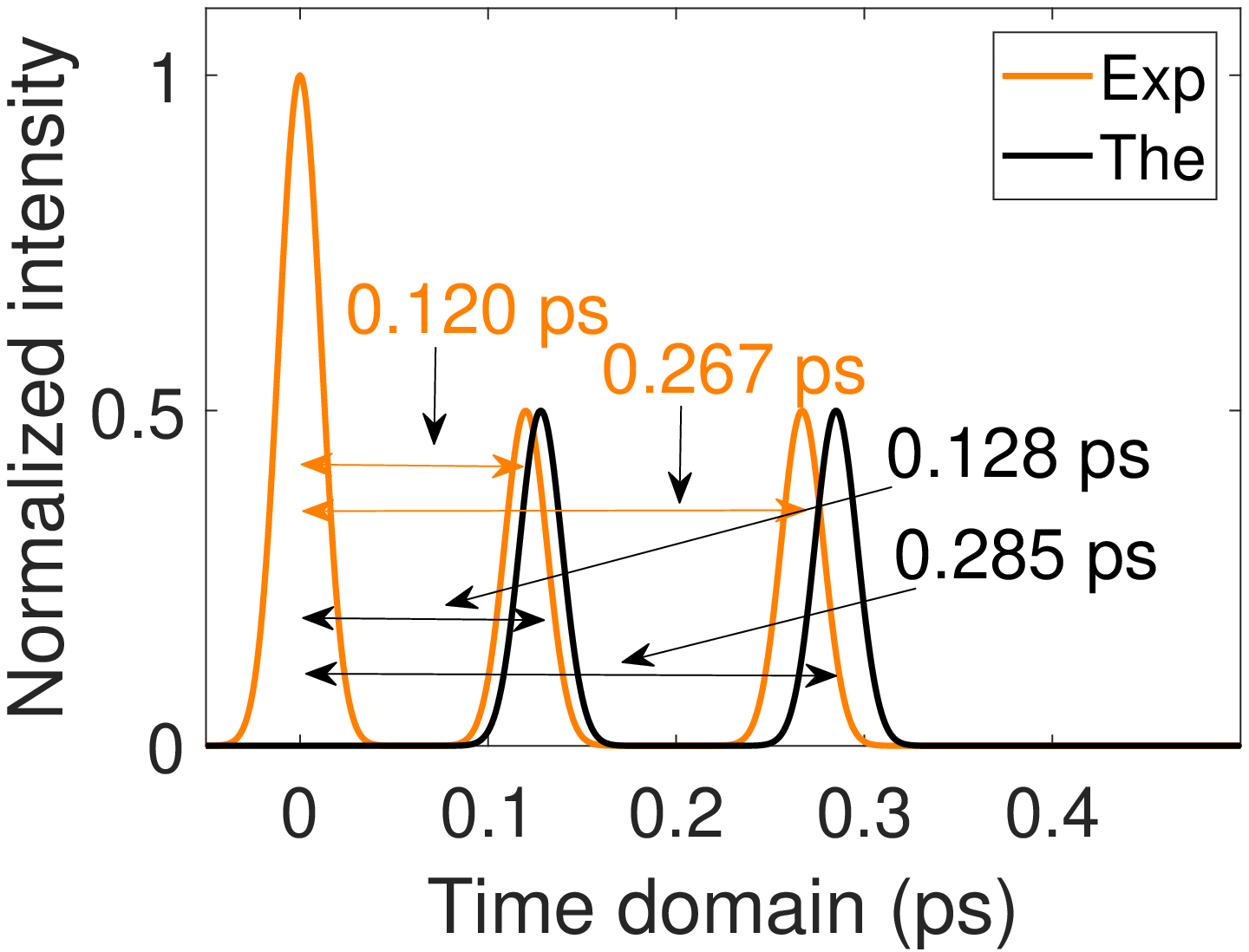}}
\subfigure[]{
\label{Fig6.sub.4}
\includegraphics[width=0.24\linewidth]{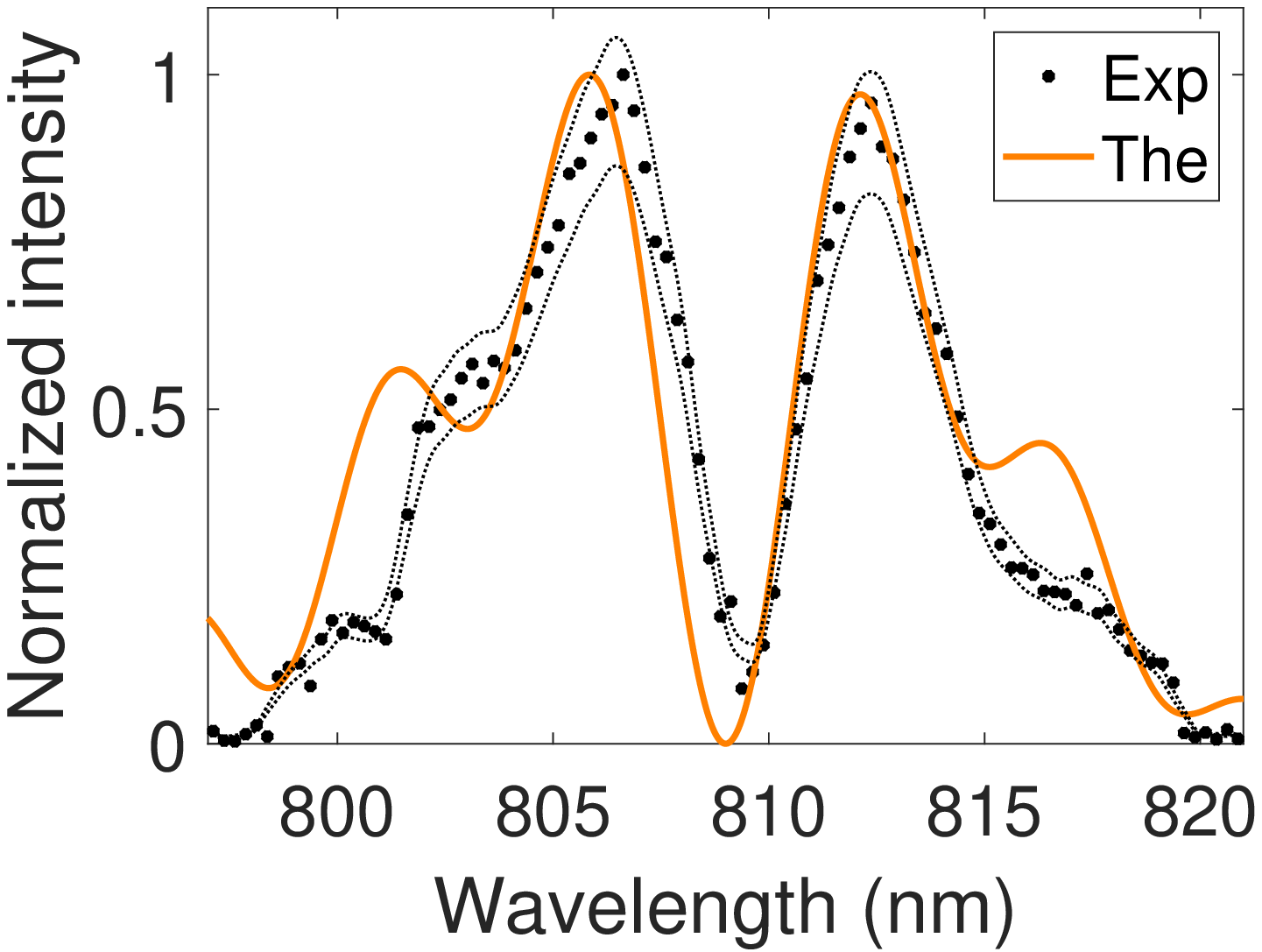}}
\subfigure[]{
\label{Fig6.sub.5}
\includegraphics[width=0.24\linewidth]{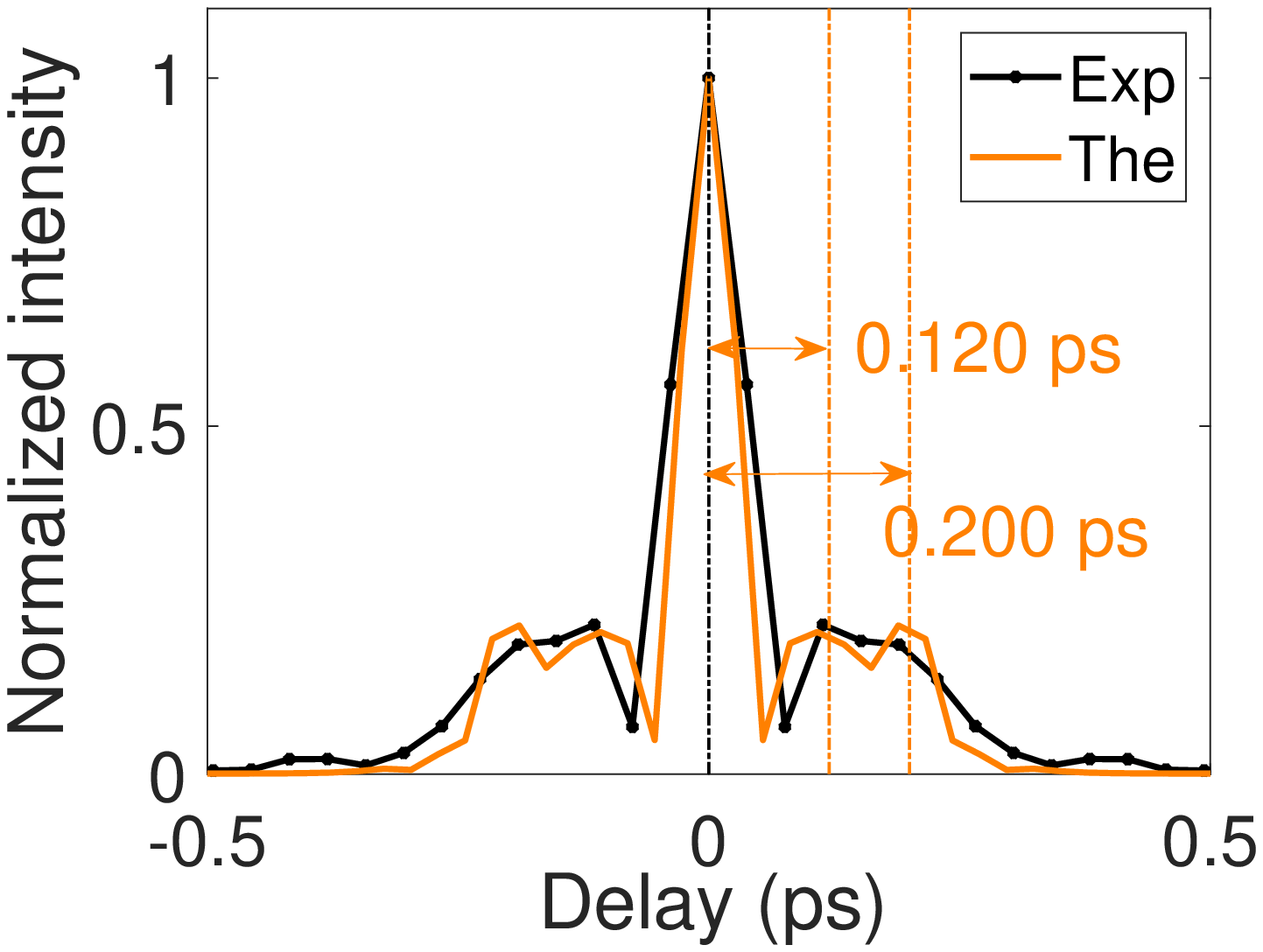}}
\subfigure[]{
\label{Fig6.sub.6}
\includegraphics[width=0.24\linewidth]{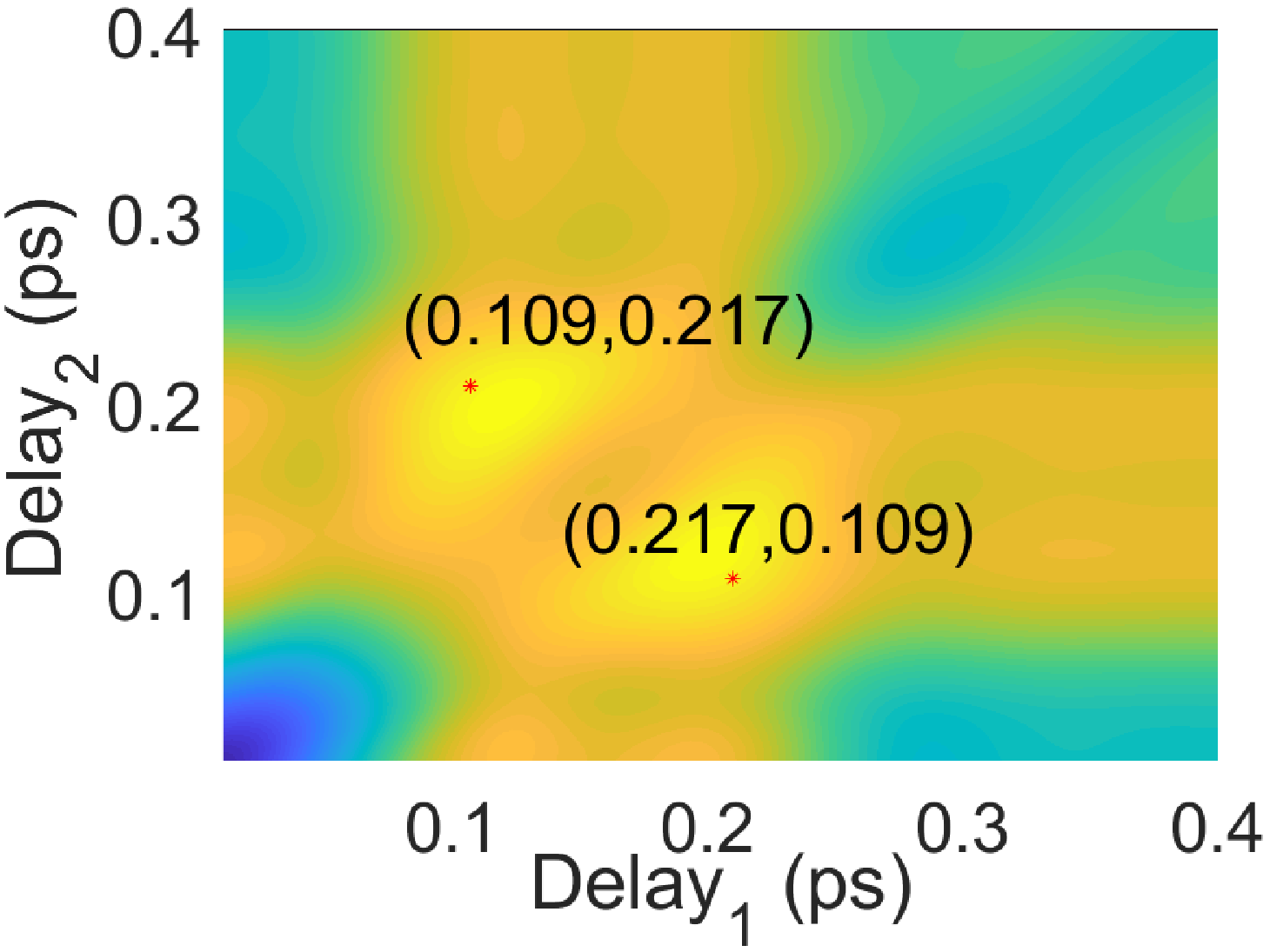}}
\subfigure[]{
\label{Fig6.sub.6}
\includegraphics[width=0.24\linewidth]{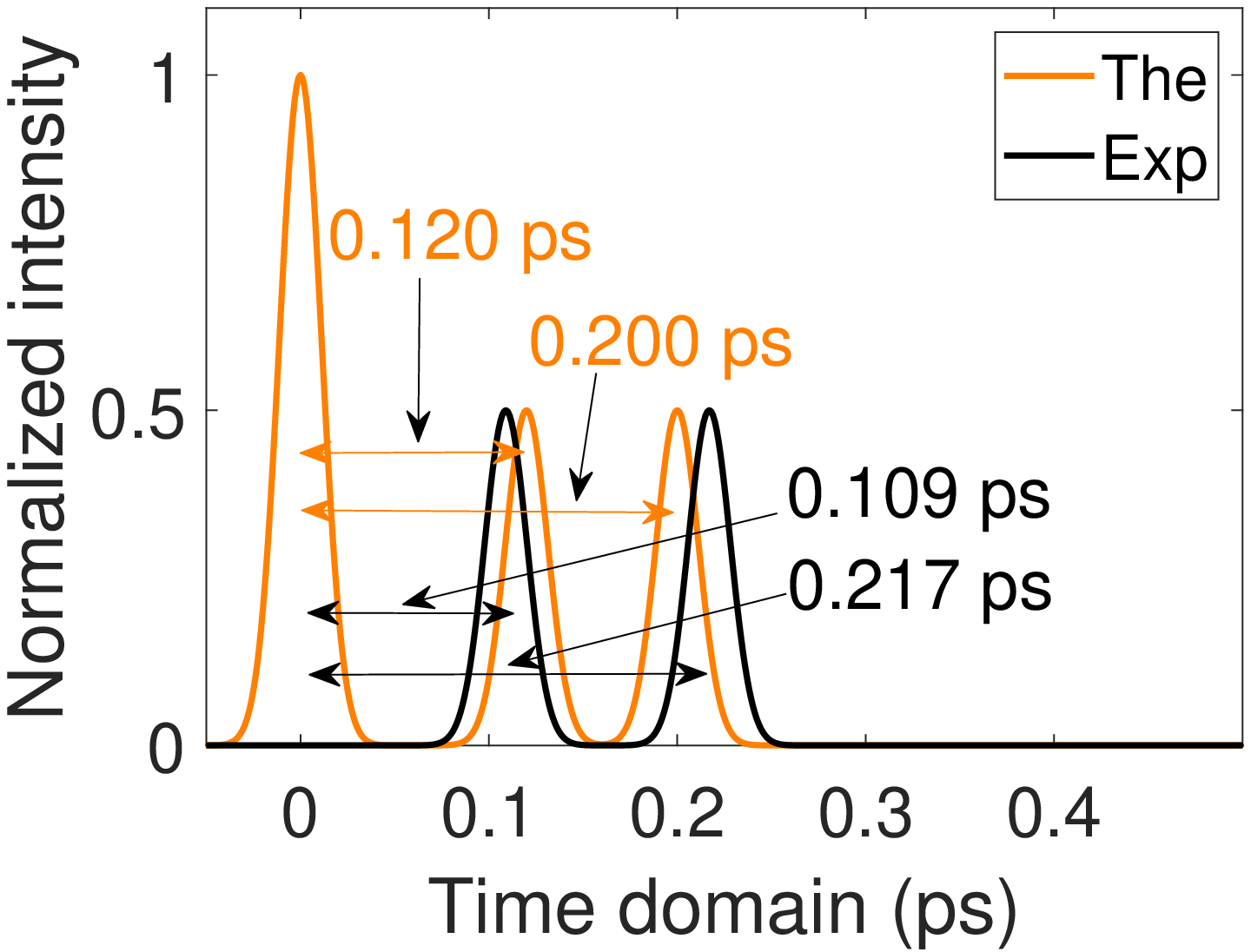}}
\caption{Experimental measurement and theoretical simulation of (a,e) two-photon joint spectral intensities, where the orange lines represent the theoretical predictions, and the black points represent the experimental results that are bounded by the standard deviation estimated by statistical methods assuming a Poisson distribution. (b,f) the cross-correlation functions by making an inverse Fourier transform, (c,g) the optimal results by using maximum likelihood estimation and (d,h) the predicted two-photon relative delays by setting as (a-d) $\Delta T_1=\unit{0.120}{ps}$ and $\Delta T_2=\unit{0.267}{ps}$, (e-h) $\Delta T_1=\unit{0.120}{ps}$ and $\Delta T_2=\unit{0.200}{ps}$.}
\label{figure_6}
\end{figure*}

In our proof-of-concept experiment, we build a home-made single-photon monochromator to scan the frequency correlation. The monochromator is constituted by a holographic grating that spread the spectrum in space and a plano-convex lens that calibrates the spatial distribution of the spectrum. Assisted by a detection array, we are able to obtain the two-photon joint spectral intensity from one experimental trial. As shown in Fig. \ref{figure_5}\textcolor{blue}{(a-d)}, the experimental results exhibit as discrete frequency-bins, wherein the frequency bandwidth of single peaks is determined by time delay. Aside from the sample with uniform thickness, we also experimentally verify the viability of our method by laying two pieces of glass partially on top of each other. Accordingly, the generated two-photon joint spectral intensity turns out to be the accumulation of multiple individual spectral distributions as shown in Fig. \ref{figure_6}\textcolor{blue}{(a,e)}.

\section{Spectral-domain quantum optical coherence tomography}
A concise yet essential application of QWKT is the SD-QOCT, where spectral pattern reveals the longitudinal structural information of test samples. In comparison with time-domain quantum optical coherence tomography \cite{huang1991optical,adrian2000three}, SD-QOCT has no requirement of precise temporal scanning such that it is superior in terms of its capturing speed, signal to noise ratio, and sensitivity \cite{leitgeb2003performance,choma2003sensitivity,siddiqui2018high,marchand2021soliton}. Here we experimentally demonstrate a spectrally-resolved HOM sensor that we use to detect delays that introduced by transparent samples, which can be considered as a time-reversed process of QWKT.

With respect to transparent samples with uniform thickness, their CCFs in QWKT are obtained by applying an inverse Fourier transform on joint spectral intensity as shown in Fig. \ref{figure_5}\textcolor{blue}{(e-h)}. Since the longitudinal information determines the separation distance between two bilateral side peaks, the extracted two-photon relative time delays are shown in Fig. \ref{figure_5}\textcolor{blue}{(i-l)}. As our experiment utilizes a continuous wave laser and the emission of down-converted photons is uncertain in time domain, we note that the start points of time are random.

We also verify that QWKT is appropriate for the exploitation of SD-QOCT in extracting depth information from the transparent samples with non-uniform thickness. A direct method to obtain the target depth parameter is to distinguish the temporal separation of two bilateral side peaks as discussed before. However, if multiple depth parameters are close enough, their side peaks maybe too ambiguous to be separated as shown in Fig. \ref{figure_6}\textcolor{blue}{(b,f)}. To tackle this issue, we use maximum likelihood estimation to obtain the estimators of target longitudinal parameters as shown in Fig. \ref{figure_6}\textcolor{blue}{(c,g)}, and extract their two-photon relative time delays as shown in Fig. \ref{figure_6}\textcolor{blue}{(d,h)}. In contrast to direct observation, maximum likelihood estimation enables us to search for the optimal estimator, which maybe a more efficient method for the statistical analysis.

\begin{figure*}[!t]
\centering
\subfigure[]{
\label{Fig7.sub.1}
\includegraphics[width=0.24\linewidth]{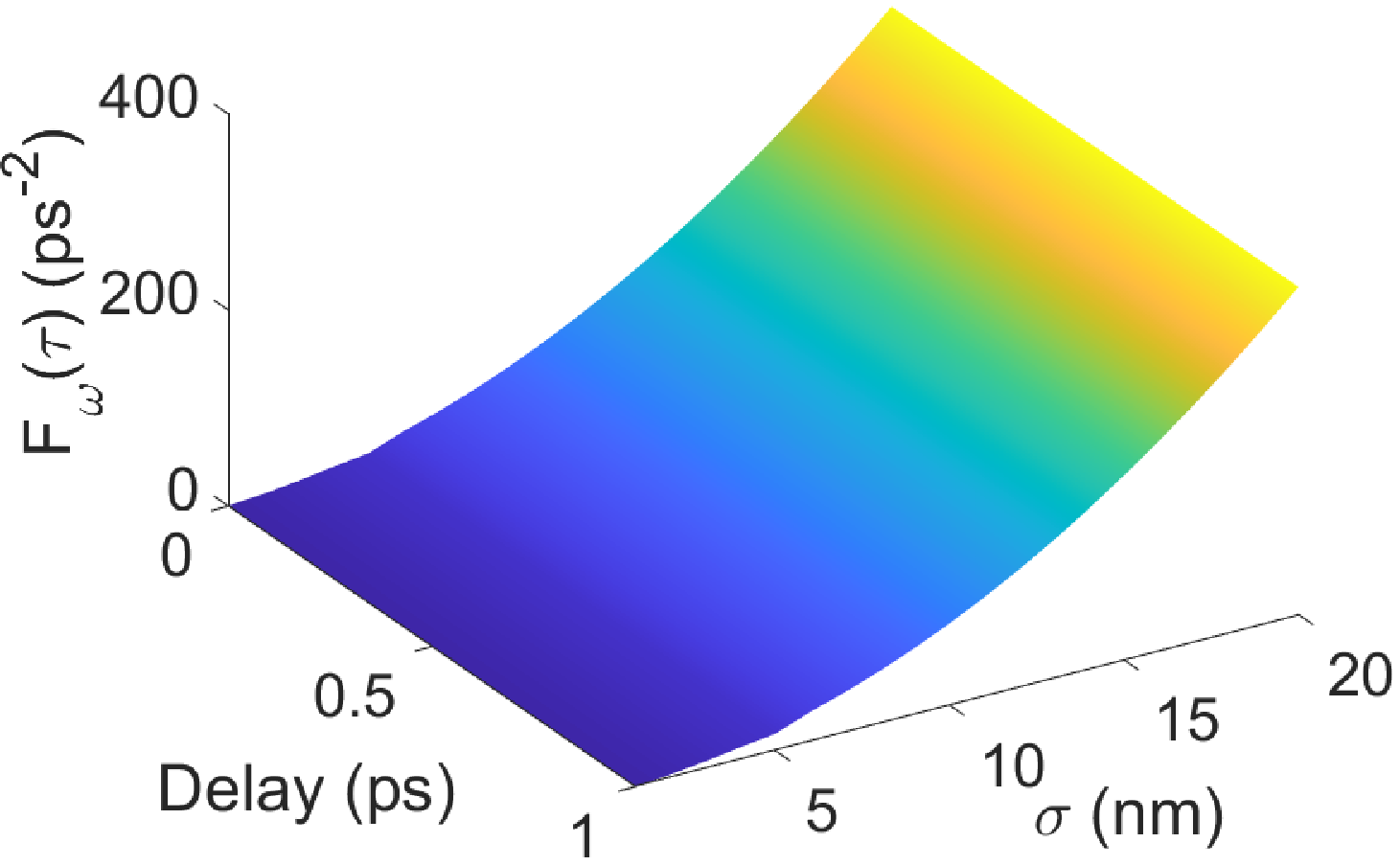}}
\subfigure[]{
\label{Fig7.sub.2}
\includegraphics[width=0.24\linewidth]{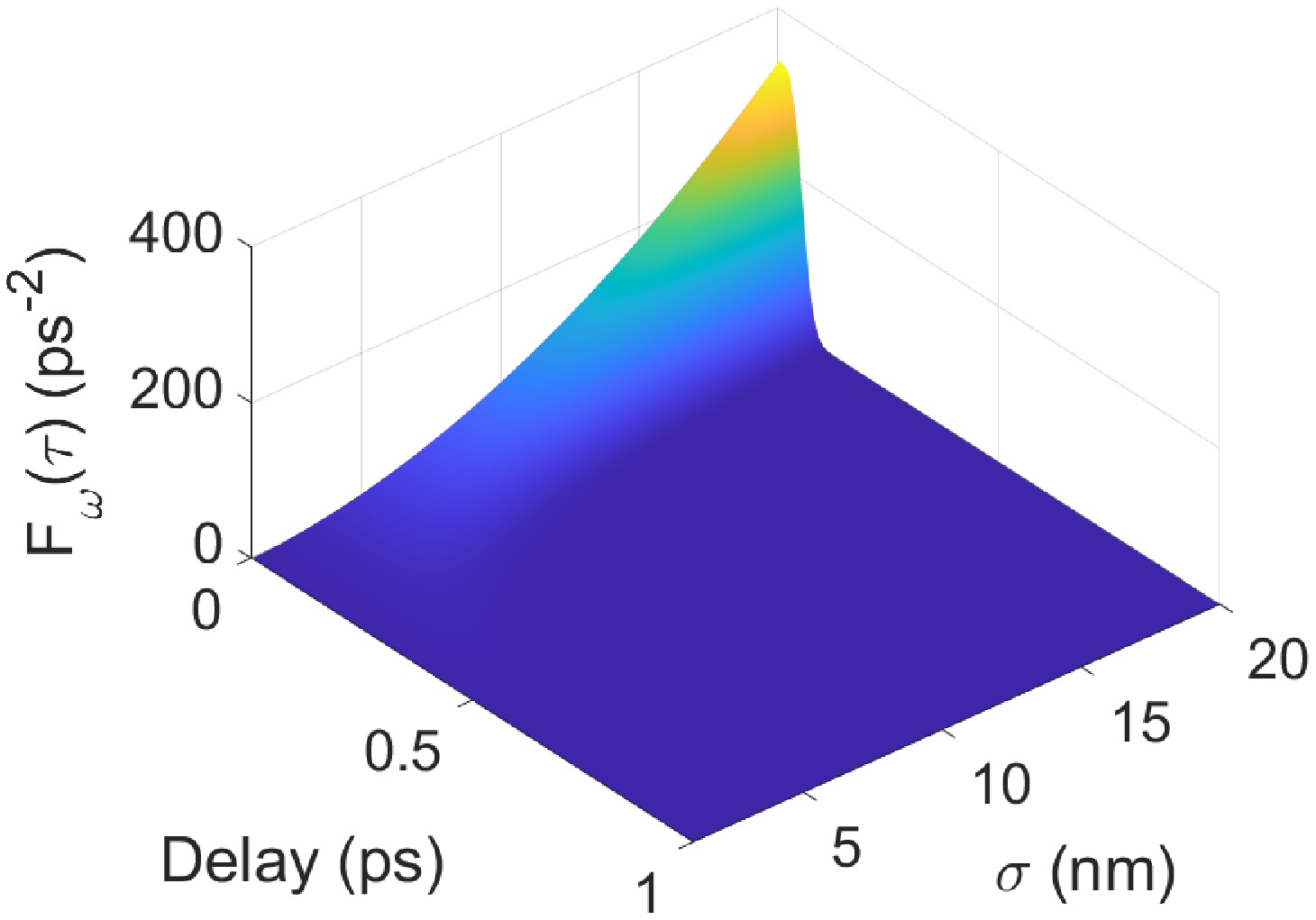}}
\subfigure[]{
\label{Fig7.sub.3}
\includegraphics[width=0.24\linewidth]{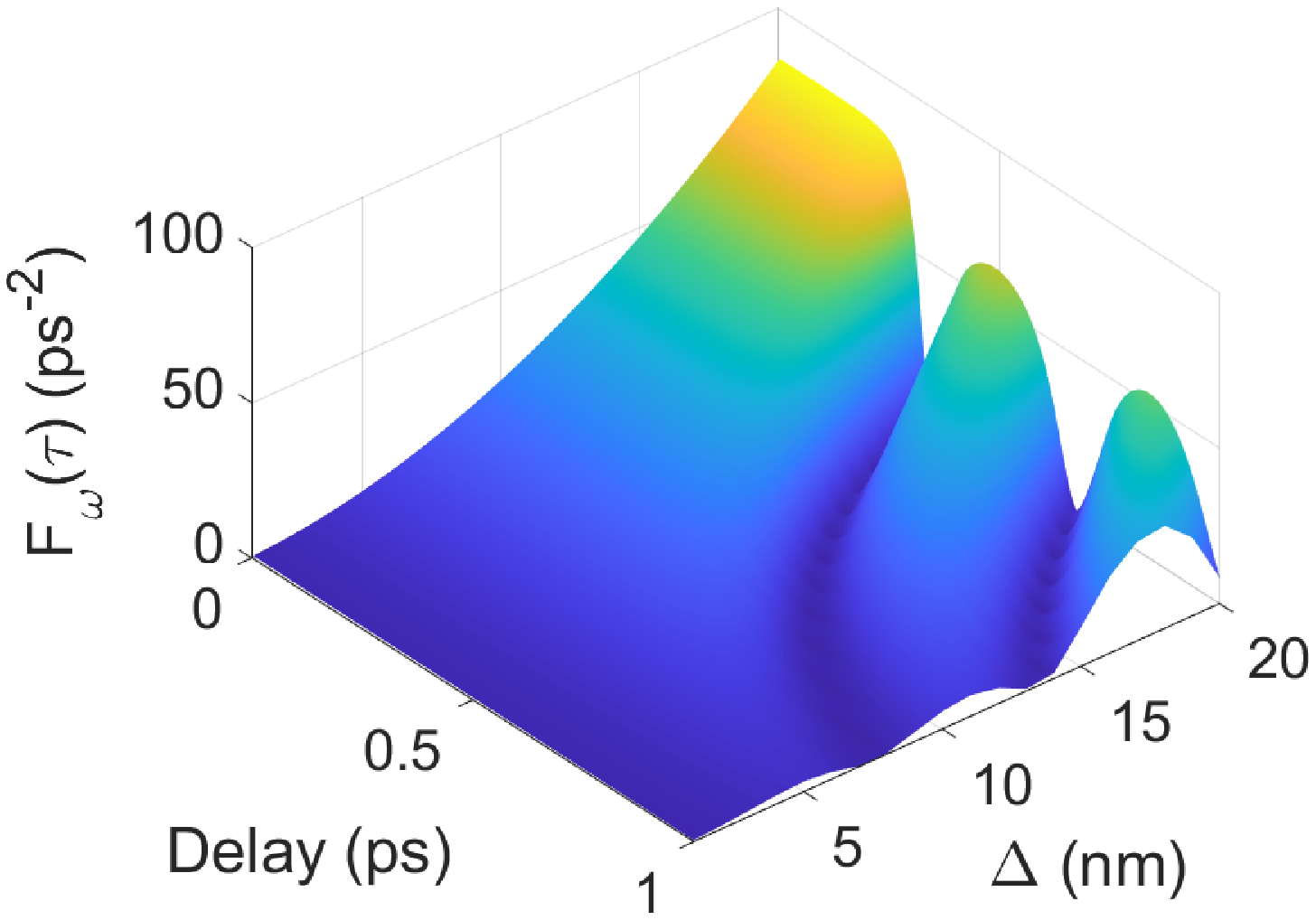}}
\subfigure[]{
\label{Fig7.sub.4}
\includegraphics[width=0.24\linewidth]{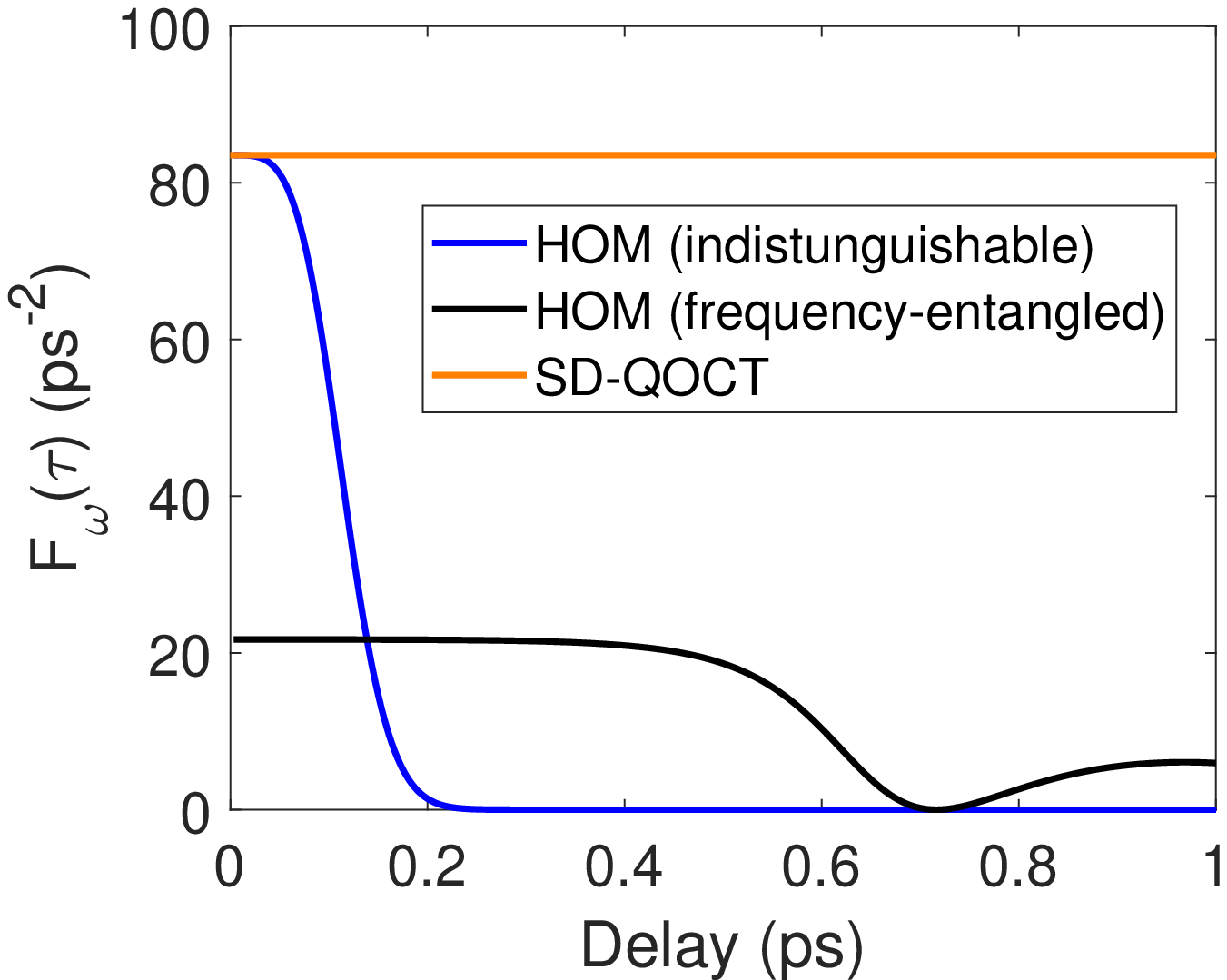}}
\subfigure[]{
\label{Fig7.sub.5}
\includegraphics[width=0.24\linewidth]{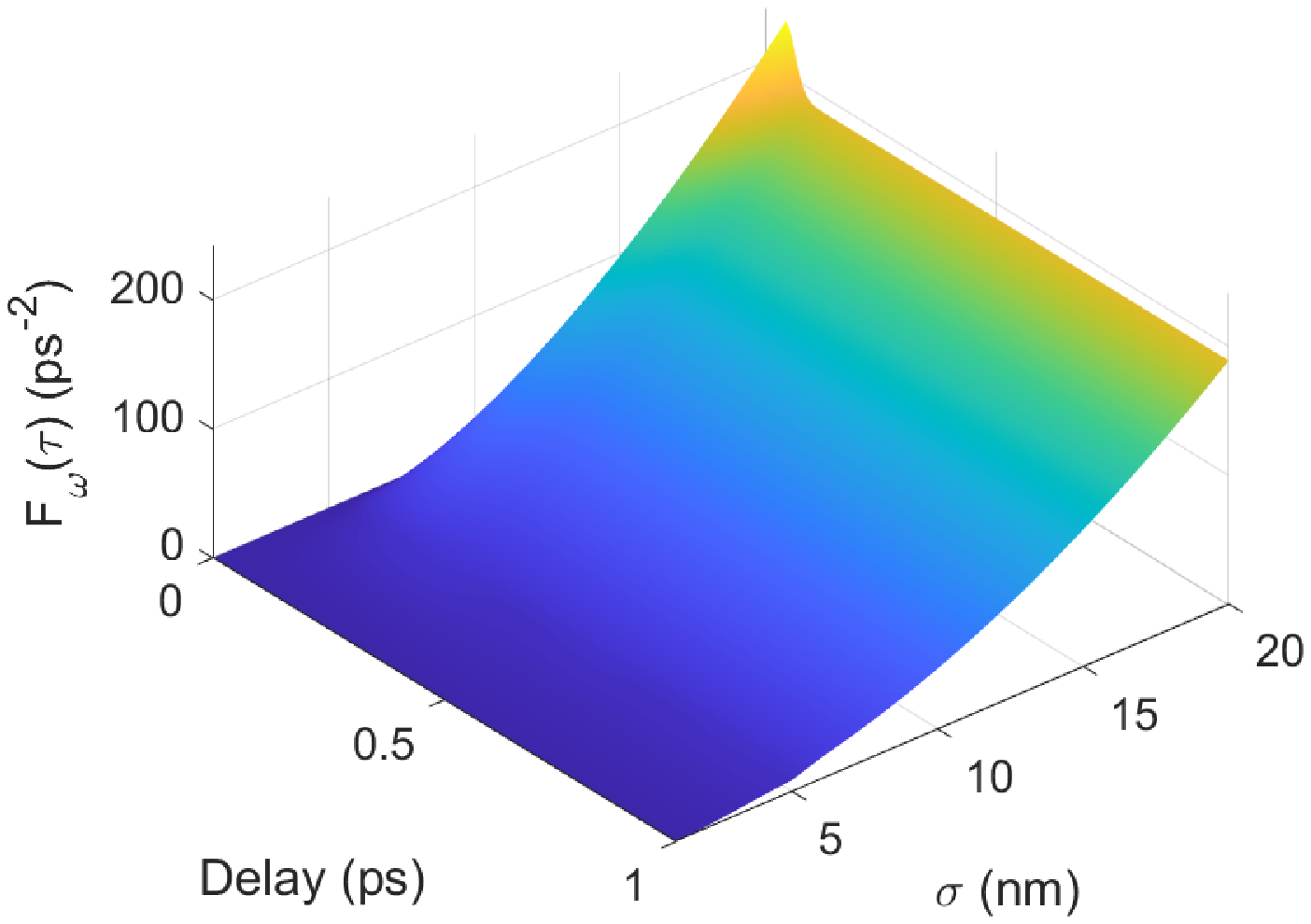}}
\subfigure[]{
\label{Fig7.sub.6}
\includegraphics[width=0.24\linewidth]{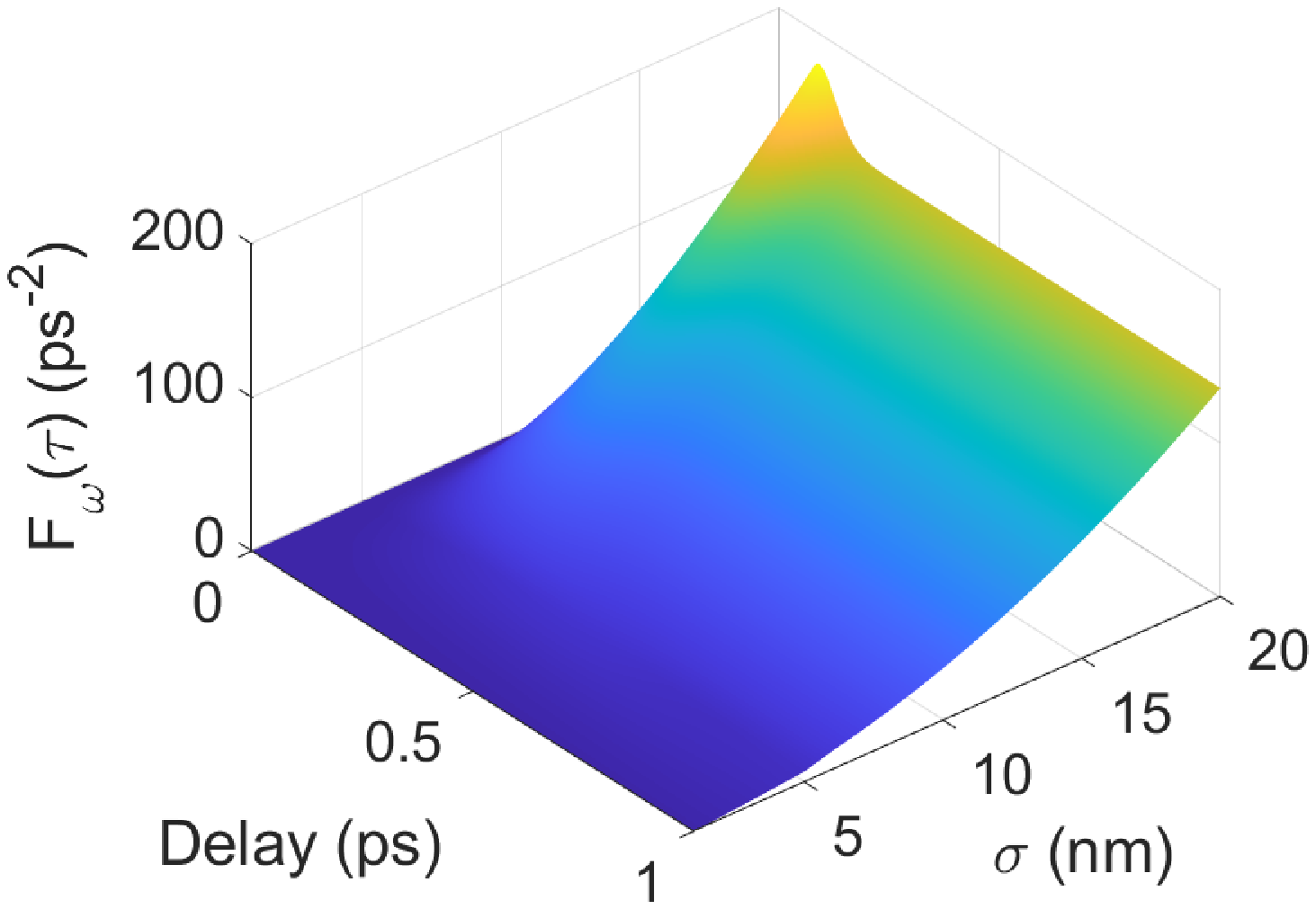}}
\subfigure[]{
\label{Fig7.sub.7}
\includegraphics[width=0.24\linewidth]{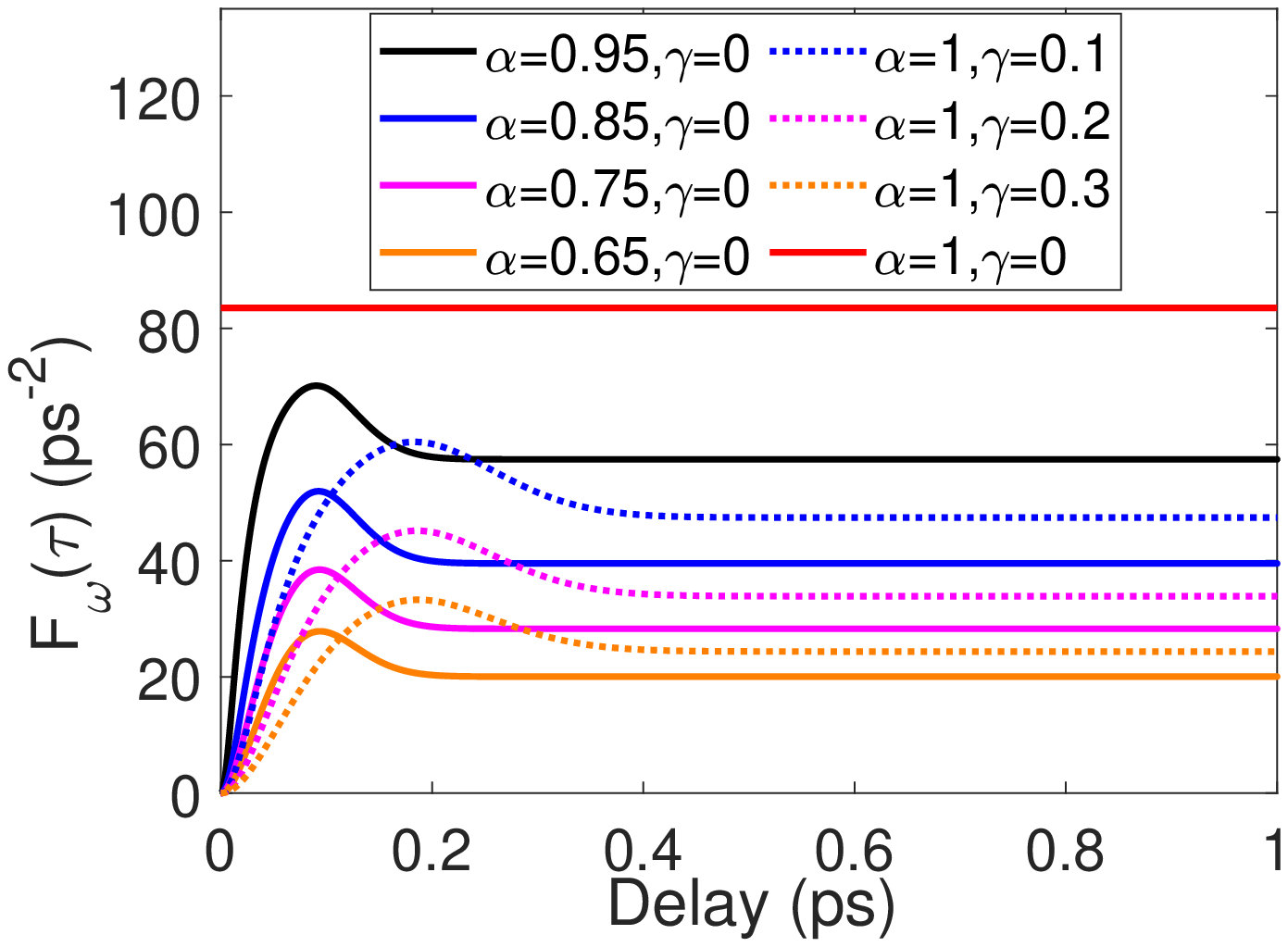}}
\subfigure[]{
\label{Fig7.sub.8}
\includegraphics[width=0.24\linewidth]{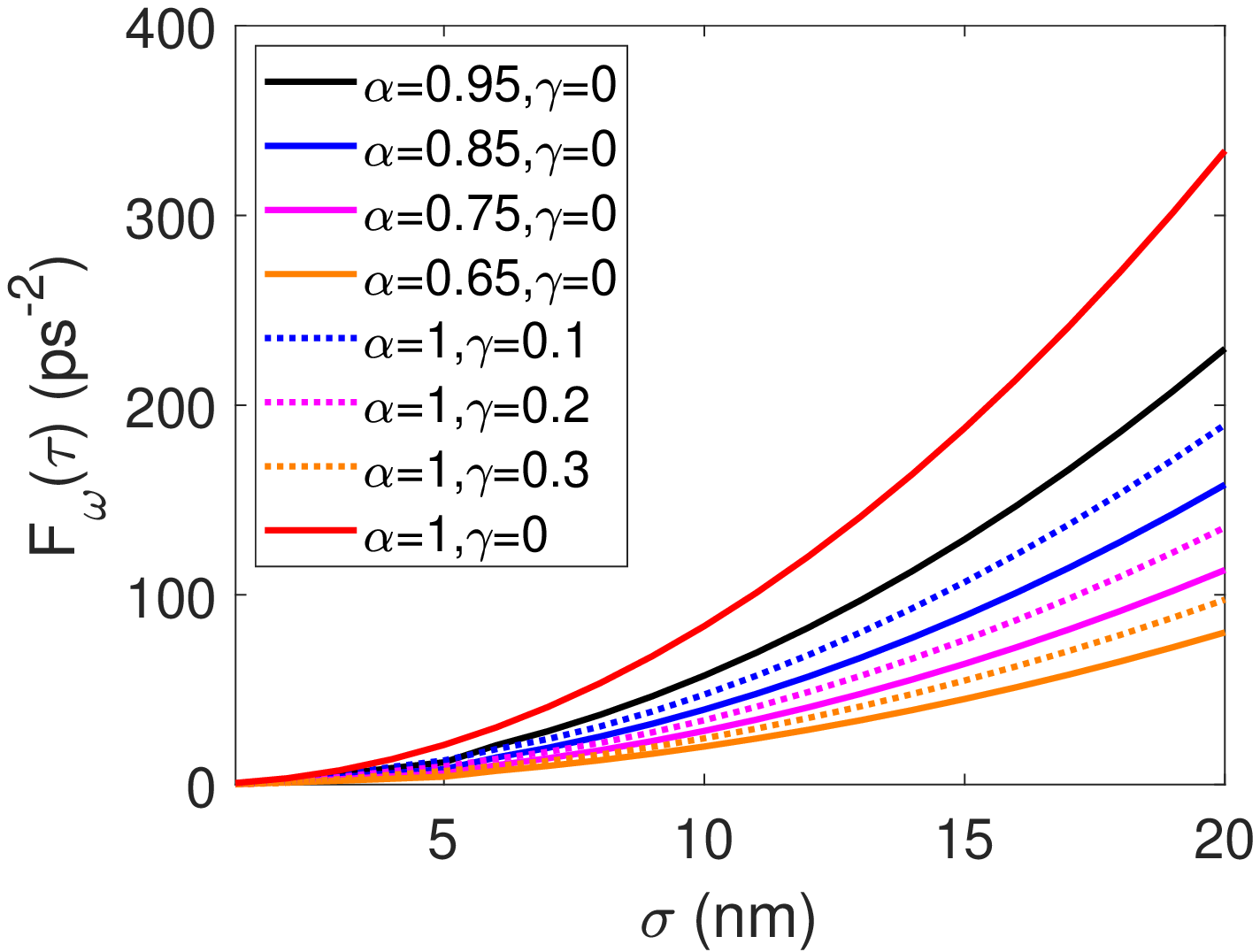}}
\caption{Theoretical prediction of Fisher information in (a) SD-QOCT, HOM interferometry based on (b) indistinguishable photons (HOM )(difference-frequency $\Delta=\unit{0}{nm}$), (c) frequency-entangled photons ($\sigma=\unit{0.5}{nm}$), and (d) their comparison in the case of setting single-photon bandwidth as $\unit{10}{nm}$ in the case of zero loss and perfect visibility. Theoretical prediction of Fisher information in SD-QOCT with (e) visibility of $0.9$, (f) loss of $0.2$. (g) Fisher information as a function of time delay ($\sigma=\unit{10}{nm}$), and (h) as a function of single-photon bandwidth ($\tau=\unit{0.5}{ps}$) for various imperfect visibilities and channel loss rates.}
\label{figure_7}
\end{figure*}

Since the probe state is shown in \eqref{eq:probe}, a fundamental limit for the precision of estimation, so-called Quantum Cram$\acute{e}$r-Rao bound, is obtained as
\begin{equation}\label{eq:cr bound}
\delta\tau\geq\frac{1}{2\sqrt{NQ}},
\end{equation}
where
\begin{equation}
Q=\braket{\frac{\partial\psi(\tau)}{\partial\tau}}{\frac{\partial\psi(\tau)}{\partial\tau}}-|\braket{\psi(\tau)}{\frac{\partial\psi(\tau)}{\partial\tau}}|^2,
\end{equation}
$N$ is the number of experimental trials. It has been proven that for an entangled state with single-photon bandwidth of $\sigma$, the Quantum Cram$\acute{e}$r-Rao bound is obtained as $\delta\tau\geq 1/(2\sigma N^{1/2})$ \cite{lyons2018attosecond,chen2019Hong}. Now we confirm that our SD-QOCT approach can experimentally saturate \eqref{eq:cr bound}. The two-photon joint spectral intensity at the output of HOM interferometer is expressed as \eqref{eq:probability}. In the case of a real HOM interferometer, that is subject to photon loss $\gamma$ and imperfect experimental visibility $\alpha$, there are three possible measurement outcomes, i.e., both photons are detected, one photon is detected, or no photon detected, with corresponding probability distributions as
\begin{equation}
\begin{split}
P_2(\tau,\omega)&=\frac{(1-\gamma)^2}{2}\frac{exp(-\omega^2/8\sigma^2)[1+\alpha cos(\omega\tau)]}{\sqrt{8\pi\sigma^2}}\\
P_1(\tau,\omega)&=\frac{(1-\gamma)^2}{2}[\frac{2(1+\gamma)}{1-\gamma}-\frac{exp(-\omega^2/8\sigma^2)(1+\alpha cos(\omega\tau))}{\sqrt{8\pi\sigma^2}}]\\
P_0(\tau,\omega)&=\gamma^2,
\end{split}
\end{equation}
where subscripts 2, 1, and 0 represent the number of detectors that click as direct results of coincidence, bunching and total loss, respectively. Explicitly, an estimator is a function of the experimental data that allows us to infer the value of the unknown time delay using a particular statistical model for the probability distribution of spectrum. For any such estimator \cite{lyons2018attosecond,chen2019Hong}, classical estimation theory states that standard deviation is lower bounded by $\delta\tau_{CR}=1/[NG_\omega(\tau)]^{1/2}$, where the Fisher information $G_\omega(\tau)$ quantifies the information can be extracted from a particular measurement as
\begin{equation}
\begin{split}
G_\omega(\tau)=&\int_{-\infty}^\infty\frac{[\partial_\tau P_2(\tau,\omega)]^2}{P_2(\tau,\omega)}+\frac{[\partial_\tau P_1(\tau,\omega)]^2}{P_1(\tau,\omega)}\\
&+\frac{[\partial_\tau P_0(\tau,\omega)]^2}{P_0(\tau,\omega)}d\omega.
\end{split}
\end{equation}
It indicates that the limit of Fisher information depends on both of a particular quantum state and a specific measurement strategy. In the case of zero loss and perfect visibility, its upper bound is achieved as $G_\omega(\tau)=4\sigma^2$, which indicates that we can recover the Quantum Cram$\acute{e}$r-Rao bound, and thus confirming that the SD-QOCT is an optimal measurement strategy. While this ultimate limit of $G_\omega(\tau)$ is independent of time delay, the maximal Fisher information that can be obtained in practical measurements is severely limited by experimental imperfections, which makes the optimal sensing position is closely relevant to the time delay (see Fig.\ \ref{Fig7.sub.7}). Note that the estimation that involves observable quantities $N$, $\gamma$, $\alpha$ and $\sigma$ need to be separately estimated before the measurements begin.
The theoretical simulation of Fisher information as functions of single-photon bandwidth $\sigma$ and target delay to be estimated is shown in Fig. \ref{Fig7.sub.1}. Since broader bandwidth leads to clearer distinguishment in spectral distribution, $G_\omega(\tau)$ increases with respect to single-photon bandwidth.

While HOM interferometry holds great promise for sensing schemes that require precise knowledge of optical delays, its dynamic range is limited by single-photon bandwidth in practical experiment (see Fig. \ref{Fig7.sub.2}). It is generally assumed that great precision in the measurement requires that photons contain a large bandwidth \cite{lyons2018attosecond}. Nevertheless, this well-known assumption is challenged and show that the use of well-separated color entanglement suffices to achieve great precision (see Fig. \ref{Fig7.sub.3}). It shows that the precision with which the delays can be measured is mainly determined not only by the coherence time of single photons, but also by the separation distance of the center frequencies of the frequency-entangled state \cite{chen2019Hong}.

As shown in Fig.\ \ref{Fig7.sub.4}, $G_\omega(\tau)$ obtained in SD-QOCT can reach the ultimate bound in measurement precision that is allowed by the probe state. With comparison to the conventional HOM measurement strategy, SD-QOCT offers an avenue to detect longitudinal structural information within a wide dynamic range and with provable advantages in the precision and sensitivity. This enhancement can be attributed to the exploitation of spectral analysis that is mainly determined by single-photon spectrometer, whereas temporal measurement in HOM interferometry is severely limited by single-photon bandwidth. Additionally, both of the HOM interferometric schemes based on indistinguishable photons and frequency-entangled photons require the precise scanning in time domain, our SD-QOCT has a great advantage in shortening the capturing time that is realized through simultaneous measurement of spectral distributions by using detection array and without the strict requirement of temporal scanning.

In the case of non-zero loss (see Fig.\ \ref{Fig7.sub.5}) and imperfect visibility (see Fig.\ \ref{Fig7.sub.6}), SD-QOCT is unable to reach the Quantum Cram$\acute{e}$r-Rao bound, and the specific Fisher information depends on both of single-photon bandwidth and time delay. As shown in Fig.\ \ref{Fig7.sub.7}), the dynamic range of SD-QOCT is still quite wide, but the precision would drops to zero in the case of $\tau\rightarrow0$. However, the ultimate precision still increases as a function of single-photon bandwidth as shown in Fig.\ \ref{Fig7.sub.8}). Thus, SD-QOCT still can offer provable advantages in precision and sensitivity in measuring the longitudinal length of transparent samples within a wide dynamic range even in the practical experimental applications.
\section{Discussion}
Backed by a mathematical definition of quantum Wiener-Khinchin theorem that explains the connection between two-photon relative temporal signal and two-photon joint spectral intensity, we report on its experimental implementation in the frequency-entangled two-photon interference by using a spectrally-resolved HOM interferometer. In addition, a time-reversed process of QWKT can be used to implement a resource-efficient SD-QOCT, including but not limited to the precise measurement of depth information of transparent samples with uniform or non-uniform thickness.

This connection in HOM interference may be extended to other degrees of freedom, such as angle and orbital angular momentum that are linked by discrete Fourier series \cite{Franke2004uncertainty} and radial position and radial momentum that are linked by quantum Mellin transform \cite{twamley2006quantum}. We hope this study spurs further investigation into the translation between different degrees of freedom of photons, which can be exploited in quantum metrology for greater performance even in the presence of excess noise.

In conclusion, we believe that fully harnessing QWKT and its experimental demonstration in frequency-entangled two-photon HOM interference will provide additional tools, e.g., quantum interferometric spectroscopy, ultimately broadening the path towards practical quantum information processing and quantum metrology. For example, with respect to quantum information processing, temporal distinguishability in HOM interference can be harnessed to prepare and characterize high-dimensional frequency entanglement \cite{chen2020Temporal,jin2016simple,xie2015Harnessing}. With respect to quantum metrology, entanglement-assisted absorption spectroscopy based on quantum interference has been used to detect the absorptive properties of materials and molecules, which has the potential to provide quantum advantages in robustness against noise and loss \cite{jin2018extended,chen2022entanglement,shi2020entanglement}. In addition, entanglement-based optical coherence tomography have inspired many unique applications such as quantum sensing with undetected photons, which enables the manipulation and detection of photons by harnessing its parter photons with a completely different wavelength, in particular for commercial near-IR-OCT systems \cite{aron2020frequency}.

\section{Acknowledgements}
This work is supported by the National Natural Science Foundation of China (NSFC) (12034016, 12004318, 61975169), the Fundamental Research Funds for the Central Universities at Xiamen University (20720190057, 20720210096), the Natural Science Foundation of Fujian Province of China (2020J05004), the Natural Science Foundation of Fujian Province of China for Distinguished Young Scientists (2015J06002), and the program for New Century Excellent Talents in University of China (NCET-13-0495).

\appendix
\section{HOM interference with frequency-entangled states}
In the ideal case, the HOM interference measurement is accomplished by using a lossless and balanced beam splitter. The beam splitter transformation on the input modes can be expressed by
\begin{equation}
\begin{split}
\hat{a}_s^\dag(\omega_s)=\frac{1}{\sqrt{2}}[\hat{a}_1^\dag(\omega_s)+\hat{a}_2^\dag(\omega_s)]\\
\hat{a}_i^\dag(\omega_s)=\frac{1}{\sqrt{2}}[\hat{a}_1^\dag(\omega_i)-\hat{a}_2^\dag(\omega_i)].\\
\end{split}
\end{equation}
As a relative time delay $\tau$ is introduced by a transparent sample, it results in a phase shift $exp(-i\omega_i\tau_1)$ that transmits the two-photon state as $\ket{\psi(\tau)}\rightarrow\ket{\psi_A(\tau)}+\ket{\psi_B(\tau)}$
\begin{equation}
\begin{split}
\ket{\psi(\tau)}=&\frac{1}{2}\int_0^\infty\int_0^\infty d\omega_sd\omega_if(\omega_s,\omega_i)e^{-i\omega_i\tau_1}[i\hat{a}_1^\dag(\omega_s)\hat{a}_1^\dag(\omega_i)\\
&+i\hat{a}_2^\dag(\omega_s)\hat{a}_2^\dag(\omega_i)+\hat{a}_1^\dag(\omega_i)\hat{a}_2^\dag(\omega_s)-\hat{a}_1^\dag(\omega_s)\hat{a}_2^\dag(\omega_i)]\ket{0},
\end{split}
\end{equation}
where subscript 1/2 represent two output modes of the beam splitter, and
\begin{equation}
\begin{split}
\ket{\psi_A(\tau)}=&\frac{1}{2}\int_0^\infty\int_0^\infty d\omega_sd\omega_i[f(\omega_s,\omega_i)e^{-i\omega_i\tau}\\
&-f(\omega_i,\omega_s)e^{-i\omega_s\tau}]\hat{a}_1^\dag(\omega_s)\hat{a}_2^\dag(\omega_i)\ket{0},\\
\ket{\psi_B(\tau)}=&\frac{1}{2}\int_0^\infty\int_0^\infty d\omega_sd\omega_i[f(\omega_s,\omega_i)e^{-i\omega_i\tau}-f(\omega_i,\omega_s)\\
&e^{-i\omega_s\tau}](\hat{a}_1^\dag(\omega_s)\hat{a}_1^\dag(\omega_i)+\hat{a}_2^\dag(\omega_s)\hat{a}_2^\dag(\omega_i))\ket{0}.\\
\end{split}
\end{equation}
Since the spectrally-resolved HOM interference pattern is identify in opposite spatial modes resulting from anti-bunching effect, we focus on the calculation of by $P_c(\tau)=|\braket{\psi(\tau)}{\psi_A(\tau)}|^2$. It is obviously that two photons after the beam splitter are indistinguishable, we substitute $\omega_s$ and $\omega_i$ with $\omega_1$ and $\omega_2$. By multiplying $e^{-i\omega_i\tau}$ to cancel the global phase, we obtain
\begin{equation}
\begin{split}
\ket{\psi_A(\tau)}=&\frac{1}{2}\int_0^\infty\int_0^\infty d\omega_1d\omega_2[f(\omega_1,\omega_2)-f(\omega_2,\omega_1)\\
&e^{-i(\omega_1-\omega_2)\tau}]\hat{a}_1^\dag(\omega_1)\hat{a}_2^\dag(\omega_2)\ket{0}.
\end{split}
\end{equation}
The detection operators of two detectors in different output modes are
\begin{equation}
\begin{split}
\hat{E}_1^{(+)}=\frac{1}{\sqrt{2\pi}}\int_0^\infty d\omega_1\hat{a}_1(\omega_1)e^{-i\omega_1t_1},\\
\hat{E}_2^{(+)}=\frac{1}{\sqrt{2\pi}}\int_0^\infty d\omega_2\hat{a}_2(\omega_2)e^{-i\omega_2t_2}.\\
\end{split}
\end{equation}
Thus the normalized coincidence probability $P(\tau)$ as a function of time delay can be expressed as
\begin{equation}
\begin{split}
P(\tau)=&\langle\psi_A(\tau)|\hat{E}_1^{(-)}\hat{E}_2^{(-)}\hat{E}_2^{(+)}\hat{E}_1^{(+)}\ket{\psi_A(\tau)},
\end{split}
\end{equation}
and the simplified result is shown in \eqref{eq: hom interference}.

\newpage
\bibliography{apssamp}

\begin{thebibliography}{38}%
\makeatletter
\providecommand \@ifxundefined [1]{%
 \@ifx{#1\undefined}
}%
\providecommand \@ifnum [1]{%
 \ifnum #1\expandafter \@firstoftwo
 \else \expandafter \@secondoftwo
 \fi
}%
\providecommand \@ifx [1]{%
 \ifx #1\expandafter \@firstoftwo
 \else \expandafter \@secondoftwo
 \fi
}%
\providecommand \natexlab [1]{#1}%
\providecommand \enquote  [1]{``#1''}%
\providecommand \bibnamefont  [1]{#1}%
\providecommand \bibfnamefont [1]{#1}%
\providecommand \citenamefont [1]{#1}%
\providecommand \href@noop [0]{\@secondoftwo}%
\providecommand \href [0]{\begingroup \@sanitize@url \@href}%
\providecommand \@href[1]{\@@startlink{#1}\@@href}%
\providecommand \@@href[1]{\endgroup#1\@@endlink}%
\providecommand \@sanitize@url [0]{\catcode `\\12\catcode `\$12\catcode
  `\&12\catcode `\#12\catcode `\^12\catcode `\_12\catcode `\%12\relax}%
\providecommand \@@startlink[1]{}%
\providecommand \@@endlink[0]{}%
\providecommand \url  [0]{\begingroup\@sanitize@url \@url }%
\providecommand \@url [1]{\endgroup\@href {#1}{\urlprefix }}%
\providecommand \urlprefix  [0]{URL }%
\providecommand \Eprint [0]{\href }%
\providecommand \doibase [0]{http://dx.doi.org/}%
\providecommand \selectlanguage [0]{\@gobble}%
\providecommand \bibinfo  [0]{\@secondoftwo}%
\providecommand \bibfield  [0]{\@secondoftwo}%
\providecommand \translation [1]{[#1]}%
\providecommand \BibitemOpen [0]{}%
\providecommand \bibitemStop [0]{}%
\providecommand \bibitemNoStop [0]{.\EOS\space}%
\providecommand \EOS [0]{\spacefactor3000\relax}%
\providecommand \BibitemShut  [1]{\csname bibitem#1\endcsname}%
\let\auto@bib@innerbib\@empty
\bibitem [{\citenamefont {Wiener}(1930)}]{wiener1930generalized}%
  \BibitemOpen
  \bibfield  {author} {\bibinfo {author} {\bibfnamefont {Norbert}\ \bibnamefont
  {Wiener}},\ }\bibfield  {title} {\enquote {\bibinfo {title} {Generalized
  harmonic analysis},}\ }\href@noop {} {\bibfield  {journal} {\bibinfo
  {journal} {Acta mathematica}\ }\textbf {\bibinfo {volume} {55}},\ \bibinfo
  {pages} {117--258} (\bibinfo {year} {1930})}\BibitemShut {NoStop}%
\bibitem [{\citenamefont {Khintchine}(1934)}]{khinchin1934korrelationstheorie}%
  \BibitemOpen
  \bibfield  {author} {\bibinfo {author} {\bibfnamefont {Alexander}\
  \bibnamefont {Khintchine}},\ }\bibfield  {title} {\enquote {\bibinfo {title}
  {Korrelationstheorie der station{\"a}ren stochastischen prozesse},}\
  }\href@noop {} {\bibfield  {journal} {\bibinfo  {journal} {Mathematische
  Annalen}\ }\textbf {\bibinfo {volume} {109}},\ \bibinfo {pages} {604--615}
  (\bibinfo {year} {1934})}\BibitemShut {NoStop}%
\bibitem [{\citenamefont {Kubo}\ \emph {et~al.}(2012)\citenamefont {Kubo},
  \citenamefont {Toda},\ and\ \citenamefont
  {Hashitsume}}]{kubo1985statistical}%
  \BibitemOpen
  \bibfield  {author} {\bibinfo {author} {\bibfnamefont {Ryogo}\ \bibnamefont
  {Kubo}}, \bibinfo {author} {\bibfnamefont {Morikazu}\ \bibnamefont {Toda}}, \
  and\ \bibinfo {author} {\bibfnamefont {Natsuki}\ \bibnamefont {Hashitsume}},\
  }\href@noop {} {\emph {\bibinfo {title} {Statistical physics II:
  nonequilibrium statistical mechanics}}},\ Vol.~\bibinfo {volume} {31}\
  (\bibinfo  {publisher} {Springer Science \& Business Media},\ \bibinfo {year}
  {2012})\BibitemShut {NoStop}%
\bibitem [{\citenamefont {Davis}\ \emph {et~al.}(2013)\citenamefont {Davis},
  \citenamefont {Abrams},\ and\ \citenamefont {Brault}}]{davis2013fourier}%
  \BibitemOpen
  \bibfield  {author} {\bibinfo {author} {\bibfnamefont {S.~P.}\ \bibnamefont
  {Davis}}, \bibinfo {author} {\bibfnamefont {M.~C.}\ \bibnamefont {Abrams}}, \
  and\ \bibinfo {author} {\bibfnamefont {J.~W.}\ \bibnamefont {Brault}},\
  }\bibfield  {title} {\enquote {\bibinfo {title} {Fourier transform
  spectrometry},}\ }\href@noop {} {\bibfield  {journal} {\bibinfo  {journal}
  {ELSEVIER}\ } (\bibinfo {year} {2013})}\BibitemShut {NoStop}%
\bibitem [{\citenamefont {Griffiths}(1983)}]{griffiths1983fourier}%
  \BibitemOpen
  \bibfield  {author} {\bibinfo {author} {\bibfnamefont {Peter~R}\ \bibnamefont
  {Griffiths}},\ }\bibfield  {title} {\enquote {\bibinfo {title} {Fourier
  transform infrared spectrometry},}\ }\href {\doibase 10.1126/science.6623077}
  {\bibfield  {journal} {\bibinfo  {journal} {Science}\ }\textbf {\bibinfo
  {volume} {222}},\ \bibinfo {pages} {297--302} (\bibinfo {year}
  {1983})}\BibitemShut {NoStop}%
\bibitem [{\citenamefont {Yu}\ \emph {et~al.}(2018)\citenamefont {Yu},
  \citenamefont {Lu}, \citenamefont {Tang}, \citenamefont {Yuan}, \citenamefont
  {Yuan},\ and\ \citenamefont {Yu}}]{yu2018application}%
  \BibitemOpen
  \bibfield  {author} {\bibinfo {author} {\bibfnamefont {Shasha}\ \bibnamefont
  {Yu}}, \bibinfo {author} {\bibfnamefont {Chengzhe}\ \bibnamefont {Lu}},
  \bibinfo {author} {\bibfnamefont {Xin}\ \bibnamefont {Tang}}, \bibinfo
  {author} {\bibfnamefont {Xiaoyong}\ \bibnamefont {Yuan}}, \bibinfo {author}
  {\bibfnamefont {Bo}~\bibnamefont {Yuan}}, \ and\ \bibinfo {author}
  {\bibfnamefont {Zhe}\ \bibnamefont {Yu}},\ }\bibfield  {title} {\enquote
  {\bibinfo {title} {Application of spectral domain optical coherence
  tomography to objectively evaluate posterior capsular opacity in vivo},}\
  }\href {\doibase 10.1155/2018/5461784} {\bibfield  {journal} {\bibinfo
  {journal} {Journal of ophthalmology}\ }\textbf {\bibinfo {volume} {2018}}
  (\bibinfo {year} {2018}),\ 10.1155/2018/5461784}\BibitemShut {NoStop}%
\bibitem [{\citenamefont {Adhi}\ and\ \citenamefont
  {Duker}(2013)}]{adhi2013optical}%
  \BibitemOpen
  \bibfield  {author} {\bibinfo {author} {\bibfnamefont {Mehreen}\ \bibnamefont
  {Adhi}}\ and\ \bibinfo {author} {\bibfnamefont {Jay~S}\ \bibnamefont
  {Duker}},\ }\bibfield  {title} {\enquote {\bibinfo {title} {Optical coherence
  tomography--current and future applications},}\ }\href {\doibase
  10.1097/ICU.0b013e32835f8bf8} {\bibfield  {journal} {\bibinfo  {journal}
  {Current opinion in ophthalmology}\ }\textbf {\bibinfo {volume} {24}},\
  \bibinfo {pages} {213} (\bibinfo {year} {2013})}\BibitemShut {NoStop}%
\bibitem [{\citenamefont {Arute}\ \emph {et~al.}(2019)\citenamefont {Arute},
  \citenamefont {Arya}, \citenamefont {Babbush}, \citenamefont {Bacon},
  \citenamefont {Bardin}, \citenamefont {Barends}, \citenamefont {Biswas},
  \citenamefont {Boixo}, \citenamefont {Brandao}, \citenamefont {Buell} \emph
  {et~al.}}]{2019Quantum}%
  \BibitemOpen
  \bibfield  {author} {\bibinfo {author} {\bibfnamefont {Frank}\ \bibnamefont
  {Arute}}, \bibinfo {author} {\bibfnamefont {Kunal}\ \bibnamefont {Arya}},
  \bibinfo {author} {\bibfnamefont {Ryan}\ \bibnamefont {Babbush}}, \bibinfo
  {author} {\bibfnamefont {Dave}\ \bibnamefont {Bacon}}, \bibinfo {author}
  {\bibfnamefont {Joseph~C}\ \bibnamefont {Bardin}}, \bibinfo {author}
  {\bibfnamefont {Rami}\ \bibnamefont {Barends}}, \bibinfo {author}
  {\bibfnamefont {Rupak}\ \bibnamefont {Biswas}}, \bibinfo {author}
  {\bibfnamefont {Sergio}\ \bibnamefont {Boixo}}, \bibinfo {author}
  {\bibfnamefont {Fernando~GSL}\ \bibnamefont {Brandao}}, \bibinfo {author}
  {\bibfnamefont {David~A}\ \bibnamefont {Buell}},  \emph {et~al.},\ }\bibfield
   {title} {\enquote {\bibinfo {title} {Quantum supremacy using a programmable
  superconducting processor},}\ }\href {\doibase 10.1038/s41586-019-1666-5}
  {\bibfield  {journal} {\bibinfo  {journal} {Nature}\ }\textbf {\bibinfo
  {volume} {574}},\ \bibinfo {pages} {505--510} (\bibinfo {year}
  {2019})}\BibitemShut {NoStop}%
\bibitem [{\citenamefont {Lo}\ and\ \citenamefont
  {Chau}(1999)}]{lo1999Unconditional}%
  \BibitemOpen
  \bibfield  {author} {\bibinfo {author} {\bibfnamefont {Hoi-Kwong}\
  \bibnamefont {Lo}}\ and\ \bibinfo {author} {\bibfnamefont {Hoi~Fung}\
  \bibnamefont {Chau}},\ }\bibfield  {title} {\enquote {\bibinfo {title}
  {Unconditional security of quantum key distribution over arbitrarily long
  distances},}\ }\href {\doibase 10.1126/science.283.5410.2050} {\bibfield
  {journal} {\bibinfo  {journal} {Science}\ }\textbf {\bibinfo {volume}
  {283}},\ \bibinfo {pages} {2050--2056} (\bibinfo {year} {1999})}\BibitemShut
  {NoStop}%
\bibitem [{\citenamefont {Untern\"{a}hrer}\ \emph {et~al.}(2018)\citenamefont
  {Untern\"{a}hrer}, \citenamefont {Bessire}, \citenamefont {Gasparini},
  \citenamefont {Perenzoni},\ and\ \citenamefont {Stefanov}}]{manuel2018super}%
  \BibitemOpen
  \bibfield  {author} {\bibinfo {author} {\bibfnamefont {Manuel}\ \bibnamefont
  {Untern\"{a}hrer}}, \bibinfo {author} {\bibfnamefont {B\"{a}nz}\ \bibnamefont
  {Bessire}}, \bibinfo {author} {\bibfnamefont {Leonardo}\ \bibnamefont
  {Gasparini}}, \bibinfo {author} {\bibfnamefont {Matteo}\ \bibnamefont
  {Perenzoni}}, \ and\ \bibinfo {author} {\bibfnamefont {Andr\'{e}}\
  \bibnamefont {Stefanov}},\ }\bibfield  {title} {\enquote {\bibinfo {title}
  {Super-resolution quantum imaging at the heisenberg limit},}\ }\href
  {\doibase 10.1364/OPTICA.5.001150} {\bibfield  {journal} {\bibinfo  {journal}
  {Optica}\ }\textbf {\bibinfo {volume} {5}},\ \bibinfo {pages} {1150--1154}
  (\bibinfo {year} {2018})}\BibitemShut {NoStop}%
\bibitem [{\citenamefont {Jin}\ and\ \citenamefont
  {Shimizu}(2018)}]{jin2018extended}%
  \BibitemOpen
  \bibfield  {author} {\bibinfo {author} {\bibfnamefont {Rui-Bo}\ \bibnamefont
  {Jin}}\ and\ \bibinfo {author} {\bibfnamefont {Ryosuke}\ \bibnamefont
  {Shimizu}},\ }\bibfield  {title} {\enquote {\bibinfo {title} {Extended
  wiener--khinchin theorem for quantum spectral analysis},}\ }\href {\doibase
  10.1364/OPTICA.5.000093} {\bibfield  {journal} {\bibinfo  {journal} {Optica}\
  }\textbf {\bibinfo {volume} {5}},\ \bibinfo {pages} {93--98} (\bibinfo {year}
  {2018})}\BibitemShut {NoStop}%
\bibitem [{\citenamefont {Hong}\ \emph {et~al.}(1987)\citenamefont {Hong},
  \citenamefont {Ou},\ and\ \citenamefont {Mandel}}]{hong1987measurement}%
  \BibitemOpen
  \bibfield  {author} {\bibinfo {author} {\bibfnamefont {C.~K.}\ \bibnamefont
  {Hong}}, \bibinfo {author} {\bibfnamefont {Z.~Y.}\ \bibnamefont {Ou}}, \ and\
  \bibinfo {author} {\bibfnamefont {L.}~\bibnamefont {Mandel}},\ }\bibfield
  {title} {\enquote {\bibinfo {title} {Measurement of subpicosecond time
  intervals between two photons by interference},}\ }\href {\doibase
  10.1103/PhysRevLett.59.2044} {\bibfield  {journal} {\bibinfo  {journal}
  {Phys. Rev. Lett.}\ }\textbf {\bibinfo {volume} {59}},\ \bibinfo {pages}
  {2044--2046} (\bibinfo {year} {1987})}\BibitemShut {NoStop}%
\bibitem [{\citenamefont {Nasr}\ \emph {et~al.}(2003)\citenamefont {Nasr},
  \citenamefont {Saleh}, \citenamefont {Sergienko},\ and\ \citenamefont
  {Teich}}]{nasr2003demonstration}%
  \BibitemOpen
  \bibfield  {author} {\bibinfo {author} {\bibfnamefont {Magued~B.}\
  \bibnamefont {Nasr}}, \bibinfo {author} {\bibfnamefont {Bahaa E.~A.}\
  \bibnamefont {Saleh}}, \bibinfo {author} {\bibfnamefont {Alexander~V.}\
  \bibnamefont {Sergienko}}, \ and\ \bibinfo {author} {\bibfnamefont
  {Malvin~C.}\ \bibnamefont {Teich}},\ }\bibfield  {title} {\enquote {\bibinfo
  {title} {Demonstration of dispersion-canceled quantum-optical coherence
  tomography},}\ }\href {\doibase 10.1103/PhysRevLett.91.083601} {\bibfield
  {journal} {\bibinfo  {journal} {Phys. Rev. Lett.}\ }\textbf {\bibinfo
  {volume} {91}},\ \bibinfo {pages} {083601} (\bibinfo {year}
  {2003})}\BibitemShut {NoStop}%
\bibitem [{\citenamefont {Yepiz-Graciano}\ \emph {et~al.}(2020)\citenamefont
  {Yepiz-Graciano}, \citenamefont {Mart\'{i}nez}, \citenamefont {Lopez-Mago},
  \citenamefont {Cruz-Ramirez},\ and\ \citenamefont
  {U'Ren}}]{pablo2020spectrally}%
  \BibitemOpen
  \bibfield  {author} {\bibinfo {author} {\bibfnamefont {Pablo}\ \bibnamefont
  {Yepiz-Graciano}}, \bibinfo {author} {\bibfnamefont {Al\'{i} Michel~Angulo}\
  \bibnamefont {Mart\'{i}nez}}, \bibinfo {author} {\bibfnamefont {Dorilian}\
  \bibnamefont {Lopez-Mago}}, \bibinfo {author} {\bibfnamefont {Hector}\
  \bibnamefont {Cruz-Ramirez}}, \ and\ \bibinfo {author} {\bibfnamefont
  {Alfred~B.}\ \bibnamefont {U'Ren}},\ }\bibfield  {title} {\enquote {\bibinfo
  {title} {Spectrally resolved hong-ou-mandel interferometry for
  quantum-optical coherence tomography},}\ }\href {\doibase 10.1364/PRJ.388693}
  {\bibfield  {journal} {\bibinfo  {journal} {Photon. Res.}\ }\textbf {\bibinfo
  {volume} {8}},\ \bibinfo {pages} {1023--1034} (\bibinfo {year}
  {2020})}\BibitemShut {NoStop}%
\bibitem [{\citenamefont {Lyons}\ \emph {et~al.}(2018)\citenamefont {Lyons},
  \citenamefont {Knee}, \citenamefont {Bolduc}, \citenamefont {Roger},
  \citenamefont {Leach}, \citenamefont {Gauger},\ and\ \citenamefont
  {Faccio}}]{lyons2018attosecond}%
  \BibitemOpen
  \bibfield  {author} {\bibinfo {author} {\bibfnamefont {Ashley}\ \bibnamefont
  {Lyons}}, \bibinfo {author} {\bibfnamefont {George~C}\ \bibnamefont {Knee}},
  \bibinfo {author} {\bibfnamefont {Eliot}\ \bibnamefont {Bolduc}}, \bibinfo
  {author} {\bibfnamefont {Thomas}\ \bibnamefont {Roger}}, \bibinfo {author}
  {\bibfnamefont {Jonathan}\ \bibnamefont {Leach}}, \bibinfo {author}
  {\bibfnamefont {Erik~M}\ \bibnamefont {Gauger}}, \ and\ \bibinfo {author}
  {\bibfnamefont {Daniele}\ \bibnamefont {Faccio}},\ }\bibfield  {title}
  {\enquote {\bibinfo {title} {Attosecond-resolution hong-ou-mandel
  interferometry},}\ }\href {\doibase 10.1126/sciadv.aap9416} {\bibfield
  {journal} {\bibinfo  {journal} {Sci. Adv.}\ }\textbf {\bibinfo {volume}
  {4}},\ \bibinfo {pages} {eaap9416} (\bibinfo {year} {2018})}\BibitemShut
  {NoStop}%
\bibitem [{\citenamefont {Pan}\ \emph {et~al.}(2012)\citenamefont {Pan},
  \citenamefont {Chen}, \citenamefont {Lu}, \citenamefont {Weinfurter},
  \citenamefont {Zeilinger},\ and\ \citenamefont {\ifmmode~\dot{Z}\else
  \.{Z}\fi{}ukowski}}]{pan2012multiphoton}%
  \BibitemOpen
  \bibfield  {author} {\bibinfo {author} {\bibfnamefont {Jian-Wei}\
  \bibnamefont {Pan}}, \bibinfo {author} {\bibfnamefont {Zeng-Bing}\
  \bibnamefont {Chen}}, \bibinfo {author} {\bibfnamefont {Chao-Yang}\
  \bibnamefont {Lu}}, \bibinfo {author} {\bibfnamefont {Harald}\ \bibnamefont
  {Weinfurter}}, \bibinfo {author} {\bibfnamefont {Anton}\ \bibnamefont
  {Zeilinger}}, \ and\ \bibinfo {author} {\bibfnamefont {Marek}\ \bibnamefont
  {\ifmmode~\dot{Z}\else \.{Z}\fi{}ukowski}},\ }\bibfield  {title} {\enquote
  {\bibinfo {title} {Multiphoton entanglement and interferometry},}\ }\href
  {\doibase 10.1103/RevModPhys.84.777} {\bibfield  {journal} {\bibinfo
  {journal} {Rev. Mod. Phys.}\ }\textbf {\bibinfo {volume} {84}},\ \bibinfo
  {pages} {777--838} (\bibinfo {year} {2012})}\BibitemShut {NoStop}%
\bibitem [{\citenamefont {Xie}\ \emph {et~al.}(2015)\citenamefont {Xie},
  \citenamefont {Tian}, \citenamefont {Shrestha}, \citenamefont {Xu},
  \citenamefont {Liang}, \citenamefont {Gong}, \citenamefont {Bienfang},
  \citenamefont {Restelli}, \citenamefont {Shapiro},\ and\ \citenamefont
  {Wong}}]{xie2015Harnessing}%
  \BibitemOpen
  \bibfield  {author} {\bibinfo {author} {\bibfnamefont {Z.}~\bibnamefont
  {Xie}}, \bibinfo {author} {\bibfnamefont {Z.}~\bibnamefont {Tian}}, \bibinfo
  {author} {\bibfnamefont {S.}~\bibnamefont {Shrestha}}, \bibinfo {author}
  {\bibfnamefont {X.~A.}\ \bibnamefont {Xu}}, \bibinfo {author} {\bibfnamefont
  {J.}~\bibnamefont {Liang}}, \bibinfo {author} {\bibfnamefont {Y.~X.}\
  \bibnamefont {Gong}}, \bibinfo {author} {\bibfnamefont {J.~C.}\ \bibnamefont
  {Bienfang}}, \bibinfo {author} {\bibfnamefont {A.}~\bibnamefont {Restelli}},
  \bibinfo {author} {\bibfnamefont {J.~H.}\ \bibnamefont {Shapiro}}, \ and\
  \bibinfo {author} {\bibfnamefont {Fnc}\ \bibnamefont {Wong}},\ }\bibfield
  {title} {\enquote {\bibinfo {title} {Harnessing high-dimensional
  hyperentanglement through a biphoton frequency comb},}\ }\href {\doibase
  10.1038/nphoton.2015.110} {\bibfield  {journal} {\bibinfo  {journal} {Nat.
  Photonics}\ }\textbf {\bibinfo {volume} {9}},\ \bibinfo {pages} {536}
  (\bibinfo {year} {2015})}\BibitemShut {NoStop}%
\bibitem [{\citenamefont {Chen}\ \emph
  {et~al.}(2021{\natexlab{a}})\citenamefont {Chen}, \citenamefont {Ecker},
  \citenamefont {Chen}, \citenamefont {Steinlechner}, \citenamefont {Huber},\
  and\ \citenamefont {Ursin}}]{chen2021temporal}%
  \BibitemOpen
  \bibfield  {author} {\bibinfo {author} {\bibfnamefont {Yuanyuan}\
  \bibnamefont {Chen}}, \bibinfo {author} {\bibfnamefont {Sebastian}\
  \bibnamefont {Ecker}}, \bibinfo {author} {\bibfnamefont {Lixiang}\
  \bibnamefont {Chen}}, \bibinfo {author} {\bibfnamefont {Fabian}\ \bibnamefont
  {Steinlechner}}, \bibinfo {author} {\bibfnamefont {Marcus}\ \bibnamefont
  {Huber}}, \ and\ \bibinfo {author} {\bibfnamefont {Rupert}\ \bibnamefont
  {Ursin}},\ }\bibfield  {title} {\enquote {\bibinfo {title} {Temporal
  distinguishability in hong-ou-mandel interference for harnessing
  high-dimensional frequency entanglement},}\ }\href {\doibase
  10.1038/s41534-021-00504-0} {\bibfield  {journal} {\bibinfo  {journal} {npj
  Quantum Inf.}\ }\textbf {\bibinfo {volume} {7}},\ \bibinfo {pages} {1--7}
  (\bibinfo {year} {2021}{\natexlab{a}})}\BibitemShut {NoStop}%
\bibitem [{\citenamefont {Thompson}\ \emph {et~al.}(2017)\citenamefont
  {Thompson}, \citenamefont {Moran},\ and\ \citenamefont
  {Swenson}}]{thompson2017analysis}%
  \BibitemOpen
  \bibfield  {author} {\bibinfo {author} {\bibfnamefont {A.~R.}\ \bibnamefont
  {Thompson}}, \bibinfo {author} {\bibfnamefont {J.~M.}\ \bibnamefont {Moran}},
  \ and\ \bibinfo {author} {\bibfnamefont {George~W.}\ \bibnamefont
  {Swenson}},\ }\enquote {\bibinfo {title} {Analysis of the interferometer
  response},}\ in\ \href {\doibase 10.1007/978-3-319-44431-4_3} {\emph
  {\bibinfo {booktitle} {Interferometry and Synthesis in Radio Astronomy}}}\
  (\bibinfo  {publisher} {Springer International Publishing},\ \bibinfo {year}
  {2017})\ pp.\ \bibinfo {pages} {89--108}\BibitemShut {NoStop}%
\bibitem [{\citenamefont {Devaux}\ \emph {et~al.}(2020)\citenamefont {Devaux},
  \citenamefont {Mosset}, \citenamefont {Moreau},\ and\ \citenamefont
  {Lantz}}]{devaux2020imaging}%
  \BibitemOpen
  \bibfield  {author} {\bibinfo {author} {\bibfnamefont {Fabrice}\ \bibnamefont
  {Devaux}}, \bibinfo {author} {\bibfnamefont {Alexis}\ \bibnamefont {Mosset}},
  \bibinfo {author} {\bibfnamefont {Paul-Antoine}\ \bibnamefont {Moreau}}, \
  and\ \bibinfo {author} {\bibfnamefont {Eric}\ \bibnamefont {Lantz}},\
  }\bibfield  {title} {\enquote {\bibinfo {title} {Imaging spatiotemporal
  hong-ou-mandel interference of biphoton states of extremely high schmidt
  number},}\ }\href {\doibase 10.1103/PhysRevX.10.031031} {\bibfield  {journal}
  {\bibinfo  {journal} {Phys. Rev. X}\ }\textbf {\bibinfo {volume} {10}},\
  \bibinfo {pages} {031031} (\bibinfo {year} {2020})}\BibitemShut {NoStop}%
\bibitem [{\citenamefont {Kolenderska}\ \emph {et~al.}(2020)\citenamefont
  {Kolenderska}, \citenamefont {Vanholsbeeck},\ and\ \citenamefont
  {Kolenderski}}]{sylwia2020fourier}%
  \BibitemOpen
  \bibfield  {author} {\bibinfo {author} {\bibfnamefont {Sylwia~M.}\
  \bibnamefont {Kolenderska}}, \bibinfo {author} {\bibfnamefont
  {Fr\'{e}d\'{e}rique}\ \bibnamefont {Vanholsbeeck}}, \ and\ \bibinfo {author}
  {\bibfnamefont {Piotr}\ \bibnamefont {Kolenderski}},\ }\bibfield  {title}
  {\enquote {\bibinfo {title} {Fourier domain quantum optical coherence
  tomography},}\ }\href {\doibase 10.1364/OE.399913} {\bibfield  {journal}
  {\bibinfo  {journal} {Opt. Express}\ }\textbf {\bibinfo {volume} {28}},\
  \bibinfo {pages} {29576--29589} (\bibinfo {year} {2020})}\BibitemShut
  {NoStop}%
\bibitem [{\citenamefont {Chen}\ \emph {et~al.}(2022)\citenamefont {Chen},
  \citenamefont {Shen}, \citenamefont {Luo}, \citenamefont {Zhang},
  \citenamefont {Chen},\ and\ \citenamefont {Chen}}]{chen2022entanglement}%
  \BibitemOpen
  \bibfield  {author} {\bibinfo {author} {\bibfnamefont {Yuanyuan}\
  \bibnamefont {Chen}}, \bibinfo {author} {\bibfnamefont {Qian}\ \bibnamefont
  {Shen}}, \bibinfo {author} {\bibfnamefont {Song}\ \bibnamefont {Luo}},
  \bibinfo {author} {\bibfnamefont {Long}\ \bibnamefont {Zhang}}, \bibinfo
  {author} {\bibfnamefont {Zhanghai}\ \bibnamefont {Chen}}, \ and\ \bibinfo
  {author} {\bibfnamefont {Lixiang}\ \bibnamefont {Chen}},\ }\bibfield  {title}
  {\enquote {\bibinfo {title} {Entanglement-assisted absorption spectroscopy by
  hong-ou-mandel interference},}\ }\href {\doibase
  10.1103/PhysRevApplied.17.014010} {\bibfield  {journal} {\bibinfo  {journal}
  {Phys. Rev. Appl.}\ }\textbf {\bibinfo {volume} {17}},\ \bibinfo {pages}
  {014010} (\bibinfo {year} {2022})}\BibitemShut {NoStop}%
\bibitem [{\citenamefont {Chen}\ \emph {et~al.}(2019)\citenamefont {Chen},
  \citenamefont {Fink}, \citenamefont {Steinlechner}, \citenamefont {Torres},\
  and\ \citenamefont {Ursin}}]{chen2019Hong}%
  \BibitemOpen
  \bibfield  {author} {\bibinfo {author} {\bibfnamefont {Yuanyuan}\
  \bibnamefont {Chen}}, \bibinfo {author} {\bibfnamefont {Matthias}\
  \bibnamefont {Fink}}, \bibinfo {author} {\bibfnamefont {Fabian}\ \bibnamefont
  {Steinlechner}}, \bibinfo {author} {\bibfnamefont {Juan~P}\ \bibnamefont
  {Torres}}, \ and\ \bibinfo {author} {\bibfnamefont {Rupert}\ \bibnamefont
  {Ursin}},\ }\bibfield  {title} {\enquote {\bibinfo {title} {Hong-ou-mandel
  interferometry on a biphoton beat note},}\ }\href {\doibase
  10.1038/s41534-019-0161-z} {\bibfield  {journal} {\bibinfo  {journal} {npj
  Quantum Inf.}\ }\textbf {\bibinfo {volume} {5}},\ \bibinfo {pages} {1--6}
  (\bibinfo {year} {2019})}\BibitemShut {NoStop}%
\bibitem [{\citenamefont {Giovannini}\ \emph {et~al.}(2015)\citenamefont
  {Giovannini}, \citenamefont {Romero}, \citenamefont {Poto{\v{c}}ek},
  \citenamefont {Ferenczi}, \citenamefont {Speirits}, \citenamefont {Barnett},
  \citenamefont {Faccio},\ and\ \citenamefont
  {Padgett}}]{giovannini2015spatially}%
  \BibitemOpen
  \bibfield  {author} {\bibinfo {author} {\bibfnamefont {Daniel}\ \bibnamefont
  {Giovannini}}, \bibinfo {author} {\bibfnamefont {Jacquiline}\ \bibnamefont
  {Romero}}, \bibinfo {author} {\bibfnamefont {V{\'a}clav}\ \bibnamefont
  {Poto{\v{c}}ek}}, \bibinfo {author} {\bibfnamefont {Gergely}\ \bibnamefont
  {Ferenczi}}, \bibinfo {author} {\bibfnamefont {Fiona}\ \bibnamefont
  {Speirits}}, \bibinfo {author} {\bibfnamefont {Stephen~M}\ \bibnamefont
  {Barnett}}, \bibinfo {author} {\bibfnamefont {Daniele}\ \bibnamefont
  {Faccio}}, \ and\ \bibinfo {author} {\bibfnamefont {Miles~J}\ \bibnamefont
  {Padgett}},\ }\bibfield  {title} {\enquote {\bibinfo {title} {Spatially
  structured photons that travel in free space slower than the speed of
  light},}\ }\href {\doibase 10.1126/science.aaa3035} {\bibfield  {journal}
  {\bibinfo  {journal} {Science}\ }\textbf {\bibinfo {volume} {347}},\ \bibinfo
  {pages} {857--860} (\bibinfo {year} {2015})}\BibitemShut {NoStop}%
\bibitem [{\citenamefont {Chen}\ \emph {et~al.}(2018)\citenamefont {Chen},
  \citenamefont {Ecker}, \citenamefont {Wengerowsky}, \citenamefont {Bulla},
  \citenamefont {Joshi}, \citenamefont {Steinlechner},\ and\ \citenamefont
  {Ursin}}]{chen2018polarization}%
  \BibitemOpen
  \bibfield  {author} {\bibinfo {author} {\bibfnamefont {Yuanyuan}\
  \bibnamefont {Chen}}, \bibinfo {author} {\bibfnamefont {Sebastian}\
  \bibnamefont {Ecker}}, \bibinfo {author} {\bibfnamefont {S\"oren}\
  \bibnamefont {Wengerowsky}}, \bibinfo {author} {\bibfnamefont {Lukas}\
  \bibnamefont {Bulla}}, \bibinfo {author} {\bibfnamefont {Siddarth~Koduru}\
  \bibnamefont {Joshi}}, \bibinfo {author} {\bibfnamefont {Fabian}\
  \bibnamefont {Steinlechner}}, \ and\ \bibinfo {author} {\bibfnamefont
  {Rupert}\ \bibnamefont {Ursin}},\ }\bibfield  {title} {\enquote {\bibinfo
  {title} {Polarization entanglement by time-reversed hong-ou-mandel
  interference},}\ }\href {\doibase 10.1103/PhysRevLett.121.200502} {\bibfield
  {journal} {\bibinfo  {journal} {Phys. Rev. Lett.}\ }\textbf {\bibinfo
  {volume} {121}},\ \bibinfo {pages} {200502} (\bibinfo {year}
  {2018})}\BibitemShut {NoStop}%
\bibitem [{\citenamefont {Chen}\ \emph {et~al.}(2020)\citenamefont {Chen},
  \citenamefont {Ecker}, \citenamefont {Bavaresco}, \citenamefont {Scheidl},
  \citenamefont {Chen}, \citenamefont {Steinlechner}, \citenamefont {Huber},\
  and\ \citenamefont {Ursin}}]{chen2020verification}%
  \BibitemOpen
  \bibfield  {author} {\bibinfo {author} {\bibfnamefont {Yuanyuan}\
  \bibnamefont {Chen}}, \bibinfo {author} {\bibfnamefont {Sebastian}\
  \bibnamefont {Ecker}}, \bibinfo {author} {\bibfnamefont {Jessica}\
  \bibnamefont {Bavaresco}}, \bibinfo {author} {\bibfnamefont {Thomas}\
  \bibnamefont {Scheidl}}, \bibinfo {author} {\bibfnamefont {Lixiang}\
  \bibnamefont {Chen}}, \bibinfo {author} {\bibfnamefont {Fabian}\ \bibnamefont
  {Steinlechner}}, \bibinfo {author} {\bibfnamefont {Marcus}\ \bibnamefont
  {Huber}}, \ and\ \bibinfo {author} {\bibfnamefont {Rupert}\ \bibnamefont
  {Ursin}},\ }\bibfield  {title} {\enquote {\bibinfo {title} {Verification of
  high-dimensional entanglement generated in quantum interference},}\ }\href
  {\doibase 10.1103/PhysRevA.101.032302} {\bibfield  {journal} {\bibinfo
  {journal} {Phys. Rev. A}\ }\textbf {\bibinfo {volume} {101}},\ \bibinfo
  {pages} {032302} (\bibinfo {year} {2020})}\BibitemShut {NoStop}%
\bibitem [{\citenamefont {Huang}\ \emph {et~al.}(1991)\citenamefont {Huang},
  \citenamefont {Swanson}, \citenamefont {Lin}, \citenamefont {Schuman},
  \citenamefont {Stinson}, \citenamefont {Chang}, \citenamefont {Hee},
  \citenamefont {Flotte}, \citenamefont {Gregory}, \citenamefont {Puliafito}
  \emph {et~al.}}]{huang1991optical}%
  \BibitemOpen
  \bibfield  {author} {\bibinfo {author} {\bibfnamefont {David}\ \bibnamefont
  {Huang}}, \bibinfo {author} {\bibfnamefont {Eric~A}\ \bibnamefont {Swanson}},
  \bibinfo {author} {\bibfnamefont {Charles~P}\ \bibnamefont {Lin}}, \bibinfo
  {author} {\bibfnamefont {Joel~S}\ \bibnamefont {Schuman}}, \bibinfo {author}
  {\bibfnamefont {William~G}\ \bibnamefont {Stinson}}, \bibinfo {author}
  {\bibfnamefont {Warren}\ \bibnamefont {Chang}}, \bibinfo {author}
  {\bibfnamefont {Michael~R}\ \bibnamefont {Hee}}, \bibinfo {author}
  {\bibfnamefont {Thomas}\ \bibnamefont {Flotte}}, \bibinfo {author}
  {\bibfnamefont {Kenton}\ \bibnamefont {Gregory}}, \bibinfo {author}
  {\bibfnamefont {Carmen~A}\ \bibnamefont {Puliafito}},  \emph {et~al.},\
  }\bibfield  {title} {\enquote {\bibinfo {title} {Optical coherence
  tomography},}\ }\href {\doibase 10.1126/science.1957169} {\bibfield
  {journal} {\bibinfo  {journal} {Science}\ }\textbf {\bibinfo {volume}
  {254}},\ \bibinfo {pages} {1178--1181} (\bibinfo {year} {1991})}\BibitemShut
  {NoStop}%
\bibitem [{\citenamefont {Podoleanu}\ \emph {et~al.}(2000)\citenamefont
  {Podoleanu}, \citenamefont {Rogers}, \citenamefont {Jackson},\ and\
  \citenamefont {Dunne}}]{adrian2000three}%
  \BibitemOpen
  \bibfield  {author} {\bibinfo {author} {\bibfnamefont {Adrian~Gh.}\
  \bibnamefont {Podoleanu}}, \bibinfo {author} {\bibfnamefont {John~A.}\
  \bibnamefont {Rogers}}, \bibinfo {author} {\bibfnamefont {David~A.}\
  \bibnamefont {Jackson}}, \ and\ \bibinfo {author} {\bibfnamefont {Shane}\
  \bibnamefont {Dunne}},\ }\bibfield  {title} {\enquote {\bibinfo {title}
  {Three dimensional oct images from retina and skin},}\ }\href {\doibase
  10.1364/OE.7.000292} {\bibfield  {journal} {\bibinfo  {journal} {Opt.
  Express}\ }\textbf {\bibinfo {volume} {7}},\ \bibinfo {pages} {292--298}
  (\bibinfo {year} {2000})}\BibitemShut {NoStop}%
\bibitem [{\citenamefont {Leitgeb}\ \emph {et~al.}(2003)\citenamefont
  {Leitgeb}, \citenamefont {Hitzenberger},\ and\ \citenamefont
  {Fercher}}]{leitgeb2003performance}%
  \BibitemOpen
  \bibfield  {author} {\bibinfo {author} {\bibfnamefont {R}~\bibnamefont
  {Leitgeb}}, \bibinfo {author} {\bibfnamefont {CK}~\bibnamefont
  {Hitzenberger}}, \ and\ \bibinfo {author} {\bibfnamefont {Adolf~F}\
  \bibnamefont {Fercher}},\ }\bibfield  {title} {\enquote {\bibinfo {title}
  {Performance of fourier domain vs. time domain optical coherence
  tomography},}\ }\href {\doibase 10.1364/OE.11.000889} {\bibfield  {journal}
  {\bibinfo  {journal} {Opt. express}\ }\textbf {\bibinfo {volume} {11}},\
  \bibinfo {pages} {889--894} (\bibinfo {year} {2003})}\BibitemShut {NoStop}%
\bibitem [{\citenamefont {Choma}\ \emph {et~al.}(2003)\citenamefont {Choma},
  \citenamefont {Sarunic}, \citenamefont {Yang},\ and\ \citenamefont
  {Izatt}}]{choma2003sensitivity}%
  \BibitemOpen
  \bibfield  {author} {\bibinfo {author} {\bibfnamefont {Michael~A}\
  \bibnamefont {Choma}}, \bibinfo {author} {\bibfnamefont {Marinko~V}\
  \bibnamefont {Sarunic}}, \bibinfo {author} {\bibfnamefont {Changhuei}\
  \bibnamefont {Yang}}, \ and\ \bibinfo {author} {\bibfnamefont {Joseph~A}\
  \bibnamefont {Izatt}},\ }\bibfield  {title} {\enquote {\bibinfo {title}
  {Sensitivity advantage of swept source and fourier domain optical coherence
  tomography},}\ }\href {\doibase 10.1364/OE.11.002183} {\bibfield  {journal}
  {\bibinfo  {journal} {Opt. express}\ }\textbf {\bibinfo {volume} {11}},\
  \bibinfo {pages} {2183--2189} (\bibinfo {year} {2003})}\BibitemShut {NoStop}%
\bibitem [{\citenamefont {Siddiqui}\ \emph {et~al.}(2018)\citenamefont
  {Siddiqui}, \citenamefont {Nam}, \citenamefont {Tozburun}, \citenamefont
  {Lippok}, \citenamefont {Blatter},\ and\ \citenamefont
  {Vakoc}}]{siddiqui2018high}%
  \BibitemOpen
  \bibfield  {author} {\bibinfo {author} {\bibfnamefont {Meena}\ \bibnamefont
  {Siddiqui}}, \bibinfo {author} {\bibfnamefont {Ahhyun~S}\ \bibnamefont
  {Nam}}, \bibinfo {author} {\bibfnamefont {Serhat}\ \bibnamefont {Tozburun}},
  \bibinfo {author} {\bibfnamefont {Norman}\ \bibnamefont {Lippok}}, \bibinfo
  {author} {\bibfnamefont {Cedric}\ \bibnamefont {Blatter}}, \ and\ \bibinfo
  {author} {\bibfnamefont {Benjamin~J}\ \bibnamefont {Vakoc}},\ }\bibfield
  {title} {\enquote {\bibinfo {title} {High-speed optical coherence tomography
  by circular interferometric ranging},}\ }\href {\doibase
  10.1038/s41566-017-0088-x} {\bibfield  {journal} {\bibinfo  {journal} {Nat.
  photonics}\ }\textbf {\bibinfo {volume} {12}},\ \bibinfo {pages} {111--116}
  (\bibinfo {year} {2018})}\BibitemShut {NoStop}%
\bibitem [{\citenamefont {Marchand}\ \emph {et~al.}(2021)\citenamefont
  {Marchand}, \citenamefont {Riemensberger}, \citenamefont {Skehan},
  \citenamefont {Ho}, \citenamefont {Pfeiffer}, \citenamefont {Liu},
  \citenamefont {Hauger}, \citenamefont {Lasser},\ and\ \citenamefont
  {Kippenberg}}]{marchand2021soliton}%
  \BibitemOpen
  \bibfield  {author} {\bibinfo {author} {\bibfnamefont {Paul~J}\ \bibnamefont
  {Marchand}}, \bibinfo {author} {\bibfnamefont {Johann}\ \bibnamefont
  {Riemensberger}}, \bibinfo {author} {\bibfnamefont {J~Connor}\ \bibnamefont
  {Skehan}}, \bibinfo {author} {\bibfnamefont {Jia-Jung}\ \bibnamefont {Ho}},
  \bibinfo {author} {\bibfnamefont {Martin~HP}\ \bibnamefont {Pfeiffer}},
  \bibinfo {author} {\bibfnamefont {Junqiu}\ \bibnamefont {Liu}}, \bibinfo
  {author} {\bibfnamefont {Christoph}\ \bibnamefont {Hauger}}, \bibinfo
  {author} {\bibfnamefont {Theo}\ \bibnamefont {Lasser}}, \ and\ \bibinfo
  {author} {\bibfnamefont {Tobias~J}\ \bibnamefont {Kippenberg}},\ }\bibfield
  {title} {\enquote {\bibinfo {title} {Soliton microcomb based spectral domain
  optical coherence tomography},}\ }\href {\doibase 10.1038/s41467-020-20404-9}
  {\bibfield  {journal} {\bibinfo  {journal} {Nat. Commun.}\ }\textbf {\bibinfo
  {volume} {12}},\ \bibinfo {pages} {1--9} (\bibinfo {year}
  {2021})}\BibitemShut {NoStop}%
\bibitem [{\citenamefont {Franke-Arnold}\ \emph {et~al.}(2004)\citenamefont
  {Franke-Arnold}, \citenamefont {Barnett}, \citenamefont {Yao}, \citenamefont
  {Leach}, \citenamefont {Courtial},\ and\ \citenamefont
  {Padgett}}]{Franke2004uncertainty}%
  \BibitemOpen
  \bibfield  {author} {\bibinfo {author} {\bibfnamefont {Sonja}\ \bibnamefont
  {Franke-Arnold}}, \bibinfo {author} {\bibfnamefont {Stephen~M}\ \bibnamefont
  {Barnett}}, \bibinfo {author} {\bibfnamefont {Eric}\ \bibnamefont {Yao}},
  \bibinfo {author} {\bibfnamefont {Jonathan}\ \bibnamefont {Leach}}, \bibinfo
  {author} {\bibfnamefont {Johannes}\ \bibnamefont {Courtial}}, \ and\ \bibinfo
  {author} {\bibfnamefont {Miles}\ \bibnamefont {Padgett}},\ }\bibfield
  {title} {\enquote {\bibinfo {title} {Uncertainty principle for angular
  position and angular momentum},}\ }\href {\doibase 10.1088/1367-2630/6/1/103}
  {\bibfield  {journal} {\bibinfo  {journal} {New J. Phys.}\ }\textbf {\bibinfo
  {volume} {6}},\ \bibinfo {pages} {103--103} (\bibinfo {year}
  {2004})}\BibitemShut {NoStop}%
\bibitem [{\citenamefont {Twamley}\ and\ \citenamefont
  {Milburn}(2006)}]{twamley2006quantum}%
  \BibitemOpen
  \bibfield  {author} {\bibinfo {author} {\bibfnamefont {J}~\bibnamefont
  {Twamley}}\ and\ \bibinfo {author} {\bibfnamefont {G~J}\ \bibnamefont
  {Milburn}},\ }\bibfield  {title} {\enquote {\bibinfo {title} {The quantum
  mellin transform},}\ }\href {\doibase 10.1088/1367-2630/8/12/328} {\bibfield
  {journal} {\bibinfo  {journal} {New J. Phys.}\ }\textbf {\bibinfo {volume}
  {8}},\ \bibinfo {pages} {328--328} (\bibinfo {year} {2006})}\BibitemShut
  {NoStop}%
\bibitem [{\citenamefont {Chen}\ \emph
  {et~al.}(2021{\natexlab{b}})\citenamefont {Chen}, \citenamefont {Ecker},
  \citenamefont {Chen}, \citenamefont {Steinlechner}, \citenamefont {Huber},\
  and\ \citenamefont {Ursin}}]{chen2020Temporal}%
  \BibitemOpen
  \bibfield  {author} {\bibinfo {author} {\bibfnamefont {Yuanyuan}\
  \bibnamefont {Chen}}, \bibinfo {author} {\bibfnamefont {Sebastian}\
  \bibnamefont {Ecker}}, \bibinfo {author} {\bibfnamefont {Lixiang}\
  \bibnamefont {Chen}}, \bibinfo {author} {\bibfnamefont {Fabian}\ \bibnamefont
  {Steinlechner}}, \bibinfo {author} {\bibfnamefont {Marcus}\ \bibnamefont
  {Huber}}, \ and\ \bibinfo {author} {\bibfnamefont {Rupert}\ \bibnamefont
  {Ursin}},\ }\bibfield  {title} {\enquote {\bibinfo {title} {Temporal
  distinguishability in hong-ou-mandel interference for harnessing
  high-dimensional frequency entanglement},}\ }\href {\doibase
  10.1038/s41534-021-00504-0} {\bibfield  {journal} {\bibinfo  {journal} {npj
  Quantum Inf.}\ }\textbf {\bibinfo {volume} {7}},\ \bibinfo {pages} {1--7}
  (\bibinfo {year} {2021}{\natexlab{b}})}\BibitemShut {NoStop}%
\bibitem [{\citenamefont {Jin}\ \emph {et~al.}(2016)\citenamefont {Jin},
  \citenamefont {Shimizu}, \citenamefont {Fujiwara}, \citenamefont {Takeoka},
  \citenamefont {Wakabayashi}, \citenamefont {Yamashita}, \citenamefont {Miki},
  \citenamefont {Terai}, \citenamefont {Gerrits},\ and\ \citenamefont
  {Sasaki}}]{jin2016simple}%
  \BibitemOpen
  \bibfield  {author} {\bibinfo {author} {\bibfnamefont {Rui-Bo}\ \bibnamefont
  {Jin}}, \bibinfo {author} {\bibfnamefont {Ryosuke}\ \bibnamefont {Shimizu}},
  \bibinfo {author} {\bibfnamefont {Mikio}\ \bibnamefont {Fujiwara}}, \bibinfo
  {author} {\bibfnamefont {Masahiro}\ \bibnamefont {Takeoka}}, \bibinfo
  {author} {\bibfnamefont {Ryota}\ \bibnamefont {Wakabayashi}}, \bibinfo
  {author} {\bibfnamefont {Taro}\ \bibnamefont {Yamashita}}, \bibinfo {author}
  {\bibfnamefont {Shigehito}\ \bibnamefont {Miki}}, \bibinfo {author}
  {\bibfnamefont {Hirotaka}\ \bibnamefont {Terai}}, \bibinfo {author}
  {\bibfnamefont {Thomas}\ \bibnamefont {Gerrits}}, \ and\ \bibinfo {author}
  {\bibfnamefont {Masahide}\ \bibnamefont {Sasaki}},\ }\bibfield  {title}
  {\enquote {\bibinfo {title} {Simple method of generating and distributing
  frequency-entangled qudits},}\ }\href {\doibase 10.1088/2058-9565/1/1/015004}
  {\bibfield  {journal} {\bibinfo  {journal} {Quantum Sci. Technol.}\ }\textbf
  {\bibinfo {volume} {1}},\ \bibinfo {pages} {015004} (\bibinfo {year}
  {2016})}\BibitemShut {NoStop}%
\bibitem [{\citenamefont {Shi}\ \emph {et~al.}(2020)\citenamefont {Shi},
  \citenamefont {Zhang}, \citenamefont {Pirandola},\ and\ \citenamefont
  {Zhuang}}]{shi2020entanglement}%
  \BibitemOpen
  \bibfield  {author} {\bibinfo {author} {\bibfnamefont {Haowei}\ \bibnamefont
  {Shi}}, \bibinfo {author} {\bibfnamefont {Zheshen}\ \bibnamefont {Zhang}},
  \bibinfo {author} {\bibfnamefont {Stefano}\ \bibnamefont {Pirandola}}, \ and\
  \bibinfo {author} {\bibfnamefont {Quntao}\ \bibnamefont {Zhuang}},\
  }\bibfield  {title} {\enquote {\bibinfo {title} {Entanglement-assisted
  absorption spectroscopy},}\ }\href {\doibase 10.1103/PhysRevLett.125.180502}
  {\bibfield  {journal} {\bibinfo  {journal} {Phys. Rev. Lett.}\ }\textbf
  {\bibinfo {volume} {125}},\ \bibinfo {pages} {180502} (\bibinfo {year}
  {2020})}\BibitemShut {NoStop}%
\bibitem [{\citenamefont {Vanselow}\ \emph {et~al.}(2020)\citenamefont
  {Vanselow}, \citenamefont {Kaufmann}, \citenamefont {Zorin}, \citenamefont
  {Heise}, \citenamefont {Chrzanowski},\ and\ \citenamefont
  {Ramelow}}]{aron2020frequency}%
  \BibitemOpen
  \bibfield  {author} {\bibinfo {author} {\bibfnamefont {Aron}\ \bibnamefont
  {Vanselow}}, \bibinfo {author} {\bibfnamefont {Paul}\ \bibnamefont
  {Kaufmann}}, \bibinfo {author} {\bibfnamefont {Ivan}\ \bibnamefont {Zorin}},
  \bibinfo {author} {\bibfnamefont {Bettina}\ \bibnamefont {Heise}}, \bibinfo
  {author} {\bibfnamefont {Helen~M.}\ \bibnamefont {Chrzanowski}}, \ and\
  \bibinfo {author} {\bibfnamefont {Sven}\ \bibnamefont {Ramelow}},\ }\bibfield
   {title} {\enquote {\bibinfo {title} {Frequency-domain optical coherence
  tomography with undetected mid-infrared photons},}\ }\href {\doibase
  10.1364/OPTICA.400128} {\bibfield  {journal} {\bibinfo  {journal} {Optica}\
  }\textbf {\bibinfo {volume} {7}},\ \bibinfo {pages} {1729--1736} (\bibinfo
  {year} {2020})}\BibitemShut {NoStop}%
\end{thebibliography}%
\end{document}